%% file: main.tex
\documentclass[11pt]{article}

\input{macro.tex}

\begin{document}

\title{Efficient Quantum Algorithm for Port-based Teleportation}

\def\sphys{%
Stanford Institute for Theoretical Physics \\Stanford, CA
}%


\author{Jiani Fei}
\author{Sydney Timmerman}
\author{Patrick Hayden}
\affil{\sphys}
\date{\today}
\maketitle

\begin{quotation}
In this paper, we provide the first efficient algorithm for port-based teleportation, a unitarily equivariant version of teleportation useful for constructing programmable quantum processors and performing instantaneous nonlocal computation (NLQC).
The latter connection is important in AdS/CFT, where bulk computations are realized as boundary NLQC. Our algorithm yields an exponential improvement to the known relationship between the amount of entanglement available and the complexity of the nonlocal part of any unitary that can be implemented using NLQC. Similarly, our algorithm provides the first nontrivial efficient algorithm for an approximate universal programmable quantum processor.
The key to our approach is a generalization of Schur-Weyl duality we call \textit{twisted Schur-Weyl duality}, as well as an efficient algorithm we develop for the \textit{twisted Schur transform}, which transforms to a subgroup-reduced irrep basis of the partially transposed permutation algebra, whose dual is the $U^{\otimes n-k} \otimes (U^*)^{\otimes k}$ representation of the unitary group. 
\end{quotation}

\newpage
\setcounter{tocdepth}{2}
\tableofcontents

\input{sections/intro.tex}

\input{sections/schur.tex}

\input{sections/reptheory.tex}
\input{sections/circuit.tex}
\input{sections/complexity.tex}
\input{sections/discussion.tex}

\section*{Acknowledgments}
The authors thank Kfir Dolev, Eric Chitambar, Felix Leditzky and Alex May for helpful discussions. We especially thank Felix for sharing his notes on the representation theory of $A_n^{t_n}(d)$. This work was supported by ARO (award W911NF2120214), DOE (Q-NEXT), CIFAR and the Simons Foundation.

\newpage
\bibliographystyle{alpha}
\bibliography{main}
\newpage
\input{sections/appendixblock.tex}
\newpage
\input{sections/appendixcomplexity.tex}

\end{document}

%% file: macro.tex
\usepackage[utf8]{inputenc}
\usepackage[T1]{fontenc}

\usepackage{fullpage}
\usepackage{mdframed}
\usepackage{euscript}
\usepackage[dvipsnames]{xcolor}

\usepackage[
pagebackref,
colorlinks=true,
urlcolor=blue,
linkcolor=blue,
citecolor=OliveGreen,
]{hyperref}

\usepackage{amsmath}
\DeclareMathOperator{\Tr}{Tr}

\usepackage{graphicx}
\graphicspath{{./graphs/}}
\usepackage{caption}
\usepackage{subcaption}

\usepackage{dcolumn}
\usepackage{bm}
\usepackage{hyperref}
\usepackage{mathrsfs}
\usepackage{color}
\usepackage{geometry}
\usepackage{amsthm}
\usepackage{tikz}
\usepackage{mathtools}
\usetikzlibrary{arrows}
\usetikzlibrary{quantikz}
\usepackage{float}
\usepackage{empheq}
\usepackage{cases}
\usepackage[most]{tcolorbox}

\usepackage{makecell}

\usepackage{booktabs}

\usepackage{youngtab}
\usepackage{enumitem}
\newlist{primenumerate}{enumerate}{1}
\setlist[primenumerate,1]{label={\arabic*$'$.}}

\usepackage[export]{adjustbox}

\usepackage{amssymb, verbatim, xcolor,bbm}
\usepackage{doi}
\usepackage{authblk}

\usepackage{etoolbox}
\patchcmd{\section}
  {\centering}
  {\raggedright}
  {}
  {}
\patchcmd{\subsection}
  {\centering}
  {\raggedright}
  {}
  {}
\patchcmd{\subsubsection}
  {\centering}
  {\raggedright}
  {}
  {}

\newtheorem*{problem*}{Problem}

\newcommand{\bcdot}{{\boldsymbol{\cdot}}}%
\renewcommand\bra[1]{{\langle{#1}|}}
\renewcommand\ket[1]{{|{#1}\rangle}}
\newcommand{\norm}[1]{\Big\lVert #1 \Big\rVert}%
\newcommand{\mycong}[1]{\stackrel{\mathclap{\scriptsize\mbox{#1}}}{\cong}}
\newcommand{\poly}[1]{\text{poly}(#1)}%
\newcommand{\polylog}[1]{\text{polylog}(#1)}%
\newcommand{\thd}[1]{\thead{#1}}%
\newcommand{\pp}[1]{\left({#1}\right)}%

\newcommand{\tikzscale}[2]{\begin{center}\begin{tikzpicture}\node[scale=#1] {#2};\end{tikzpicture}\end{center}}

\makeatletter
\@addtoreset{equation}{section}
\makeatother

\allowdisplaybreaks

%% file: sections/intro.tex
\section{Introduction}
\label{sec:intro}

\subsection{Port-based teleportation}

Port-based teleportation is a variant of quantum teleportation first introduced by Ishizaka and Hiroshima~\cite{ishizaka2008asymptotic,ishizaka2009quantum} building on a teleportation scheme in~\cite{Knill2001}.  Alice and Bob each hold $n-1$ pairs of $d$-dimensional quantum systems called \textit{ports}, which begin in a joint entangled resource state, and Alice has an additional system in a state she wishes to teleport to Bob.  She performs a joint POVM on the state and her ports and classically sends the measurement outcome $i$ to Bob, who learns that his $i$th port now holds the approximately teleported state.  In the limit $n \rightarrow \infty$, the fidelity of the teleportation goes to $1$ for some choices of resource state and POVM~\cite{ishizaka2008asymptotic, ishizaka2009quantum, Beigi_2011, Mozrzymas_2018, Walter21}.  Port-based teleportation consumes a large amount of entanglement to achieve only approximate teleportation, a necessary tradeoff to achieve \textit{unitary equivariance}~\cite{nielsen1997programmable} --- Bob can perform a unitary operation $U$ on the teleported state, before he has learned which port holds it from the classical communication, by simply performing $U$ on all ports.  The unitary equivariance of port-based teleportation makes it useful for a variety of applications, including to universal programmable quantum processors and nonlocal quantum computation~\cite{ishizaka2008asymptotic, Beigi_2011}.  

In this work, we develop the first efficient $\poly{n,d}$ quantum algorithm for a port-based teleportation protocol.  We consider the protocol where the entangled resource state is $n-1$ Bell pairs and the joint POVM is the \textit{pretty good measurement}, which is the optimal POVM for this resource state as well as for an optimized resource state \cite{Leditzky20}.  The measurement operators for the pretty good measurement are given by 
\begin{align}
		\Pi_i &= \tilde{\Pi}_i + \Delta \text{, for } i = 1,..., n-1
\end{align}
where
\begin{align}
        \tilde{\Pi}_i &= \rho^{-1/2} \rho_i \rho^{-1/2}, \quad \Delta = \left(\mathbb{I} - \sum_{i = 1}^{n-1} \tilde{\Pi}_i\right)\Big/(n-1) \\
        \rho &= \sum_{i=1}^{n-1} \rho_i, \quad     \rho_i = \ket{\phi_+}\bra{\phi_+}_{A_i A_n} \otimes \frac{\mathbb{I}_{\overline{A_i}}}{d^{n-2}}.
\end{align}
Here, $\ket{\phi_+}=1/\sqrt{d}\sum_{j=1}^{d}\ket{jj}$, $A_i$ is Alice's $i$th port, $A_n$ holds the state to be teleported, and $\overline{A_i}$ denotes all of Alice's ports except $A_i$.

\subsubsection{Twisted Schur-Weyl duality}
A significant challenge to performing port-based teleportation with the pretty-good measurement is implementing $\rho^{-1/2}$, because of the matrix power.  
Our strategy to do this is to take advantage of the symmetry of the problem, in particular, a generalization of the celebrated \textit{Schur-Weyl duality} in representation theory.

The original duality states that the algebra $\mathbb{C}V[S(n)]$ of permutation operators that permute the $n$ factors of the tensor space $(\mathbb{C}^d)^{\otimes n}$, and the algebra $\mathbb{C}Q[U(d)]$ of all $n$-fold unitary tensor operators $Q(U)=U^{\otimes n}$, are each other's \textit{commutants}, i.e., any operator that commutes with all elements of one algebra lies in the other. This results in a multiplicity-free irrep decomposition of the Hilbert space,
\begin{align}
    (\mathbb{C}^d)^{\otimes n} \cong \bigoplus_{\lambda \vdash n,\ h(\lambda) \leqslant d} \mathcal{P}_\lambda \otimes \mathcal{Q}^d_{\lambda},
\end{align}
where each irrep of $V[S(n)]\times Q[U(d)]$ is labeled by a Young diagram $\lambda$, and is composed of the corresponding $S(n)$ irrep $\mathcal{P}_\lambda$ and $U(d)$ irrep $\mathcal{Q}^d_{\lambda}$. The unitary transformation from the standard basis of $(\mathbb{C}^d)^{\otimes n}$ to the irrep-decomposing basis is called the \textit{Schur transform} and can be implemented by an efficient quantum algorithm~\cite{Harrow05, Harrow06}.

Our problem exhibits a \textit{twisted} Schur-Weyl duality. This is rooted in the nature of the Bell state
\begin{align}
    \ket{\phi_+}\bra{\phi_+}_{in}=\frac{1}{d}\sum_{j,k=1}^{d}\ket{jk}\bra{kj}_{in}^{t_n}=\frac{1}{d}V[(i\ n)]^{t_n},\quad \big[\ket{\phi_+}\bra{\phi_+}_{in}, U\otimes U^{*}\big]=0
\end{align}
where $t_n$ denotes partial transposition on the $n$-th system. It implies that the constituent operators $\rho_i$ and $\rho$ of $\sqrt{\tilde{\Pi}_i}$ both live in the algebra of partially transposed permutation operators $A^{t_n}_n(d) = \mathbb{C}\{V[\sigma]^{t_n}: \forall \sigma\in S(n) \}$ and possess a $\big[\rho_{(i)}, \chi(U) \big]=0$ symmetry, for $\chi(U) = U^{\otimes n-1}\otimes U^*$. In fact, these two statements are equivalent because the algebras $A_n^{t_n}(d)$ and $\mathbb{C}\chi[U(d)]$ are mutual commutants, just as $\mathbb{C}V[S(n)]$ and $\mathbb{C}Q[U(d)]$ are.  
The corresponding twisted Schur-Weyl duality decomposes the Hilbert space as~\cite{Walter21, Leditzky20}
\begin{align}
    \label{eq:intro_H_M}(\mathbb{C}^d)^{\otimes n}\cong\mathcal{H}_M + \mathcal{H}_S,\quad 
    \mathcal{H}_M = \bigoplus_{\alpha \vdash n-2,\ h(\alpha) \leqslant d} \textcolor{red}{\left( \bigoplus_{\mu = \alpha+\Box, h(\mu) \leqslant d} \mathcal{P}_{\mu} \right)} \otimes \mathcal{Q}_{\alpha}^d.
\end{align}

In this paper, we focus on $\mathcal{H}_M$, the representation space of the ideal $M\subset A_n^{t_n}(d)$, which is generated by the elements that nontrivially permute and transpose the $n$-th system and are therefore relevant to our problem.  We present explicit formulae and a $\poly{n,d}$ quantum algorithm for a \textit{twisted} Schur transform, which maps from the standard basis of $\mathcal{H}_M$ to one that decomposes $A_n^{t_n}(d)\times \mathbb{C}\chi[U(d)]$ into irreps.  This allows us to block-diagonalize  $\rho_i$ and $\rho$, such that matrix operations can be implemented block-by-block in parallel.  

It is not enough to merely block-diagonalize $A^{t_n}_n(d)$, however.
The key is that we recognize that irreps of $A_n^{t_n}(d)$ enjoy a multiplicity-free decomposition into irreps $\mathcal{P}_\mu$ of $S(n-1)$. 
(These irreps, and their decomposition, are shown here in parentheses and highlighted in red.)
This means that we can do more than implement just any twisted Schur transform---we can use the freedom of basis choice within each irrep of $A^{t_n}_n(d)$ to transform to a basis that is additionally subalgebra-reduced on $\mathbb{C}[S(n-1)] \subset A^{t_n}_n(d)$.  
As a result, operators like $\rho$ that bear an $S(n-1)$ symmetry will be \textit{diagonalized}, making computing the matrix power trivial.  More work is needed to complete the algorithm for port-based teleportation, but this is the essence of why it is possible to implement the POVM efficiently. 

In fact, finding the subalgebra-reduced basis for the irreps of $A^{t_n}_n (d)$ is a particular instance of the general problem of block-diagonalizing \textit{generalized induced representations}.  This is because the irreps of $A^{t_n}_n (d)$ can be viewed as representations of $S(n-1)$ lifted from a representation $V$ of the subgroup $S(n-2)$; we call them \textit{generalized} because the isomorphic copies of $V$ may not necessarily be linearly independent.  Inspired by this motivating example, we have developed a quantum algorithm that solves the broader problem of block-diagonalizing induced representations in an upcoming companion paper~\cite{inducedRepPrep}. 

Independently of this work, Burchardt, Grinko, and Ozols developed another efficient implementation of PBT, again by using a twisted Schur transform~\cite{Grinkoprep}. Also independently, Nguyen has constructed an efficient algorithm for performing the the transform~\cite{Nguyen2023efficient}.
Our works can be seen as complementary as Burchardt et al. and Nguyen develop the transform from the $\mathbb{C}\{U^{\otimes n-1} \otimes U^*\}$ side of the duality, while we work from the $A^{t_n}_n (d)$ side.

\subsubsection{Performing the twisted Schur transform}

In order to sketch how to perform the subalgebra-reduced twisted Schur transform, we first need to develop a concrete picture of the irreps of $A^{t_n}(d)$.
The structure of $A^{t_n}(d)$ and its irreps for all $n$ and $d$ was first derived in~\cite{Studzinski13,Studzinski14,Studzinski18} using the nested ideal structure of semisimple algebras. 
Our work provides a complementary Hilbert space approach that constructively identifies the irrep spaces---the key is to employ the prior knowledge of Schur-Weyl duality.

Explicitly, we find the irreps of $A^{t_n}_n(d)$ within $\mathcal{H}_M$ by first identifying the irreps of $\mathbb{C}\chi[U(d)]$, the commutant of $A^{t_n}_n(d)$, then extracting the irreps of $A^{t_n}_n(d)$ as the multiplicity spaces of each inequivalent $\mathbb{C}\chi[U(d)]$ irrep. Spelling this out, the relevant subspace $\mathcal{H}_M$ turns out to be 
\begin{align}
    \mathcal{H}_M &=\mathbb{C}\{|\phi_i\rangle |\phi_+\rangle _{kn} : k=1,\dots,n-1,i=1,\dots,d^{n-2}\} 
    \\&= \mathbb{C}\{V[(k\ n-1)]|\phi_i\rangle |\phi_+\rangle _{n-1 n} :k=1,\dots,n-1,i=1,\dots,d^{n-2}\},
\end{align}
where each $\ket{\phi_i}$ is a computational basis state on $n-2$ qudits. Because $\ket{\phi_+}_{n-1 n}$ is $U\otimes U^*$-invariant, $\chi(U)$ will act on $\mathcal{H}_M$ just as $U^{\otimes n-2}$, whose irreps are directly identifiable using the Schur-Weyl duality for $U^{\otimes n-2}$. Finding the pairing spaces then gives the $A^{t_n}_n(d)$ irreps
\begin{align}
    \label{eq:intro_H_al}\mathcal{H}_{\alpha,r} &=\mathbb{C}\{ V[(k\ n-1)] |r, \alpha, k_\alpha \rangle^{(n-2)}_{\text{Sch}} |\phi_+\rangle _{n-1 n} : k=1,\dots,n-1,k_\alpha = 1,\dots,d_\alpha\}
\end{align}
where $|r, \alpha, k_\alpha \rangle^{(n-2)}_{\text{Sch}}$ are elements of the $(n-2)$-qudit Schur basis labeled by the Young diagram $\alpha \vdash n-2$, with $r$ labeling the multiplicity and $k_\alpha$ indexing within each of them.

Given this set of spanning vectors for each irrep $\mathcal{H}_{\alpha, r}$, we then only need to construct an orthonormal basis that decomposes it into irreps of $S(n-1)$.  
A result of Refs.~\cite{Studzinski14,Studzinski18} shows that this can be done using permutations in subgroup-reduced irreps of $S(n-1)$.
Altogether, this means that the irrep basis of $A^{t_n}_n (d)$ can be constructed using just permutations and the $n-2$-qudit Schur transform, which together generate the spanning vectors of $\mathcal{H}_{\alpha,r}$ in Eq.~\eqref{eq:intro_H_al}, and permutations and submatrices of the $n-1$-qudit Schur transform, which together perform the reduction.  Because the twisted Schur transform can be constructed using only these simple building blocks, it will enjoy an efficiency it inherits directly from the efficient quantum Schur transform~\cite{Harrow05, Harrow06}.

\subsubsection{Using the twisted Schur transform to execute PBT}

We have already explained that transforming to the twisted Schur basis diagonalizes $\rho$, making it trivial to execute $\rho^{-1/2}$.  To complete the algorithm, we need to treat the full Kraus operator $\sqrt{\Pi_i}$, which will be block-diagonalized in this basis.  It turns out that its matrix elements will also only depend on the same simple building blocks as the twisted Schur transform.  This is the key result of the representation theory that makes our algorithm efficient: by transforming to the twisted Schur basis, executing $\sqrt{\Pi_i}$, and transforming back, we can decompose $\sqrt{\Pi_i}$ as a linear superposition of products of $(n-1)$- and $(n-2)$-qudit Schur transform submatrices and permutations.

From this decomposition, it's relatively straightforward to efficiently implement port-based teleportation.  It's first necessary to find unitary block-encodings of the constituent non-unitary matrices, which can be combined into the superposition for $\sqrt{\Pi_i}$ using a method inspired by the Linear Combination of Unitaries algorithm given in \cite{Gilyen18, Berry15, Childs17}.  After finally block-encoding each $\sqrt{\Pi_i}$, we can use Naimark's theorem to implement the POVM.  The only remaining detail is that it's necessary to use \textit{oblivious amplitude amplification} to amplify the $0$-eigenspace of the ancillae used for the block-encodings---the correct POVM is only performed when we end up in this subspace \cite{Gilyen18,Berry14,Scott12}.
Altogether, our quantum circuit for port-based teleportation has a $\poly{n,d}$ time cost, an $O(n^{d})$ space cost, and an $O(1)$ error.

\subsection{Applications of efficient PBT} \label{subsec:applications}

Our efficient algorithm for port-based teleportation, as well as the twisted Schur transform, both have a number of consequences in quantum information. 

\medskip
\noindent\textbf{Nonlocal quantum computation and AdS/CFT.} 
One of the most enduring puzzles in AdS/CFT is how boundary degrees of freedom are organized into an extra spatial dimension in the bulk theory. Despite a great amount of work investigating the emergence of space from different points of view~\cite{bianchi2002holographic,Hamilton:2005ju,engelhardt2016towards,dong2016reconstruction,jafferis2016relative,cotler2019entanglement}, the phenomenon remains mysterious. One approach to studying emergent space is to calculate and interpret scattering in the bulk theory. Taking that idea a step further, May has developed thought experiments in which a quantum computer is placed in the bulk spacetime~\cite{may2019quantum,may2021holographic}. Inputs are sent to the computer from the boundary and outputs of the computation then sent back out to the boundary. In that setting, the emergence of space converts a locally-performed bulk quantum computation into a \textit{instantaneous nonlocal quantum computation}~\cite{beigi2011simplified} performed in communicating, segmented regions of the boundary spacetime.  Port-based teleportation is useful for nonlocal quantum computation because of its unitary equivariance---in fact, it is the bottleneck subroutine.

May recently proved that there is a quantitative relationship between the amount of entanglement consumed in a nonlocal quantum computation protocol and the complexity of the nonlocal part of the unitary transformation that can be executed~\cite{may2022complexity}. While the complexity and entanglement are bounded above and below by each other, his bound has a \textit{triple} exponential gap between the two. By improving the efficiency of nonlocal quantum computation, our result reduces the gap to double exponential. In the holographic setting, the entanglement consumed must be drawn from the boundary state so our results provide an improved understanding of the complexity of bulk computations that can be realized for a given boundary state.

\medskip
\noindent \textbf{Universal programmable quantum processor.}
The unitary equivariance of port-based teleportation means that any port-based teleportation protocol is also an approximate universal programmable quantum processor $\mathcal{P}$~\cite{ishizaka2008asymptotic}: such a device takes as input a quantum state $|\psi\rangle \in \mathcal{H}_D$ and a program state $| \xi_U \rangle \in \mathcal{H}_P$ and outputs $U | \psi \rangle$ to a good approximation~\cite{nielsen1997programmable,PhysRevA.65.022301,PhysRevA.65.012302}. The problem of finding the optimal asymptotic trade-off between the quality of approximation and the size of the program state was recently found, but disregarding the cost of implementing $\mathcal{P}$~\cite{PhysRevLett.122.170502,PhysRevLett.122.080505,yang2020optimal}. Our quantum algorithm yields the first nontrivial processors that are also efficient in the size of the program state. In both cases, good processors require $( \dim \mathcal{H}_D )^2 < \operatorname{const} \log \dim \mathcal{H}_P$ but the error scaling of our efficient algorithm is not optimal.

\subsection{How to navigate this paper}

We begin in Sec.~\ref{sec:prelim} by defining port-based teleportation and reviewing the necessary background to understand our strategy for constructing the algorithm, including the algebra $A^{t_n}_n (d)$ as well as Schur-Weyl duality.  Sec.~\ref{sec:reptheory} then dives into the structure of twisted Schur-Weyl duality and the irreps of $A^{t_n}_n (d)$, demonstrating why the relevant irreps enjoy a multiplicity-free decomposition into irreps of $S(n-1)$.  It then finds the corresponding subalgebra-reduced irrep basis of $A^{t_n}_n (d)$ that gives the appropriate twisted Schur transform.  Armed with the twisted Schur transform, Sec.~\ref{sec:circuit} then completes the algorithm for PBT and gives an explicit circuit, while Sec.~\ref{sec:complexity} discusses the complexity. Sec.~\ref{sec:discussion} discusses the potential applications of our results, in particular of the twisted Schur transform. 
 Finally, Appendix~\ref{app:circuit} contains the block-encoding of the Kraus operators of PBT, and Appendix~\ref{app:complexity} covers the complexity and error propagation analysis for the algorithm in detail.

%% file: sections/schur.tex
\section{Preliminaries}
\label{sec:prelim}

\subsection{PBT with the pretty good measurement\label{sec:pbt_prelim}}

In a general port-based teleportation protocol, Alice and Bob each hold $n-1$ $d$-dimensional quantum systems called \textit{ports}, $A = \{ A_i: i = 1,...,n-1\}$ and $B = \{B_i: i = 1,...,n-1\}$, respectively, which begin in an entangled resource state $\Phi_{AB}$.  Alice also holds an additional quantum system $A_n$ in a state $\eta_{A_n}$ she wishes to teleport to Bob.  To do this, she performs a joint POVM  on the systems $A A_n$ and classically sends the outcome $i$ to Bob.  

In the \textit{deterministic} version~\cite{ishizaka2008asymptotic,Beigi_2011, Mozrzymas_2018,Walter21} of the protocol, POVM $E = \{ E^i_{A A_n} : i = 1, ..., n-1 \}$ and the outcome $i \in \{ 1, ..., n-1 \}$ corresponds to the port $B_i$ which now holds the approximately teleported state.  Bob then simply throws away everything except the port $B_i$, which he renames $B_n$, and the protocol is complete. 
In the limit $n \rightarrow \infty$, the fidelity of the teleportation goes to $1$ for some POVMs. There is also \textit{probabilistic} PBT~\cite{ishizaka2009quantum, Mozrzymas_2018,Walter21}, which is implemented with $E = \{ E^i_{A A_n} : i = 1, ..., n \}$.  In that case, the state is teleported with perfect fidelity but the additional outcome $i = n$ corresponds to failure of the protocol.  In this paper, however, we will focus on deterministic PBT.

Notice that because Bob does not have to apply any Pauli operation to decode the state, in contrast to traditional teleportation, his portion of the protocol commutes with the application of any unitary $U^{\otimes n-1}$, for $U \in U(d)$,  to his ports---this is why PBT is unitarily equivariant~\cite{nielsen1997programmable}.

A specific deterministic PBT protocol is characterized by the choice of entangled resource state $\Phi_{AB}$ and POVM $E = \{ E^i_{A A_n} : i = 1, ..., n-1 \}$.  This choice defines a channel $\Lambda: A_n \rightarrow B_n$, given by
\begin{equation}
    \label{eq:channel_PBT}\Lambda (\eta_{A_n}) = \sum_{i=1}^{n-1} [\Tr_{A A_n \overline{B_i}} [E^i_{A A_n} (\Phi_{AB} \otimes \eta_{A_n} ) ]]_{B_i \mapsto B_n}
\end{equation}
where $\overline{B_i}$ indicates all of Bob's ports except $B_i$, and $B_i \mapsto B_n$ represents the renaming of $B_i$ to $B_n$ in each term of the sum. We can quantify how close this channel is to the identity channel, i.e. to perfectly teleporting $\eta_{A_n}$ to $B_n$, by using the entanglement fidelity $F(\Lambda)$.  Here,
\begin{equation}
F(\Lambda) = \Tr [ \phi_{+B_n R} (\Lambda \otimes 1_R) \phi_{+A_n R}] = \frac{1}{d^2} \sum_{i=1}^{n-1}\Tr [E^i_{A B_n} (\rho_i)_{A B_n} ] 
\end{equation}
which measures how well $\Lambda$ preserves correlations with an ancilla $R$.  In the second equality, the joint POVM is now interpreted to be a measurement on $A$ and $B_n$, and 
\begin{equation}
    (\rho_i)_{A B_n} = \Tr_{\overline{B_i}} [\Phi_{A B}]_{B_i \mapsto B_n},
\end{equation}
where the trace is over all systems of $B$ except the system $B_i$, denoted by $\overline{B_i}$.

This expression for the entanglement fidelity is proportional to the probability of successfully distinguishing the states $(\rho_i)_{A B_n}$ when they are drawn uniformly at random~\cite{Leditzky20}.  This suggests the use of the \textit{pretty good measurement} for state discrimination, for which the measurement operators $\{ \Pi_i : i = 1, ..., n-1\}$ are given by
\begin{align}
		\Pi_i &= \tilde{\Pi}_i + \Delta\\
        \tilde{\Pi}_i &= \rho^{-1/2} \rho_i \rho^{-1/2}\text{, for } i = 1,..., n-1
\end{align}
where
\begin{align}
\rho &= \sum_{i=1}^{n-1} \rho_i\\
\Delta &= \left(\mathbb{I} - \sum_{i = 1}^{n-1} \tilde{\Pi}_i\right)\Big/(n-1).
\end{align}
These operators satisfy $\sum_{i=1}^{n-1} \Pi_i = \mathbb{I}$ and are each positive semi-definite, ensuring that they compose a valid POVM.

In the case we will be interested in, where the entanglement resource state $\Phi_{AB}$ is given by $n-1$ copies of the maximally entangled state, the pretty good measurement is the optimal POVM.  In fact, the pretty good measurement is even optimal when the resource state is optimized to yield the best possible PBT protocol \cite{Leditzky20}.

In our case then, we have
\begin{equation}
    \label{eq:bell_pairs}\Phi_{A B} = \bigotimes_{i = 1}^{n-1} \ket{\phi_+} \bra{\phi_+}_{A_i B_i},
\end{equation}
where $|\phi_+\rangle = 1/\sqrt{d}\sum_{i}|ii\rangle$ is the maximally entangled state, and which gives
\begin{equation}
    \rho_i = \ket{\phi_+}\bra{\phi_+}_{A_i B_n} \otimes \frac{\mathbb{I}_{\overline{A_i}}}{d^{n-2}}.
\end{equation}
Here the subscript $B_n$, used when calculating fidelity, need to be replaced by $A_n$ when viewing $\rho_i$ as a part of the measurement operator $\Pi_i$.

\subsection{Symmetries of $\rho$\label{sec:rho_sym}}

Turning to the algorithmic problem of implementing port-based teleportation, we essentially want to implement $\sqrt{\Pi_i}$, $i=1,\dots,n-1$, as a quantum circuit acting on $\mathcal{H}\equiv (\mathbb{C}^d)^{\otimes n}$, the joint Hilbert space of Alice's ports $A$ and the qudit $A_n$, which holds the state to be teleported. The biggest obstacles to this are the matrix powers $\rho^{-1/2}$ and $\sqrt{\Pi_i}$. Na\"ively, they can be performed by finding eigendecompositions of $\rho$ and $\Pi_i$, but this would clearly not be efficient in the dimensional parameters $d$ and $n$. Our strategy will be to instead leverage the symmetry of the problem, in particular a twisted Schur-Weyl duality, to diagonalize $\rho$.

Recall that $\rho$ is a equal superposition of $\rho_i$, which are defined to be
\begin{equation}
    \rho_i = \ket{\phi_+}\bra{\phi_+}_{i n} \otimes \frac{\mathbb{I}_{\overline{i}}}{d^{n-2}}.
\end{equation}
where we have abbreviated the port $A_i$ as $i$, and $A_n$ or $B_n$ as $n$.  The density matrix of a maximally entangled state $\ket{\phi_+}\bra{\phi_+}_{in}$ can be rewritten as
\begin{align}
    \ket{\phi_+}\bra{\phi_+}_{in} &= \frac{1}{d} \sum_{j, k = 1}^{d} \ket{jj}\bra{kk}_{in} = \frac{1}{d} \sum_{j, k = 1}^{d} \ket{jk}\bra{kj}_{in}^{t_n} = \frac{1}{d} V[(i\ n)]^{t_n}.
\end{align}
This can also be seen using the tensor network representation of the maximally entangled state
\tikzset{every picture/.style={line width=0.75pt}} 
\begin{equation}
    \ket{\phi_+}\bra{\phi_+}_{in} = \frac{1}{d}
    \begin{tikzpicture}[baseline={([yshift=-1ex]current bounding box.center)}, x=0.75pt,y=0.75pt,yscale=-1,xscale=1]
    
    \draw    (155,190.8) .. controls (155,203.8) and (180,203.3) .. (180,190.55) ;
    \draw    (155.25,215.3) .. controls (155.25,202.3) and (180,201.55) .. (180.25,215.05) ;
    
    \draw (149.75,177) node [anchor=north west][inner sep=0.75pt]    {${\textstyle i}$};
    \draw (173,180) node [anchor=north west][inner sep=0.75pt]    {$n$};
    \end{tikzpicture} 
    = \frac{1}{d}
    \begin{tikzpicture}[baseline={([yshift=-0.5ex]current bounding box.center)},x=0.75pt,y=0.75pt,yscale=-1,xscale=1]
    
    \draw [color=blue,draw opacity=1 ]   (150.75,290.3) .. controls (150.75,277.3) and (175.25,280.05) .. (175.5,265.55) ;
    \draw [color=blue,draw opacity=1 ]   (150.5,265.8) .. controls (150,269.3) and (153.5,272.8) .. (160.5,276.55) ;
    \draw [color=blue,draw opacity=1 ]   (174.5,289.3) .. controls (174.25,282.8) and (169.25,281.8) .. (166,279.3) ;
    \draw [color=red ,draw opacity=1 ]   (175.5,265.55) .. controls (176.5,256.3) and (183.75,271.8) .. (187.5,275.55) ;
    \draw [color=red,draw opacity=1 ]   (174.5,289.3) .. controls (176.25,298.05) and (195.75,272.05) .. (196.75,265.8) ;
    \draw [color=red,draw opacity=1 ]   (191,280.25) .. controls (192.5,282.55) and (195,284.3) .. (197.5,290.05) ;
    
    \draw (145.25,252) node [anchor=north west][inner sep=0.75pt]    {${\textstyle i}$};
    \draw (189.5,255) node [anchor=north west][inner sep=0.75pt]    {${\textstyle n}$};
    \end{tikzpicture}
\end{equation}
where we have shown the permutation $V[(i\ n)]$ in blue and the partial transposition on the $n$th system in red.
This means
\begin{equation}
    \label{eq:rho_i_algebra}    \rho_i  = \frac{1}{d^{n-1}} V[(i\ n)]^{t_n} =  \frac{1}{d^{n-1}}
    \begin{tikzpicture}[baseline={([yshift=-1ex]current bounding box.center)}, x=0.75pt,y=0.75pt,yscale=-1,xscale=1]
    
    \draw    (155,190.8) .. controls (155,203.8) and (180,203.3) .. (180,190.55) ;
    \draw    (155.25,215.3) .. controls (155.25,202.3) and (180,201.55) .. (180.25,215.05) ;
    
    \draw (149.75,177) node [anchor=north west][inner sep=0.75pt]    {${\textstyle i}$};
    \draw (173,180) node [anchor=north west][inner sep=0.75pt]    {$n$};
    \end{tikzpicture}.
\end{equation}
which gives
\begin{equation}
    \label{eq:rho_algebra}
    \rho = \sum_{i=1}^{n-1} \rho_i = \frac{1}{d^{n-1}} \sum_{i=1}^{n-1} V[(i\ n)]^{t_n} = \frac{1}{d^{n-1}} \sum_{i=1}^{n-1}
    \begin{tikzpicture}[baseline={([yshift=-1ex]current bounding box.center)}, x=0.75pt,y=0.75pt,yscale=-1,xscale=1]
    
    \draw    (155,190.8) .. controls (155,203.8) and (180,203.3) .. (180,190.55) ;
    \draw    (155.25,215.3) .. controls (155.25,202.3) and (180,201.55) .. (180.25,215.05) ;
    
    \draw (149.75,177) node [anchor=north west][inner sep=0.75pt]    {${\textstyle i}$};
    \draw (173,180) node [anchor=north west][inner sep=0.75pt]    {$n$};
    \end{tikzpicture}.
\end{equation}
It is then obvious that $\rho$ is completely symmetric with respect to the first $n-1$ systems, i.e., $[\rho, V(\sigma)]=0$ for any $\sigma\in S(n-1)$.

Additionally, $\rho_i$ possesses a partially conjugated unitary group symmetry, which can be quickly seen from the tensor network representation:
\begin{align}
    d^{n-1}\left(U\otimes U^*\right) \rho_i \left(U\otimes U^*\right)^\dagger = \includegraphics[width = 1cm, valign = c]{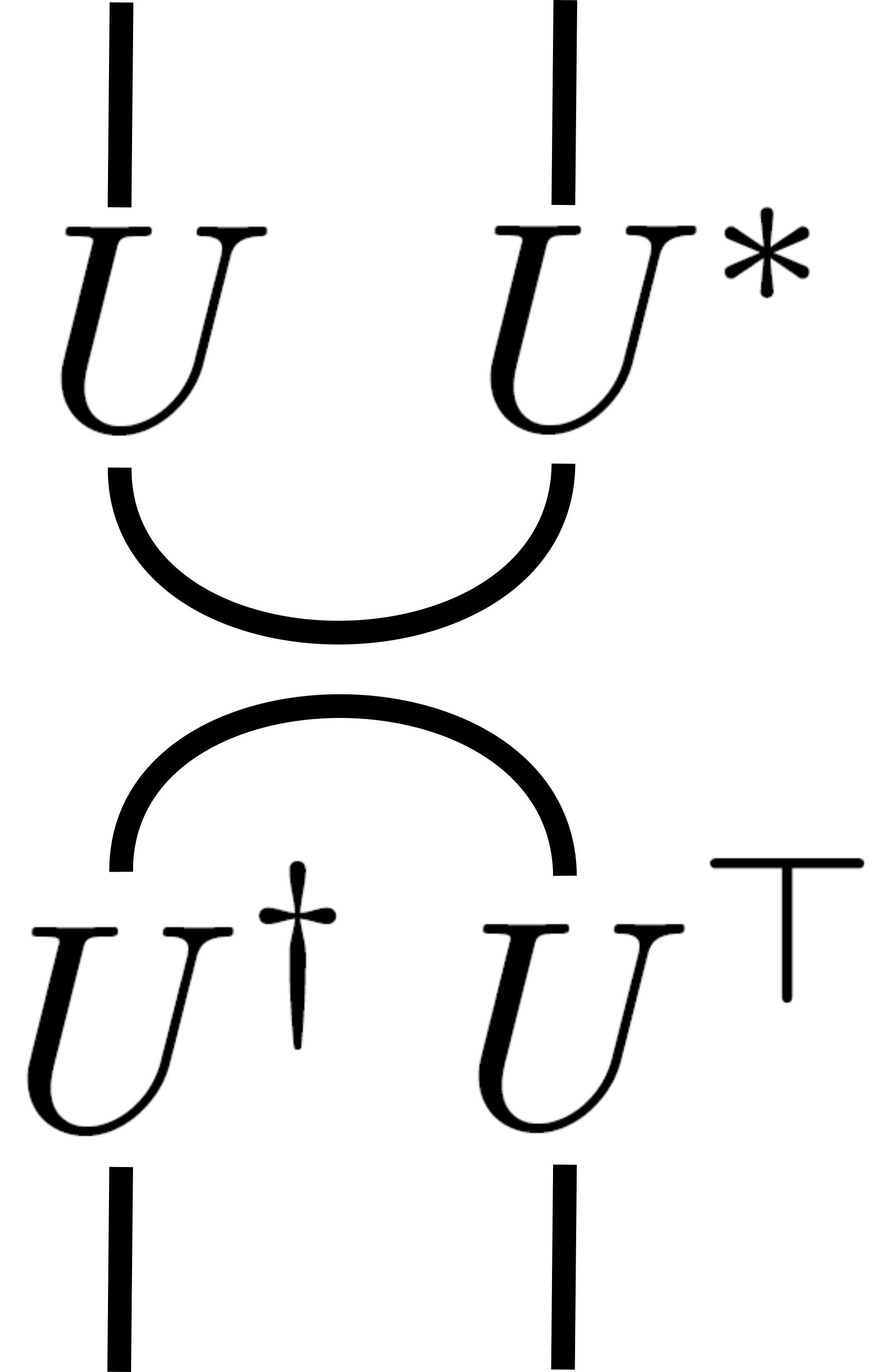} = \includegraphics[width = 0.9cm, valign = c]{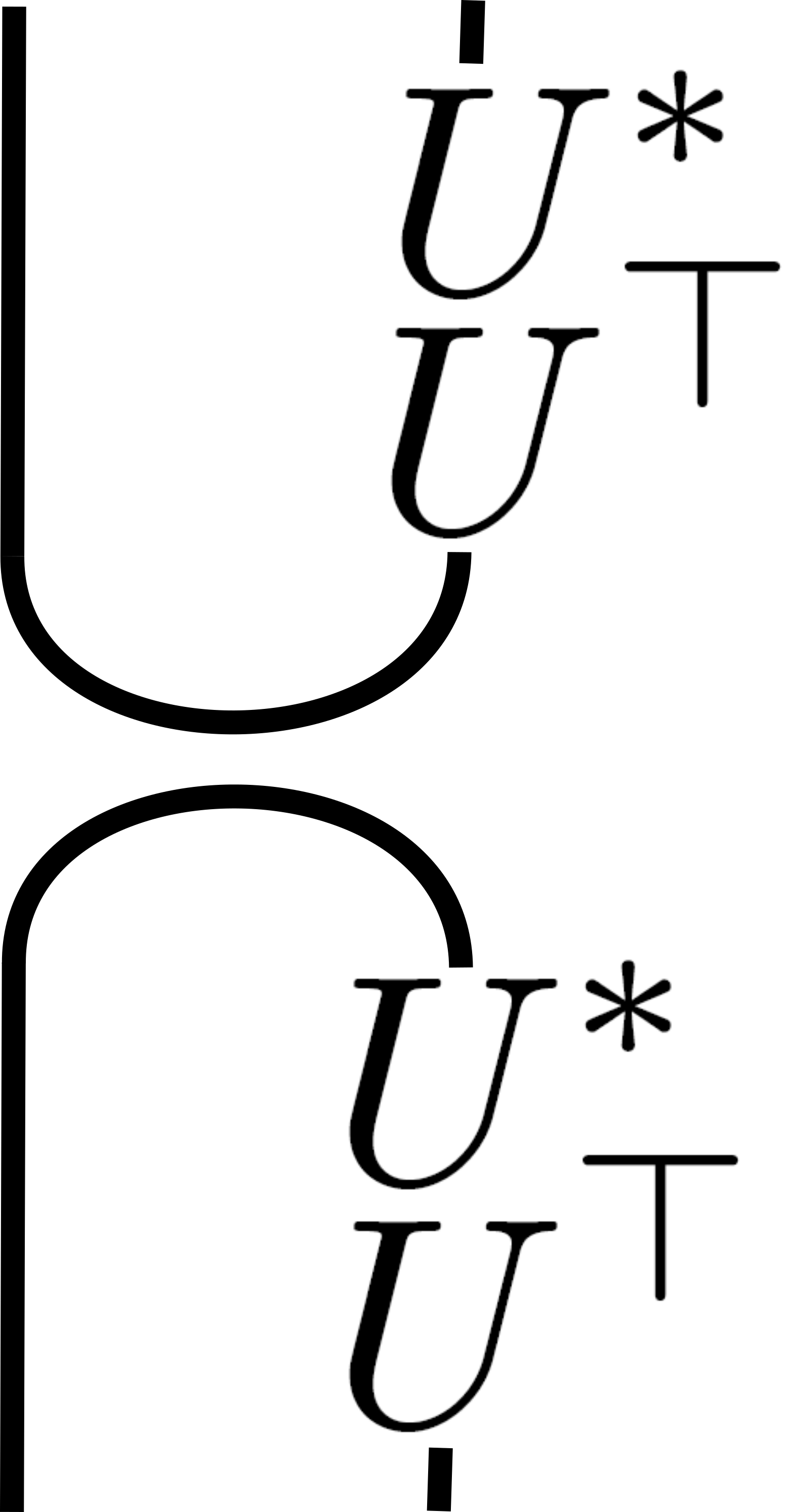} 
    = d^{n-1}\rho_i.
\end{align}
Therefore, $\rho$ satisfies $[\rho, \chi(U)]=0$ for any $U\in U(d)$.

\subsection{The algebras $A^{t_n}_n(d)$ and $A^{(k)}_n(d)$\label{sec:A_defs}}

By writing $\rho_i$ and $\rho$ in the forms shown in Eqs.~\eqref{eq:rho_i_algebra} and \eqref{eq:rho_algebra}, we have shown that they are elements of \textit{the algebra of partially transposed permutation operators} $A^{t_n}_n(d)$ where
\begin{align}
    A^{t_n}_n(d) = \mathbb{C}\{V[\sigma]^{t_n}: \forall \sigma\in S(n) \}.
\end{align}
Here and throughout this paper we use $\mathbb{C}\equiv \text{span}_{\mathbb{C}}$ as an abbreviation of a $\mathbb{C}$-vector space, and $\left(V,(\mathbb{C}^d)^{\otimes n}\right)$ to denote the natural representation of $S(n)$ on $n$ qudits, i.e. for any $\sigma \in S(n)$ we have 
\begin{align}
        V(\sigma)(\ket{i_1} \otimes ... \otimes \ket{i_n}) &= \ket{i_{\sigma^{-1}(1)}} \otimes ... \otimes \ket{i_{\sigma^{-1}(1)}}.
\end{align}
The $t_n$ superscript denotes the partial transposition on the $n$th qudit.  More generally, we can consider the partial transposition on the last $k$ qudits, denoted with a $(k)$ superscript, to define the algebra
\begin{align}
    A^{(k)}_n(d) = \mathbb{C}\{V[\sigma]^{(k)}: \forall \sigma\in S(n) \}.
\end{align}
Notice that $V[\cdot]^{t_n}$ and $V[\cdot]^{(k)}$ are not representations of $S(n)$, since, for example, 
\begin{align}
    V[(i\ n)]^{t_n}V[(i\ n)]^{t_n}\neq V[e_{S(n)}]^{t_n}.
\end{align}

As it turns out, when $d \geqslant n$, the algebra $A^{(k)}_n(d)$ is of dimension $n!$ and is isomorphic to~\cite{Benkart94} the $n!$-dimensional walled Brauer algebra $B_{n-k,k}$~\cite{Werner07,Cox08,Koike89,Benkart94,Brundan12}---a subalgebra of the Brauer algebra \cite{Gavarini97,Brauer37,Pan95}. The algebra $B_{n-k,k}(\mathbb{C})$ is a $\mathbb{C}$-span of diagrams like those illustrated below:
\begin{figure}[H]
    \centering
    \begin{subfigure}[t]{.5\textwidth}
      \centering
      \includegraphics[width=0.43\textwidth]{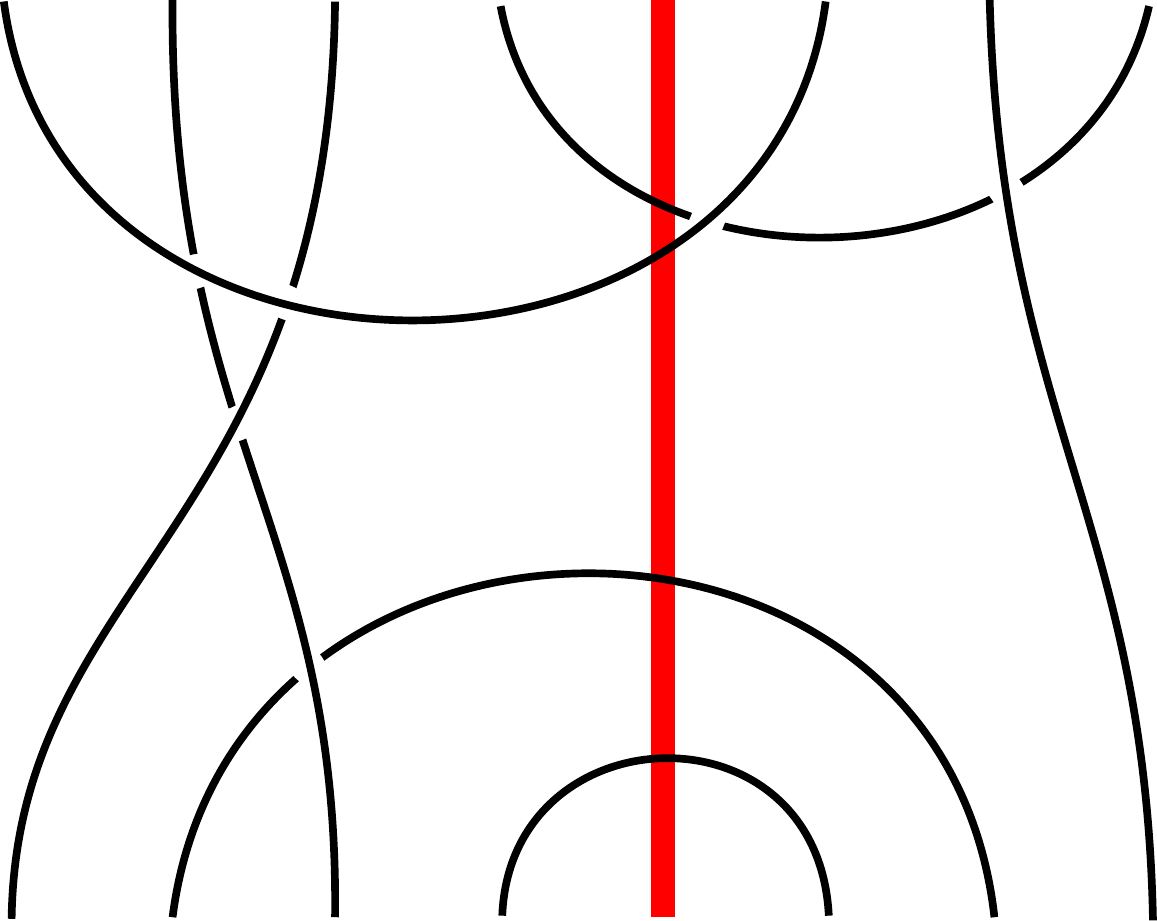}
      \captionsetup{width=.73\textwidth}
      \caption{A generator of $B_{4,3}$ with nontrivial permutations between the first $4$ and last $3$ systems. }
      \label{fig:Brauernk}
    \end{subfigure}%
    \begin{subfigure}[t]{0.5\textwidth}
      \centering
      \includegraphics[width=.43\textwidth]{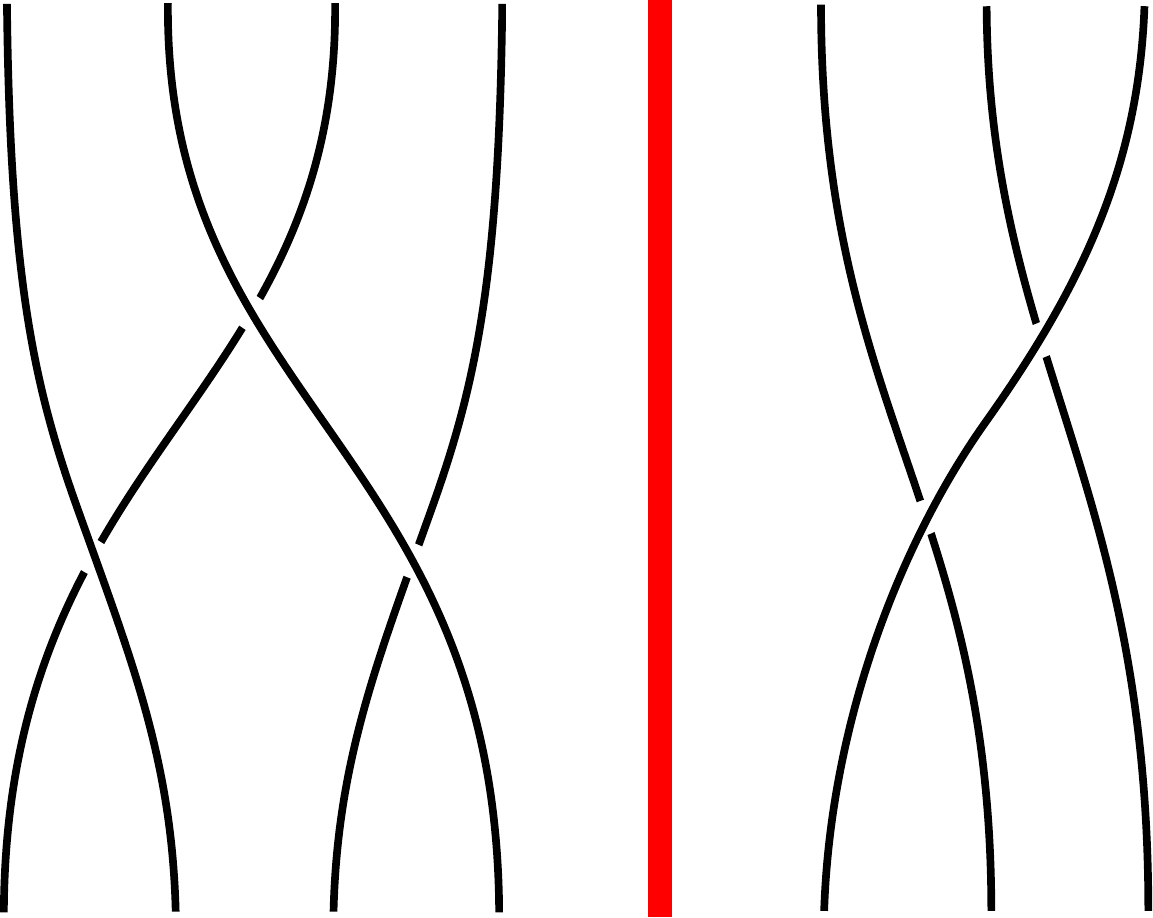}
      \captionsetup{width=.73\textwidth}
      \caption{A generator of $\mathbb{C}\{S_4\times S_3\} \subset B_{4,3}$.}
      \label{fig:Brauersub}
    \end{subfigure}
    \label{fig:Brauer}
\end{figure}
\vspace{-0.5cm}

\noindent which are essentially tensor network diagrams with the addition of a wall between the first $n-k$ and last $k$ systems.  On either side of the wall, lines can propagate from the northern to the southern side as long as they stay on the same side of the wall, representing permutations within the $n-k$ or $k$ systems.  The only lines crossing the wall are northern or southern arcs, because any permutation that crosses the wall will be followed by the partial transposition on the $k$ systems and become an arc. 

In this paper, we will primarily be interested in the less studied case $d < n$, which is the regime of interest for port-based teleportation. Our techniques work regardless of the relationship between $n$ and $d$, however.

\subsection{Twisted Schur-Weyl dualities\label{sec:dualities}}

We will now survey Schur-Weyl duality, which will play a central role in making our task of diagonalizing the PBT operator $\rho$ using representation theory viable. 

Suppose we have a vector space $L$.  We define the commutant of an algebra $A\subseteq \text{End}(L)\equiv \text{Hom}(L,L)$ as the set of all linear maps that commute with all elements of $A$:
\begin{align}
    \text{comm}(A)\equiv \text{End}_A(L) = \{b\in \text{End}(L):[b,a]=0,\forall a\in A \},
\end{align}
which is always a subalgebra of $\text{End}(L)$.

The duality theorem states the following~\cite{Goodman98}. Suppose we have two algebras $A,B\subseteq\text{End}(L)$ that are commutants of each other, i.e., $A=\text{comm}(B)$ and $B=\text{comm}(A)$.

$L$ decomposes into irreps of $A$ as
\begin{align}
    L \overset{A}{\cong} \bigoplus_{\lambda\in \hat{A}} \text{Hom}_A(F_\lambda, L) \otimes F_\lambda
\end{align}
where $F_\lambda$ is the irrep space over $A$ labeled by $\lambda$, and $\text{Hom}_A(F_\lambda, L)$ is the multiplicity space of $F_\lambda$ or equivalently the set of linear maps from $F_\lambda$ to $L$ that are invariant under the canonical action of A.  That is,
\begin{align}
    \text{Hom}_A(L_1,L_2)\equiv \{M\in\text{Hom}(L_1,L_2): R_2(a)MR_1(a)^{-1}=M, \forall a\in A\}.
\end{align}
Then, $\text{Hom}_A(F_\lambda, L)$ are in fact pairwise inequivalent irreps of $B$, which can be dubbed $E_\lambda$. The decomposition then becomes
\begin{align}
    L \overset{AB}{\cong} \bigoplus_{\lambda\in \hat{A}, \hat{B}} E_\lambda \otimes F_\lambda
\end{align}
where $\overset{AB}{\cong}$ indicates isomorphism up to a unitary transformation that respects the structure of the algebra $AB$.

There are two remarkable features of this duality we will take advantage of. 
 First, irreps of $A$ are paired with irreps of $B$ in a unique way, which means finding the irreps of one algebra is equivalent to finding the irreps of the other. Second, each $E_\lambda \otimes F_\lambda$ irrep of the algebra $AB$ is multiplicity-free, i.e., has a multiplicity of $1$ or $0$. As we will see in this paper, it is this multiplicity-free decomposition of the product of mutual commutants that will result in the diagonalization of $\rho$.

The relevant reductive dual pairs in our paper are~\cite{Goodman98,Doty03,Benkart94,Donkin14}
\begin{align}
    \mathbb{C} V[S(n)]=\text{comm}\left(\mathbb{C} Q[U(d)]\right)&\quad
    \mathbb{C} Q[U(d)]=\text{comm}\left(\mathbb{C} V[S(n)]\right)\\
    A^{t_n}_n(d)=\text{comm}\left(\mathbb{C} \chi[U(d)]\right)&\quad
    \mathbb{C} \chi[U(d)]=\text{comm}\left(A^{t_n}_n(d)\right)\\
    A^{(k)}_n(d)=\text{comm}\left(\mathbb{C} \chi^{(k)}[U(d)]\right)&\quad
    \mathbb{C} \chi^{(k)}[U(d)]=\text{comm}\left(A^{(k)}_n(d)\right).
\end{align}
Here, the first line captures standard Schur-Weyl duality, which concerns the algebras over the natural representations $\left(V,(\mathbb{C}^d)^{\otimes n}\right)$ of the symmetric group $S(n)$ and $\left(Q,(\mathbb{C}^d)^{\otimes n}\right)$ of the unitary group $U(d)$.  The defining basis is the standard basis $\{\ket{i_1} \otimes ... \otimes \ket{i_n}\}$ of $n$ qudits: for any $\sigma\in S(n)$ and any $U\in U(d)$
\begin{align}
    V(\sigma)(\ket{i_1} \otimes ... \otimes \ket{i_n}) &= \ket{i_{\sigma^{-1}(1)}} \otimes ... \otimes \ket{i_{\sigma^{-1}(1)}}\\
    Q(U) \left(\ket{i_1} \otimes ... \otimes \ket{i_n}\right) &= U^{\otimes n} \left(\ket{i_1} \otimes ... \otimes \ket{i_n}\right) = U \ket{i_1} \otimes ... \otimes U \ket{i_n}.
\end{align}
Notice that it can be easily seen that $\mathbb{C} V[S(n)]$ and $\mathbb{C} Q[U(d)]$ commute as
\begin{equation}
    Q(U) V(\sigma)(\ket{i_1} \otimes ... \otimes \ket{i_n}) = U \ket{i_{\sigma^{-1}(1)}} \otimes ... \otimes U \ket{i_{\sigma^{-1}(1)}} = V(\sigma) Q(U) (\ket{i_1} \otimes ... \otimes \ket{i_n})
\end{equation}
for all $\sigma\in S(n)$ and $U\in U(d)$.  This is a necessary but not sufficient condition for them to maximally centralize each other.  

We refer to the other reductive dual pairs, which involve the algebras of partially transposed permutation operators, $A^{t_n}_n(d)$ and $A^{(k)}_n(d)$, as \textit{twisted Schur-Weyl duality}.
Their commutants are given by the algebras over the partially conjugated representations $\left(\chi,(\mathbb{C}^d)^{\otimes n}\right)$ and $\left(\chi^{(k)},(\mathbb{C}^d)^{\otimes n}\right)$ of $U(d)$:
\begin{align}
    \chi(U)\left(\ket{i_1} \otimes ... \otimes \ket{i_n}\right) &= U^{\otimes n-1}\otimes U^* \left(\ket{i_1} \otimes ... \otimes \ket{i_n}\right) \\&=  U\ket{i_1} \otimes ... \otimes U \ket{i_{n-1}} \otimes U^* \ket{i_n}\\
    \chi^{(k)}(U)\left(\ket{i_1} \otimes ... \otimes \ket{i_n}\right) &= U^{\otimes n-k}\otimes U^{*\otimes k} \left(\ket{i_1} \otimes ... \otimes \ket{i_n}\right) \\&=  U\ket{i_1} \otimes ... \otimes U \ket{i_{n-k}} \otimes U^* \ket{i_{n-k+1}} \otimes...\otimes U^* \ket{i_n}
\end{align}
where each $U^*$ is just a complex conjugation of the matrix $U$.
Each of these representations is faithful because they preserve products of group multiplications and the identity.
It can easily be seen from the diagrams in section~\ref{sec:A_defs} that $A^{(k)}_n(d)$ and $\mathbb{C}\chi^{(k)}[U(d)]$ commute. 

In the next subsection, we will gain experience in decomposing dual pairs into irreps by decomposing $V[S(n)]Q[U(d)]$, via a unitary transformation---the Schur transform. Following that warmup, Sec.~\ref{sec:reptheory} will cover a systematic and concrete analysis of the irrep structure of $A^{t_n}_n(d)\chi[U(d)]$. 

\subsection{The Schur transform\label{sec:sch}}
\subsubsection{Defining the Schur basis and $U_{\text{Sch}}$}
Let's look at a simple case of dual pair decomposition---the decomposition of the representation $V[S(n)]Q[U(d)]$ of the direct product of groups $S(n)\times U(d)$ into irreps. The unitary change of basis that accomplishes this is the so-called Schur transform.

Applying the duality theorem, since $\mathbb{C}V[S(n)],\mathbb{C}Q[U(d)]\subseteq \text{End}((\mathbb{C}^d)^{\otimes n})$, we can decompose the representation space $(\mathbb{C}^d)^{\otimes n}$ of $\mathbb{C}V$ and $\mathbb{C}Q$ into a direct sum over irreps of $(\mathbb{C}V)(\mathbb{C}Q)$, or equivalently, of $VQ$, as
\begin{equation}
    (\mathbb{C}^d)^{\otimes n} \overset{S(n) \times U(d)}{\cong} \bigoplus_{\lambda \vdash n,\ h(\lambda) \leqslant d} \mathcal{Q}^d_{\lambda} \otimes \mathcal{P}_\lambda.
    \label{eq:schur}
\end{equation}
Here the irreps of $VQ$ are multiplicity-free, and are each labeled by a \textit{partition} $\lambda$ of an integer $n$, denoted $\lambda \vdash n$, which a non-increasing sequence of integers that sum to $n$.  That is,
\begin{equation}
    \lambda = (\lambda_1, \lambda_2, ... , \lambda_l) \text{ where } \lambda_1 \geqslant \lambda_2 \geqslant ... \geqslant \lambda_l \geqslant 0
\end{equation}
A partition can be represented as a \textit{Young diagram}, an array of boxes where the $i$th row contains $\lambda_i$ boxes. For example, $\lambda = (3,1)$ would be represented as 
\begin{equation}
    \yng(3,1)
\end{equation}
The height of the Young diagram is called $h(\lambda)$;  here $h(\lambda)=2$.

The irreps of both $S(n)$ and $U(d)$ are labeled by Young diagrams, and are called \textit{Specht modules} $\mathcal{P}_\lambda$ and the \textit{Weyl modules} $\mathcal{Q}^d_{\lambda}$, respectively. 
In this paper, we will consistently use $\alpha$ to denote a Young diagram with $n-2$ boxes, i.e., $\alpha \vdash n-2$. The expression $\alpha+\Box$ is used to mean adding a box to $\alpha$ in all legal ways such that the resulting diagram still represents a partition---we will use this to relate irreps of e.g. $S(n-2)$ to irreps of $S(n-1)$. Similarly, $\nu-\Box$ denotes removing a box.

We can make this abstract decomposition into vector spaces concrete by looking at the associated basis transformation. By construction, the left-hand side of Eq.~\eqref{eq:schur} is in the defining basis $\{\ket{i_1} \otimes ... \otimes \ket{i_n}\}$ of the representation $VQ$, while the right-hand side is implicitly in the \textit{canonical} (see next subsection) basis $\{\ket{\lambda} \ket{q_\lambda^d} \ket{p_\lambda}^{(n)}_{\text{Sch}}\}$ that simultaneously decomposes the action of $V(\sigma)$ and $Q(U)$ into irreps, which we will call the \textit{Schur basis}. For $\sigma\in S(n)$ and $U\in U(d)$, $V(\sigma)Q(U)$ acts on a Schur basis element as
\begin{align}
    V(\sigma)Q(U) \left(\ket{\lambda} \ket{q_\lambda^d} \ket{p_\lambda}^{(n)}_{\text{Sch}}\right) =  Q(U)V(\sigma)  \left(\ket{\lambda} \ket{q_\lambda^d} \ket{p_\lambda}^{(n)}_{\text{Sch}}\right) = \ket{\lambda} \Big(q^d_\lambda [U] \ket{q_\lambda^d} \Big) \Big( p_\lambda [\sigma] \ket{p_\lambda} \Big)^{(n)}_{\text{Sch}}.
\end{align}
Here, $q^d_\lambda$ is the matrix form of the $U(d)$ irrep labeled by $\lambda$, and $p_\lambda$ is the matrix form of the $S(n)$ irrep also labeled by $\lambda$. Notice that since the irreps
$q^d_\lambda$ act only on the multiplicity spaces of the irreps $p_\lambda$, and vice versa, $q^d_\lambda \otimes p_\lambda$ are the multiplicity-free irreps of $VQ$.

The unitary transformation that implements the isomorphism between $(\mathbb{C}^d)^{\otimes n}$ and the right-hand side of Eq. \eqref{eq:schur}, that is, between the computational basis and the Schur basis, is known as the Schur transform $U_{\text{Sch}}$.  Explicitly, 
\begin{equation}
    U_{\text{Sch}} V(\sigma)Q(U) U_{\text{Sch}}^{\dagger} = \sum_{\lambda \vdash n,\ h(\lambda) \leqslant d} \ket{\lambda} \bra{\lambda} \otimes q^d_\lambda [U] \otimes p_\lambda [\sigma].
\end{equation}

\subsubsection{Subgroup-reduced Schur basis}

Perhaps unsurprisingly, because of the close relationship between the algebra of permutation operators $\mathbb{C}V[S(n)]$ and the algebra of partially transposed permutations $A^{t_n}_d(d)$, our algorithm for port-based teleportation will use $U_{\text{Sch}}$ as a subroutine. 
 
Mathematically, this is because the Schur basis is subgroup-reduced. In particular, what matters for our purposes is the symmetric group portion. Each symmetric group irrep $p_\lambda$ that appears in the Schur basis is subgroup-reduced with respect to the subgroup tower $S(n) \supset S(n-1) \supset S(n-2) \supset ... \supset S(1)$. 
 The partial transposition on the $n$th qudit only affects the highest two levels of the tower, as we will see more explicitly later in this paper.  As a result, the basis that simultaneously decomposes $A^{t_n}_n(d)$ and $\mathbb{C}\chi [U(d)]$ into irreps will be closely related to the Schur bases on these levels, and the unitary that transforms to this basis will depend on the corresponding Schur transforms.
 
The Schur basis is also subgroup-reduced in another way: the $U(d)$ irreps $q^d_\lambda$ are reduced with respect to the tower $U(d) \supset U(d-1) \supset ... \supset U(1)$, where $U(i)$ is the unitary group acting on the $i$-dimensional minor of the full $d$-dimensional matrix.  Being subgroup-reduced down both of these towers fixes the Schur basis up to a complex phase so that it is a canonical basis.
This essentially results from the multiplicity-free irrep-branching of $S(n)$ and $U(d)$ down these two towers. Looking at $S(n)$ as an example, we can decompose any irrep $\lambda_i\vdash i,h(\lambda_i)\leqslant d$ of $S(i)$ as
\begin{equation}
    p_{\lambda_i} [\sigma_{i-1}] \overset{S(i-1)}{=} \bigoplus_{\lambda_{i-1} = \lambda_i - \Box \atop \lambda_{i-1}\vdash i-1, h(\lambda_{i-1})\leqslant d} p_{\lambda_{i-1}} [\sigma_{i-1}] \quad \forall  \sigma_{i-1}\in S(i-1),
\end{equation}
where each $\lambda_{i-1}$ appears only once. This means that we can unambigiously label the basis elements that $S(n)$ acts on by the irreps under which they transform at each level, i.e. by a sequence of one-box removals. As an example, for $n=5$ and an irrep $p_\lambda$ with $\lambda = (2,2,1)$, one basis element of $p_\lambda$ would be
\begin{equation}
    \yng(2,2,1) \rightarrow \yng(2,1,1) \rightarrow \yng(2,1) \rightarrow \yng(1,1) \rightarrow \yng(1).
\end{equation}
This means that if we restricted to look at $p_{\lambda} [\sigma_{4}]$ for $\sigma_{4} \in S(4)$, this basis element would be part of the $\lambda_{4} = (2,1,1)$ block rather than the $(2,2)$ block, which would correspond to a different single-block removal from $(2,2,1)$. Likewise, for $p_\lambda[\sigma_{3}], \sigma_{3} \in S(3)$ it would be part of the $\lambda_{3} = (2,1)$ block within $\lambda_{4} =(2,1,1)$, and so on. By the time we get to the bottom of the tower, $S(1)$, it must transform according to the only one-dimensional irrep, the trivial representation, so that it is fully determined.

Because the basis elements of each $S(n)$ irrep $\lambda$ are labeled by sequences of Young diagrams, the Schur basis register $\ket{p_\lambda}$, which holds these basis elements, can be split up into registers $\ket{\lambda_{n-1}}, \ket{\lambda_{n-2}} \dots,\ket{\lambda_1}$, where $\lambda_{i-1} = \lambda_i - \Box$.  In this way, the subgroup-reduced property of the basis is made manifest: for any $\sigma_k\in S(k)$ permuting the first $k$ qudits
\begin{align}
V(\sigma_k)  \ket{\lambda} \ket{q_\lambda^d} \Big(\ket{\lambda_{n-1}} \dots \ket{\lambda_k} \dots \ket{\lambda_1}\Big)^{(n)}_{\text{Sch}} &=  \ket{\lambda_k}\Big(\ket{q_\lambda^d}\ket{\lambda} \ket{\lambda_{n-1}} \dots\ket{\lambda_{k+1}}\Big) \Big( p_{\lambda_k} [\sigma_k] \ket{\lambda_{k-1}} \dots\ket{\lambda_1} \Big)^{(n)}_{\text{Sch}}
\end{align}
where on the right-hand side we have reordered the registers, such that $\lambda_k$ labels irreps of $S(k)$, $\ket{\lambda_{k-1}} \dots\ket{\lambda_1}$ contains those irreps, and $\ket{q_\lambda^d}\ket{\lambda} \ket{\lambda_{n-1}} \dots\ket{\lambda_{k+1}}$ holds the multiplicity spaces of those irreps.

The efficient quantum algorithm in~\cite{Harrow05,Harrow06} for the Schur transform takes advantage of the subgroup-reduced nature of the Schur basis to implement $U_{\text{Sch}}$ efficiently.
The algorithm is built up from the unitary group side, and essentially exploits a duality between the canonical branching of $S(n)$ and the Clebsch-Gordan transforms of $U(d)$. See Appendix~\ref{app:complexity_Sch} or the original papers for more details.

\subsection{The subalgebra-reduced twisted Schur transform \label{sec:genshc_intro}}

Ultimately, the operation to diagonalize the PBT operator $\rho$ will bear a close resemblance to the Schur transform. Let's collect the facts we have learned so far:
\begin{enumerate}
    \item $\rho$ has both $[\rho, \chi[U(d)]]=0$ and $[\rho, V[S(n-1)]]=0$ symmetries [Sec.~\ref{sec:rho_sym}]
    \item $\rho$ is an element of $A^{t_n}_n(d)$ [Sec.~\ref{sec:A_defs}]
    \item $\mathbb{C}\chi[U(d)]$ and $A^{t_n}_n(d)$ are commutants of each other [Sec.~\ref{sec:dualities}]
\end{enumerate}
 A quick consistency check is that, since $\rho \in A^{t_n}_n(d)$, it should commute with the entire commutant $\mathbb{C}\chi[U(d)]$.  This is manifestly true because of its $\chi[U(d)]$ symmetry; the additional symmetry it enjoys means that it also commutes with the subalgebra $\mathbb{C}V[S(n-1)] \subset A^{t_n}_n(d)$.

Equipped with the duality theorem from Sec.~\ref{sec:dualities} we know that we can perform a unitary change of basis from the computational basis of $(\mathbb{C}^d)^{\otimes n}$ to one that decomposes the product of commutants $A^{t_n}_n(d)\mathbb{C}\chi[U(d)]$ into irreps.  In analogy to the Schur transform in Sec.~\ref{sec:sch}, we will call this the \textit{twisted Schur transform} $U_{\text{twSch}}$.  As $\rho$ is an element of $A^{t_n}_n (d)$, performing the twisted Schur transform will block-diagonalize it.  

However, we can do even more.  The key is that we have the freedom to choose this basis such that the irreps of $A^{t_n}_n(d)$ that appear in it are reduced with respect to the two-layer subalgebra tower $A^{t_n}_n(d)\supset \mathbb{C} V[S(n-1)]$.  
By subalgebra-reduced, we mean that any irrep of $A^{t_n}_n(d)$ will be in a basis such that, when it is restricted to the subalgebra $\mathbb{C} V[S(n-1)]$, it is reducible and is block-diagonalized into a direct sum of irreps of $\mathbb{C} V[S(n-1)]$.  As we will see, for the irrep of $A^{t_n}_n (d)$ labeled by $\alpha \vdash n-2$, we will have the decomposition
\begin{align}
     \label{eq:subalg}M^{\alpha}[\sigma] \overset{S(n-1)}\cong \bigoplus_{\nu = \alpha+\Box, h(\nu) \leqslant d} p_{\nu}[\sigma], \quad \forall \sigma \in S(n-1).
\end{align}
where $M^{\alpha}[\sigma]$ denotes the matrix elements of $\sigma$ in $\alpha$.

If we construct such a subalgebra-reduced irrep-decomposing basis of $A^{t_n}_n(d)\mathbb{C}\chi[U(d)]$, it will also block-diagonalize $\chi[U(d)]V[S(n-1)]$ into irreps. Because $\rho$ is $S(n-1)$-symmetric, this means it will additionally be block-diagonalized in the multiplicity spaces of the $S(n-1)$ irreps within each $A^{t_n}_n (d)$ irrep.  But, as we have previewed in Eq.~\eqref{eq:subalg} and will see more explicitly in the next section, these multiplicity spaces are \textit{trivial} as a consequence of twisted Schur-Weyl duality.  

This means that by carrying out this twisted Schur transform, that is, by transforming to an irrep-decomposing basis of $A^{t_n}_n(d)\mathbb{C}\chi[U(d)]$ which is reduced on the subalgebra $\mathbb{C} V[S(n-1)]$ of $A^{t_n}_n(d)$, we can \textit{diagonalize} $\rho$, and easily evaluate the matrix power $\rho^{-1/2}$. This basis will also decompose each $\rho_i\in A^{t_n}_n(d)$, the other factor of each $\Pi_i$, into irreps.  As we will see in Sec.~\ref{sec:circuit}, transforming to this basis will also make it possible to evaluate $\sqrt{\Pi_i}$, solving our other problem: each $\Pi_i$ will take the form of a rank-deficient (pseudo)projector in each irrep, meaning that taking the square root can be done by scaling by its eigenvalues in each irrep, so no additional basis transformation is needed.

\medskip
\noindent A rough sketch of our algorithm is then:
\begin{enumerate}[itemsep=1.7pt,topsep=2pt]
\item Transform to the irrep/eigenbasis of $\rho$ using $U_{\text{twSch}}$.
\item Execute $\sqrt{\Pi_i}$ in each irrep block in this basis.
\item Transform the basis back using $U^\dagger_{\text{twSch}}$.
\end{enumerate}

%% file: sections/reptheory.tex
\section{Representation theory of $A_n^{t_n}(d)$}
\label{sec:reptheory}

\subsection{Identifying the irreps of $A_n^{t_n}(d)$\label{sec:characterize}}

This subsection will characterize the irrep spaces of $A_n^{t_n}(d)$ and its commutant $\mathbb{C}\chi[U(d)]$ by the relevant Specht and Weyl modules, in close analogy to the case of $V[S(n)]Q[U(d)]$ in Eq.~\eqref{eq:schur}. In particular, since $\chi[U(d)]$ is a faithful representation of the unitary group $U(d)$, we can use the Littlewood-Richardson branching rule of $U(d)$ to decompose the Hilbert space $\mathcal{H}=(\mathbb{C}^d)^{\otimes n}$ by the action of $\chi[U(d)]$~\cite{Walter21, Leditzky20}. After performing this decomposition, we will be able to identify the irreps of $A_n^{t_n}(d)$ as the multiplicity spaces of $\chi[U(d)]$ irreps.

Because $\chi[U(d)]$ is complex-conjugated on the $n$th system, let's first separate the last qudit, and decompose the first $n-1$ qudits using the Schur-Weyl duality, as in Appendix A of \cite{Walter21}. That is, decomposing $\mathcal{H}$ by the action of $V(\sigma) U_1^{\otimes n-1} \otimes U_2^*$, which is a representation of $S(n-1)\times U(d)\times U(d)$, we have
\begin{align}
    (\mathbb{C}^d)^{\otimes n} \overset{S(n-1)\times U(d)\times U(d)}{\cong} \left(\bigoplus_{\mu \vdash n-1,\ h(\mu) \leqslant d} \mathcal{P}_\mu \otimes \mathcal{Q}_{\mu}^d\right) \otimes (\mathbb{C}^d)^*,
\end{align}
where $(\mathbb{C}^d)^*$ is the dual representation of $U(d)$. Then, we can use the (dual) Pieri rule to break up the representations $\mathcal{Q}_{\mu}^d \otimes (\mathbb{C}^d)^*$ of $U(d)$ into irreps, which gives
\begin{align}
    \mathcal{Q}_{\mu}^d \otimes (\mathbb{C}^d)^* \overset{U(d)}{\cong} \bigoplus_{i\in[d], \mu_i > \mu_{i+1}} \mathcal{Q}_{\mu - e_i}^d,
\end{align}
where $e_i$ is the $i$-th standard basis vector, so that the diagrams $\mu - e_i$ have one less box than $\mu$ in the $i$-th row, and we have set $\mu_{d+1}=-\infty$.  The diagrams $\mu - e_i$ may also differ from valid Young diagrams because a box can be removed from an empty row.  Note that we are now decomposing by the action of $\chi[U] = U^{\otimes n-1} \otimes U^*$, rather than $U_1^{\otimes n-1} \otimes U_2^*$.  Combining these two steps, we find a decomposition that respects the structure of $V \chi$:
\begin{align}
    \label{eq:generalized_schur_0_orig}(\mathbb{C}^d)^{\otimes n} \overset{V\chi \atop S(n-1)\times U(d)}{\cong} \bigoplus_{\mu \vdash n-1,\ h(\mu) \leqslant d} \bigoplus_{i\in[d], \mu_i > \mu_{i+1}}  \mathcal{P}_\mu \otimes \mathcal{Q}_{\mu - e_i}^d.
\end{align}

This expression can be rewritten as
\begin{align}
    \label{eq:generalized_schur_0}(\mathbb{C}^d)^{\otimes n} \overset{V\chi \atop S(n-1)\times U(d)}{\cong} \mathcal{H}_S \oplus \mathcal{H}_M,
\end{align}
where
\begin{align}
    \label{eq:generalized_schur_1}\mathcal{H}_S &= \bigoplus_{\alpha \vdash n-1,\ h(\alpha) \leqslant d-1} \textcolor{red}{\left(\mathcal{P}_{\alpha^0}\right)} \otimes \mathcal{Q}_{\alpha^-}^d\\
    \label{eq:generalized_schur_2}\mathcal{H}_M &= \bigoplus_{\alpha \vdash n-2,\ h(\alpha) \leqslant d} \textcolor{red}{\left( \bigoplus_{\mu = \alpha+\Box, h(\mu) \leqslant d} \mathcal{P}_{\mu} \right)} \otimes \mathcal{Q}_{\alpha}^d.
\end{align}
Here $\alpha$ are all valid Young diagrams, with $\alpha^- = (\alpha_1,\dots,\alpha_{d-1},-1)$ and $\alpha^0 = (\alpha_1,\dots,\alpha_{d-1},0)$. We have separated $\mathcal{H}$ into $\mathcal{H}_M$ and $\mathcal{H}_S$ because, as we will show in the next subsection, only $\mathcal{H}_M$ is relevant to port-based teleportation.  This is particularly convenient because all of the labels appearing in $\mathcal{H}_M$ are valid Young diagrams.  
Note that, in contrast to the Weyl modules in Eq.~\eqref{eq:generalized_schur_0_orig}, which may share the same diagram label, the Weyl modules appearing in Eqs. \eqref{eq:generalized_schur_1} and \eqref{eq:generalized_schur_2} are labeled by distinct $\alpha$, with their multiplicity fully captured by the direct sum of Specht modules $\mathcal{P}_*$.

Now we can use the fact that the algebras $A_n^{t_n}(d)$ and $\mathbb{C}\chi[U(d)]$ are mutual commutants.  Because the $U(d)$ irreps $\mathcal{Q}^d_*$ appearing in  Eqs.~\eqref{eq:generalized_schur_1} and~\eqref{eq:generalized_schur_2} decompose the representation $\mathbb{C}\chi(U)$, the vector spaces shown in parenthesis and highlighted in red that they are matched with, $\mathcal{P}_{\alpha^0}$ and $\left( \bigoplus_{\mu = \alpha+\Box, h(\mu) \leqslant d} \mathcal{P}_{\mu} \right)$, must be irreps of $A_n^{t_n}(d)$.
Notice that we have found the irreps of $A_n^{t_n}(d)$ in a basis that is subalgebra-reduced with respect to the tower $A_n^{t_n}(d)\supset \mathbb{C}[S(n-1)]$, as desired; this is also indicated by the isomorphism with respect to $V[S(n-1)]\chi[U(d)]$ in Eq.~\eqref{eq:generalized_schur_0}.

Moreover, the decomposition of the irreps of $A_n^{t_n}(d)$ into irreps of $S(n-1)$ is \textit{multiplicity-free}.  This happens as a consequence of the twisted Schur-Weyl duality theorem from Sec.~\ref{sec:dualities}: each irrep of $A_n^{t_n}(d)$ in red is uniquely paired with one $\mathcal{Q}_\alpha^d$ and so can be labeled by $\alpha$, but for a fixed $\alpha$, all $\mu=\alpha+\Box$ are pairwise different.  This is the key feature of Eq.~\eqref{eq:generalized_schur_2} that will make our algorithm possible.  Because $\rho$ has a $\chi[U(d)]$ symmetry, it will be block-diagonalized in the multiplicity spaces of the $\chi[U(d)]$ irreps, i.e. in the irreps of $A_n^{t_n}(d)$.  Because it has an $S(n-1)$ symmetry, it will be further block-diagonalized in the multiplicity spaces of the $S(n-1)$ irreps within each irrep of $A_n^{t_n}(d)$.  But these multiplicity spaces are trivial, so $\rho$ will actually be automatically \textit{diagonalized} in a basis that respects the decomposition Eq.~\eqref{eq:generalized_schur_0}.

\subsection{The ideal $\mathcal{H}_M$\label{sec:identify}}
Toward the end of constructing a concrete basis of $(\mathbb{C}^d)^{\otimes n}$ that decomposes the algebra $A_n^{t_n}(d)$ and its commutant $\mathbb{C}\chi[U(d)]$ into irreps as in Eq.~\eqref{eq:generalized_schur_0}, we will first show that we can focus our attention on the subrepresentation $\mathcal{H}_M$ of $(\mathbb{C}^d)^{\otimes n}$.

It turns out~\cite{Koike89} that the subreps $\mathcal{H}_S$ and $\mathcal{H}_M$ are the defining representation spaces of two ideals $M$ and $S$ that make up the algebra $A_n^{t_n}(d)$.  That is, $M=\text{End}_{\mathbb{C}\chi(U)}(\mathcal{H}_M)$ and $S=\text{End}_{\mathbb{C}\chi(U)}(\mathcal{H}_S)$, or, in other words, $\mathcal{H}_M$ is the support of the ideal $M$ and the null space of the ideal $S$, and vice versa for $\mathcal{H}_S$.

The ideals $M$ and $S$ are constructed using the idempotents of the algebra $A_n^{t_n}(d)$ and its semisimplicity, in the following way. Following the notation in \cite{Studzinski13,Studzinski14, Studzinski18}, let $\sigma_{(a,b)}$ denote any permutation operator $\sigma\in S(n)$ with $\sigma(a)=n$ and $\sigma(n)=b$. When the $n$th object is fixed, $a = b = n$ and we use the abbreviation $\sigma_n\equiv\sigma_{(n,n)}\in S(n-1)\subset S(n)$.
Then, an obvious decomposition of $A_n^{t_n}(d)$ is
\begin{align} \label{eq:decom1}
A_n^{t_n}(d)&=\mathbb{C}\{V[\sigma_{(a,b)}]^{t_n}:a,b\neq n\}+\mathbb{C}\{V[\sigma_{n}]^{t_n}=V[\sigma_{n}]:\sigma_n\in S(n-1)\}\\
&=\mathbb{C}\{V[\sigma_{(a,b)}]^{t_n}:a,b\neq n\}+\mathbb{C}V[S(n-1)]
\end{align}
since $\{V[\sigma]^{t_n}:\sigma\in S(n)\}$ are the natural generators of the algebra. By writing out the multiplication rule of these generators [Theorem 19 in~\cite{Studzinski14}], it can be seen that the first term Eq. (\ref{eq:decom1}), which we call $M$, forms a (two-sided) ideal of $A_n^{t_n}(d)$, while the second term, $\mathbb{C}V[S(n-1)]$, does not. Because the ideal $M$ is semisimple, it can be written as a projection of $A_n^{t_n}(d)$ as $M=e A_n^{t_n}(d) e$, where $e$ is the unit element of $M$ and an idempotent of the algebra. Following from the basic properties of idempotents, $\mathbb{I} - e$ will be an idempotent as well, so we can use it to construct another ideal $S=(\mathbb{I} - e) A_n^{t_n}(d)(\mathbb{I} - e)$ such that
\begin{align}
A_n^{t_n}(d)=e A_n^{t_n}(d) e\oplus (\mathbb{I} - e) A_n^{t_n}(d)(\mathbb{I} - e)\equiv M\oplus S
\end{align}
and $MS=0$. The ideal $S$ will be generated by $\{V[\sigma_n](\mathbb{I} - e): \sigma_n\in S(n-1)\}$.

Note that the objects of interest in port-based teleportation, $\rho$ and $\rho_i$ ($i=1,\dots,n-1$), all live in the ideal $M$, because $M$ is linearly generated by the elements in $A_n^{t_n}(d)$ that permute and transpose the last system $n$. That is, they will vanish outside the rep space $\mathcal{H}_M$, i.e., in $\mathcal{H}_S$.  So in this paper, it will be sufficient to decompose the vector space $\mathcal{H}_M$ into subalgebra-reduced irreps of $A_n^{t_n}(d)$, and find the matrix forms of elements of $A_n^{t_n}(d)$, including $\rho$ and $\rho_i$, in these irreps.

\subsection{Spanning vectors of $\mathcal{H}_M$}

The first step to constructing a basis that decomposes $\mathcal{H}_M$ into subalgebra-reduced irreps of $A_n^{t_n}(d)$ in $\mathcal{H}=(\mathbb{C}^d)^{\otimes n}$ is to find a spanning set of vectors for $\mathcal{H}_M$.
Observe that each generator $V[\sigma_{(a,b)}]^{t_n}$ of $M$ can be represented as follows \cite{Studzinski13},
\begin{align}
V[\sigma_{(a,b)}]^{t_n}&=\sum_{i_1,\dots,i_n} |i_{\sigma^{-1}(1)},\dots,i_{\sigma^{-1}(b)=n},\dots,i_{\sigma^{-1}(n)=a}\rangle \langle i_1,\dots,i_a,\dots,i_n|^{t_n}\\
&=\sum_{i_1,\dots,i_n} |i_{\sigma^{-1}(1)},\dots,i_{\sigma^{-1}(b)=n},\dots,i_n\rangle \langle i_1,\dots,i_a,\dots,i_{\sigma^{-1}(n)=a}|\\
&= \sum_{i_1,\dots,i_n} \ket{i_{\sigma^{-1}(1)},\dots,i_{\sigma^{-1}(n-1)}}_{\overline{bn}}\ket{i_n, i_n}_{bn} \bra{i_1,\dots,i_{n-1}}_{\overline{an}} \bra{i_a, i_a}_{a n} \\
&=\sum_{\{|\phi_i\rangle \in (\mathbb{C}^{d})^{\otimes n-2}\}} d|\phi_{i} \rangle_{\overline{bn}}  |\phi_+\rangle_{bn} \langle \sigma_{(a,b)}(\phi_i)|_{\overline{an}} \langle \phi_+ |_{an}.\label{eq:gen1}
\end{align}
Here $\overline{bn}$ ($\overline{an})$ refers to all of the systems being acted on by the permutation except $b$ or $n$ ($a$ or $n$), and the sum runs through $\{|\phi_i\rangle \in (\mathbb{C}^{d})^{\otimes n-2}\}$, an orthonormal basis defined on these $n-2$ systems.  As usual, $|\phi_+\rangle = 1/\sqrt{d}\sum_{i}|ii\rangle$ is the maximally entangled state. Additionally, $\ket{\sigma_{(a,b)}(\phi_i)}$ permutes the subsystems containing $\phi_i$ according to the permutation rule of $\sigma_{(a,b)}$. As a result, we can recognize the rep space $\mathcal{H}_M$ of $M$, as
\begin{align}
    \mathcal{H}_M=\mathbb{C}\{|\phi_i\rangle |\phi_+\rangle _{kn}:k=1,\dots,n-1,i=1,\dots,d^{n-2}\}.
\end{align}
Alternatively, this can be quickly seen from a walled diagram of an arbitrary generator $V[\sigma_{(a,b)}]^{t_n}$ of the ideal $M$, as introduced in Sec.~\ref{sec:A_defs}: 

\begin{figure}[H]
    \centering
      \includegraphics[width=0.18\textwidth]{figures/rep_space_3.pdf}
      \label{fig:rep_space}
\end{figure}
\vspace{-0.1cm}
 \noindent 
 Here the single northern and southern arc originate from the maximally entangled state, while the permutation on the basis $\{\phi_i \}$ is captured in the other lines.
 Since the support of a maximally entangled state is only the maximally entangled state, and since the algebra of permutation operators on the remaining $n-2$ qudits is supported on the whole space $(\mathbb{C}^{d})^{\otimes n-2}$, the span of $ \mathcal{H}_M$ has been verified.

Having found the vector space $\mathcal{H}_M$, we can now identify the irrep spaces of $A_n^{t_n}(d)$ in $\mathcal{H}_M$. Again, this can be most readily done by recognizing each of them as a vector space that is paired with an irrep of the commutant, $\mathbb{C}\chi[U(d)]$, of $A_n^{t_n}(d)$. Notice that, for any $U\in U(d)$, $\chi[U]$ acts from the left on $\mathcal{H}_M$ as

\begin{figure}[H]
    \centering
      \includegraphics[width=0.47\textwidth]{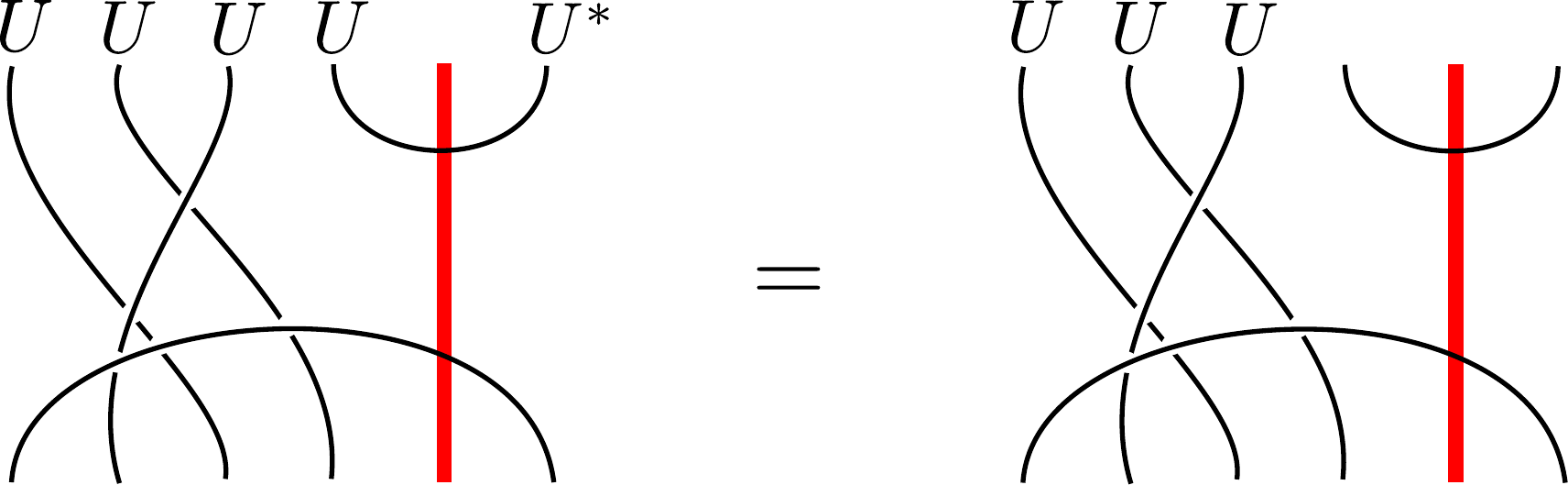}
      \label{fig:U_n}
\end{figure}
\vspace{-0.1cm}
\noindent That is, it acts as if there are only $n-1$ copies, associated with the $n-1$ maximally entangled pairs $|\phi_+\rangle_{kn}$, of $n-2$ qudits. We have already learned in Sec.~\ref{sec:sch} that the irreps of $U^{\otimes n-2}$ are the Weyl modules $\mathcal{Q}^d_{\alpha}$ labeled by Young diagrams $\alpha \vdash n-2, h(\alpha)\leqslant d$, and partition $(\mathbb{C}^{d})^{\otimes n-2}$ in the Schur basis. Thus, exchanging $\ket{\phi_i}$ for the Schur basis on $n-2$ qudits, $|\phi_{k_\alpha}(\alpha, r)\rangle\equiv \ket{\alpha, q_\alpha^d\equiv r, p_\alpha \equiv k_\alpha}_{\text{Sch}}^{(n-2)}$, and defining
\begin{align}
|\psi^k_{k_\alpha}(\alpha, r)\rangle = \sqrt{d}V[(k\ n-1)]|\phi_{k_\alpha}(\alpha, r)\rangle |\phi_+\rangle_{n-1n} \quad k=1,\dots, n-1,{k_\alpha}=1,\dots,d_\alpha, r=1,\dots, m_\alpha, 
\end{align}
the irreps of $\chi[U(d)]$ are then $\mathbb{C}\{ |\psi^k_{k_\alpha}(\alpha, r)\rangle: \forall r \}$ for all fixed tuples $(\alpha, k, {k_\alpha})$. Here $\alpha$ label pairwise inequivalent irreps, $k_\alpha$ is a multiplicity label inherited from the Schur basis, and $k$ is an additional multiplicity label because the action of $\chi[U]$ doesn't mix the copy information.

The irreps for $A_n^{t_n}(d)$ should be paired with the $\chi[U(d)]$ irreps, that is, also labeled by $\alpha$ but featuring the complementary degrees of freedom. Thus, they are
\begin{align}
    \mathcal{H}_{\alpha,r} \equiv \mathbb{C}\{ |\psi^k_{k_\alpha}(\alpha, r)\rangle: k=1\dots,n-1,\quad {k_\alpha}=1,\dots,d_\alpha \}
\end{align}
for all fixed pairs $(\alpha,r)$. That is,
\begin{align} \label{eq:H_ar_decomp}
    \mathcal{H}_M = \bigoplus_{\alpha,r} \mathcal{H}_{\alpha,r}
\end{align}
is the decomposition of $\mathcal{H}_M$ into a direct sum of $A_n^{t_n}(d)$ irrep spaces.  For a fixed $\alpha$ and varying $r$, all $\mathcal{H}_{\alpha,r}$ are isomorphic as irreps.  They are also linearly independent because the corresponding $\chi[U(d)]$ irreps are known to be Weyl modules and indeed have dimensions $m_\alpha$.

\subsection{Reducing the irreps $\mathcal{H}_{\alpha,r}$ with respect to $S(n-1)$\label{sec:induced}}

So far, we successfully divided $\mathcal{H}_M$ into the explicit $A_n^{t_n}(d)$ irreps $\mathcal{H}_{\alpha,r}$.  Putting this together with what we found in Eq.~\eqref{eq:generalized_schur_2}, we expect each $\mathcal{H}_{\alpha,r}$ to further decompose as
\begin{align}
     \label{eq:Halpha_decomp}\mathcal{H}_{\alpha,r} \overset{S(n-1)}\cong \bigoplus_{\nu = \alpha+\Box, h(\nu) \leqslant d} \mathcal{P}_{\nu}.
\end{align}
To implement this decomposition in practice, we need to transform the spanning vectors $\{ \ket{\psi^k_{k_\alpha} (\alpha,r) } \}$ of each $\mathcal{H}_{\alpha, r}$ into an orthonormal basis that decomposes it into irreps of $S(n-1)$.  

In fact, implementing this decomposition is a special case of a more general problem.  Each $\mathcal{H}_{\alpha,r}$ is an example of a \textit{generalized induced representation}, which is a representation $W$ of a group $G$ that is created by lifting a representation $V$ of a subgroup $H \subset G$.  Explicitly, we define
\begin{equation}
    \label{eq:induced_def}W = \sum_{k=1}^{[G:H]} \tau^k \cdot V
\end{equation}
where $\{ \tau^k \}$ is a complete set of representatives of the cosets of $H$ in $G$.  In a standard induced representation, the isomorphic copies $\tau^k V$ must be linearly independent, giving $W = \bigoplus_k \tau^k \cdot V$; to treat each $\mathcal{H}_{\alpha,r}$ it's necessary to generalize by relaxing this condition.  In the case of $\mathcal{H}_{\alpha,r}$, we have $G=S(n-2)$, $H = S(n-2)$, and $\phi \equiv \phi^\alpha \otimes \mathbb{I}_{n-1,n}$.  The transversals $\tau^k \equiv (k\ n-1)$ act on the base representation
\begin{align}
    V\equiv \text{span}_{\mathbb{C}}\{ |\phi_{k_\alpha}(\alpha, r) \rangle |\phi_+\rangle_{n-1n} : {k_\alpha}=1,\dots,d_\alpha \}.
\end{align}

Performing the basis transformation shown in Eq.~\eqref{eq:Halpha_decomp} is then in fact a particular instance of the general problem of block-diagonalizing generalized induced representations.  In a forthcoming companion paper, we explore this general problem and find a quantum algorithm that solves it~\cite{inducedRepPrep}.  The solution relies heavily on irreps $\psi_R^\nu$ of $G$ being in a basis that is subgroup-reduced down the tower $G\supset H$, such that every element of $H$ is block-diagonalized. 
That is,
\begin{align}
    \psi_R^\nu(h) = \bigoplus_{\xi_{\nu}\in \hat{H}(\nu)} \mathbb{I}_{m_{\xi_{\nu}}^{(\nu)}} \otimes \varphi^{\xi_{\nu}}(h)\quad \forall h\in H
\end{align}
where $\hat{H}(\nu)$ is the set of inequivalent irreps $\xi_{\nu}$ of $H$ in the irrep $\nu$ of $G$ and $m_{\xi_{\nu}}^{(\nu)}$ is the multiplicity of $\xi_{\nu}$ in $\nu$. 

However, for the particular case of $G = S(n-1)$ and $H = S(n-2)$ needed for this application, Refs.~\cite{Studzinski14,Studzinski18} have already constructed a unitary that performs the block-diagonalization by delicately manipulating projectors and sum rules for these subgroup-reduced irreps.  For an irrep $\mathcal{H}_{\alpha,r}$ labeled by a fixed $\alpha$ and any multiplicity label $r$ of $\alpha$, the resulting change of basis matrix they find is given by
\begin{align}
z_R(\alpha)^{i\xi_\nu}_{k_\alpha j_{\xi_\nu}}=\frac{1}{\sqrt{n-1}}\frac{\sqrt{d_\nu}}{\sqrt{d_\alpha}}\psi_R^\nu[(i\ n-1)]^{\alpha\xi_\nu}_{k_\alpha j_{\xi_\nu}}\quad i=1,\dots,n-1,\ \forall\nu=\alpha+\Box.
\end{align}
Here we have followed the notation in~\cite{Studzinski18}: 
$\mathcal{H}_{\alpha,r}$ is divided into $S(n-1)$ irrep blocks labeled by $\nu = \alpha + \Box$ for all $\nu$ with height less than $d$.  
Within a single fixed $\nu$ block, $\xi_\nu$ labels a reduced $S(n-2)$ irrep block, which, according to the canonical branching rule from $S(n-1)$ to $S(n-2)$, should have the form $\xi_\nu=\nu-\Box$, and $j_{\xi_\nu}$ indexes within $\xi_\nu$. Notice that $\alpha$ is one among these $\xi_\nu$ blocks, which the left-upper index of $\psi_R^\nu$ is fixed to. Additionally, we use an upper index $i$ to denote the representative $\tau=(i\ n-1)$.

The corresponding orthonormal and $S(n-1)$ irrep-decomposing basis of $\mathcal{H}_{\alpha,r}$ then has the elements
\begin{align}
\label{eq:f_vec_def}|f^{\xi_\nu}_{j_{\xi_\nu}}(\alpha, r)\rangle=\frac{1}{\sqrt{\lambda_\nu(\alpha)}}\sum_{k=1}^{n-1}\sum_{k_\alpha=1}^{d_\alpha}z_R(\alpha)_{k_\alpha j_{\xi_\nu}}^{k\  \xi_\nu}|\psi^k_{k_\alpha}(\alpha, r)\rangle\quad \nu\neq\theta,
\end{align}
where $\theta$ denotes the single possible $S(n-1)$ irrep of the form $\alpha + \Box$ with height greater than $d$ that therefore does not contribute to $\mathcal{H}_{\alpha, r}$.  
The normalization factors are the eigenvalues $\lambda_\nu(\alpha)$ of the Gram matrix of the spanning vectors $\{ \ket{\psi^k_{k_\alpha} (\alpha,r)} \}$ and are given by [Corollary 37 of \cite{Studzinski18}]
\begin{align}
    \label{eq:lambda_values}\lambda_\nu(\alpha)=(n-1)\frac{m_\nu d_\alpha}{m_\alpha d_\nu}.
\end{align}
In the above, $m_\nu$ and $m_\alpha$ are the multiplicities of the irrep $\psi^\nu_R$ in a representation $(V,(\mathbb{C}^d)^{\otimes n-1})$ of $S(n-1)$, and of $\varphi^\alpha$ in a representation $(V,(\mathbb{C}^d)^{\otimes n-2})$ of $S(n-2)$, respectively; $d_\nu$ and $d_\alpha$ are, as before, the dimensions of $\psi^\nu$ and $\varphi^\alpha$.

From Eq.~\eqref{eq:f_vec_def}, we already know that we have found an efficient subalgebra-reduced twisted Schur transform.  Each component of the $\ket{f}$ basis can be produced using permutations and submatrices of the $(n-1)$- and $(n-2)$-qudit Schur transforms: the states $\{ \ket{\psi^{k}_{k_\alpha}(\alpha,r)}\}$ are composed of permutations and $(n-2)$-qudit Schur basis states, while the matrix $z_R$ is just made up of permutations in the $(n-1)$-qudit Schur basis.  In the next section, we will find $U_{\text{twSch}}$ explicitly as a quantum circuit.

Finally, having an orthonormal basis $\{|f(\alpha, r)\rangle\}$ of each $A^{t_n}_n(d)$ irrep $\mathcal{H}_{\alpha,r}$ enables us to find the matrix form of all elements $V[\sigma]^{t_n}$ of $A^{t_n}_n(d)$ in this basis.  For a fixed $\alpha$ and arbitrary $r$, these are denoted by
\begin{equation}
M_f^\alpha[V(\sigma)^{t_n}]_{j_{\xi_\omega}j_{\xi_\nu}}^{\xi_\omega\ \xi_\nu} \equiv \langle f^{\xi_\omega}_{j_{\xi_\omega}}(\alpha, r) | V(\sigma)^{t_n} | f^{\xi_\nu}_{j_{\xi_\nu}}(\alpha, r) \rangle \quad \omega,\nu\neq\theta,\ \forall\sigma\in S(n).
\end{equation}
In fact, we only need to evaluate $M_f^\alpha$ on a minimal set of multiplicative generators of $A^{t_n}_n(d)$. This can be done in a straightforward way by calculating $\bra{\psi} V^{t_n} \ket{\psi}$ followed by a change of basis to $\ket{f}$. As a result~\cite{Studzinski18},
\begin{align}
\label{eq:Mf_1}M_f^\alpha[V[(i\ n)]^{t_n}]_{j_{\xi_\omega}j_{\xi_\nu}}^{\xi_\omega\ \xi_\nu}=\frac{1}{n-1}\frac{\sqrt{d_\nu d_\omega}}{d_\alpha}\sum_{k_\alpha}\sqrt{\lambda_\omega(\alpha)}\psi_{R\ j_{\xi_\omega}k_\alpha}^{\omega\ \xi_\omega\ \alpha}[(i\ n&-1)]\psi_{R\ k_\alpha j_{\xi_\nu}}^{\nu\ \alpha\ \xi_\nu}[(i\ n-1)]\sqrt{\lambda_\nu(\alpha)}\nonumber\\
&\quad\omega,\nu\neq\theta,\ i=1,\dots,n-1.
\end{align}
For $i = n-1$, this expression simplifies to 
\begin{align}
M_f^\alpha[V[(n-1\ n)]^{t_n}]_{j_{\xi_\omega}j_{\xi_\nu}}^{\xi_\omega\ \xi_\nu}=\frac{1}{n-1}\frac{\sqrt{d_\nu d_\omega}}{d_\alpha}\sum_{k_\alpha}\sqrt{\lambda_\omega(\alpha)\lambda_\nu(\alpha)}&\delta^{\xi_\omega\alpha}\delta^{\xi_\nu\alpha}\delta_{j_{\xi_\omega}j_{\xi_\nu}}\quad\omega,\nu\neq\theta,
\end{align}
and, for permutations that fix the last system, we have
\begin{align}
M_f^\alpha[V(\sigma_n)]_{j_{\xi_\omega}j_{\xi_\nu}}^{\xi_\omega\ \xi_\nu}=\delta^{\omega\nu}\psi_R^\nu[\sigma_n]_{j_{\xi_\omega}j_{\xi_\nu}}^{\xi_\omega\ \xi_\nu}\quad
\omega,\nu\neq\theta,\ \forall\sigma_n\in S(n-1).
\end{align}

%% file: sections/circuit.tex
\section{Quantum circuit for PBT}
\label{sec:circuit}

\subsection{Overview of the construction of the algorithm}

In this section, we'll construct an efficient quantum algorithm for port-based teleportation.  Recall that we are focused on the case where port-based teleportation is performed using the pretty good measurement, and where the entangled resource state is given by $n-1$ Bell pairs.  As we saw in Sections \ref{sec:pbt_prelim} and \ref{sec:A_defs}, in this case the Kraus operators $\sqrt{\Pi_i}$ for port-based teleportation are composed of the operators $\rho$ and $\rho_i$, which are elements of $A^{t_n}_n(d)$, the algebra of partially transposed permutation operators. Our strategy will be to take advantage of this symmetry, as well as the close relationship between $A^{t_n}_n(d)$ and the symmetric group $S(n)$, to find an efficient implementation of each $\sqrt{\Pi_i}$.

First, in this section, we observe that 
\begin{align}
    \sqrt{\Pi_i} = \sqrt{\tilde{\Pi}_i + \Delta} = \sqrt{\tilde{\Pi}_i} + \sqrt{\Delta}.
\end{align}
The key difficulty in the implementation is performing the matrix powers in, e.g.,  $\sqrt{\tilde{\Pi}_i} = \sqrt{\rho^{-1/2} \rho_i \rho^{-1/2}}$.  To get around this issue, we look at the irreducible representations of $A^{t_n}_n(d)$ in $\mathcal{H}_M$, which is the only subspace of $(\mathbb{C}^d)^{\otimes n}$ where $\rho, \rho_i \in M$ are nontrivial.  The irreps $\mathcal{H}_{\alpha,r}$ correspond to copies of $S(n-2)$ irreps ($\alpha,r$), where $\alpha \vdash n-2$ and $r$ is the multiplicity label.  In Section~\ref{sec:reptheory}, we explicitly found an orthonormal basis $\{|f^{\xi_\nu}_{j_{\xi_\nu}}(\alpha, r)\rangle \}$ that block-diagonalizes the elements of $M$ by their action in each $(\alpha, r)$ irrep, as well as their explicit form.  As it turns out, $\rho$ is diagonal in this basis, making it easy to compute $\rho^{-1/2}$.  Moreover, in this section we will show that the $\tilde{\Pi}_i$ are (pseudo)projectors in each irrep $\mathcal{H}_{\alpha,r}$, making it straightforward to scale their representations in each irrep block to find $\sqrt{\tilde{\Pi}_i}$.

Most importantly, we were able to find the $\{ \ket{f} \}$ basis, as well as the matrix elements of the generators, in terms of just $S(n-2)$ Schur basis states, $S(n-1)$ irreps, two-cycle permutations, and coefficients--all simple objects that can be produced by the $(n-1)$ and $(n-2)$ Schur transforms.  This is a consequence both of the close relationship between $A^{t_n}_n(d)$ and the symmetric group, as well as the subgroup-reduced nature of the Schur basis: the partial transposition only affects the basis at the highest two levels of the subgroup tower.

This is the key result of the representation theory that makes our algorithm efficient.  By transforming to the $\{ \ket{f} \}$ basis, which can be done with combinations of efficient Schur transforms, we can execute the Kraus operators $\sqrt{\Pi_i}$ efficiently as combinations of Schur transforms, and then transform back to the computational basis.  Explicitly, in this section we will find that 
\begin{alignat}{2}
    \sqrt{\tilde{\Pi}_i} &= \sum_{\alpha, r} U_{\text{twSch}}^\dagger(\alpha, r&&) \cdot M^\alpha_f [\sqrt{\tilde{\Pi}_i}] \cdot U_{\text{twSch}}(\alpha,r) \\
    &= \sum_{\substack{\alpha,r \\ \nu_l,\nu_r=\alpha+\Box\neq\theta \\ k_l,k_r=1,\dots,n-1}}&&[(n-1)d_\alpha-d_\theta]^{-\frac{1}{2}} [(n-1)d_\alpha]^{-\frac{3}{2}} \frac{d_{\nu_l}d_{\nu_r}}{\sqrt{\lambda_{\nu_l}\lambda_{\nu_r}}} V_L(\pi_{k_l}) \cdot \Phi^\dagger(\alpha,r)\nonumber\\ 
    & &&\cdot U_{\nu_l,\alpha}\cdot V(\pi_{k_l})\cdot V(\pi_i)\cdot U^\dagger_{\nu_l,\alpha}\cdot U_{\nu_r,\alpha} \cdot V(\pi_i)\cdot V(\pi_{k_r})\cdot U^\dagger_{\nu_r,\alpha} \cdot\Phi(\alpha,r)\cdot V_L(\pi_{k_r}),
    \label{eq:Pidecomp}
\end{alignat}
and
\begin{alignat}{2}
    \sqrt{\Delta} 
    &=\label{eq:deltabasis}\mathbb{I} \big/\sqrt{n-1} - 1\big/\sqrt{n-1}\cdot
    \sum_{\alpha,r} U^\dagger_{\text{twSch}}(\alpha, r&&) \cdot U_{\text{twSch}}(\alpha,r) \\
    &=\mathbb{I}\big/\sqrt{n-1}-1\big/\sqrt{n-1}\cdot\sum_{\substack{\alpha,r \\ \nu_l=\alpha+\Box\neq\theta \\ k_l,k_r=1,\dots,n-1}} && [(n-1)d_\alpha]^{-1} \frac{d_{\nu_l}}{\lambda_{\nu_l}} V_L(\pi_{k_l}) \cdot \Phi^\dagger(\alpha,r)\cdot\nonumber\\ 
    & &&U_{\nu_l,\alpha}\cdot V(\pi_{k_l})\cdot V(\pi_{k_r})\cdot U^\dagger_{\nu_l,\alpha} \cdot\Phi(\alpha,r)\cdot V_L(\pi_{k_r}).
    \label{eq:Deltadecomp}
\end{alignat}
Here, $\pi_k$ denotes the two-cycle $(k\ n-1)$ and $V_L$ is the natural representation of $S(n-1)$ on $n$ qudits, while $V$ is the natural representation on $n-1$ qudits.  The matrix $U_{\nu, \alpha}$ is a submatrix of the standard Schur transform on $n-1$ qudits: $U_{\nu, \alpha}$ is given by the rows of the Schur transform corresponding to the $\alpha$ (an $S(n-2)$ irrep) portion of $\nu$ (an $S(n-1)$ irrep).  Finally, $\Phi(\alpha,r)$ is constructed from the standard Schur transform on $n-2$ qudits by tensoring a Bell state to the rows corresponding to $S(n-2)$ irrep copy $(\alpha,r)$. 

Each of these relatively simple building blocks originates from the decomposition of $M^\alpha_f [\sqrt{\tilde{\Pi}_i}]$, the matrix elements of $\sqrt{\tilde{\Pi}_i}$ in the $\{ \ket{f} \}$ basis of an irrep $(\alpha,r)$, and $U_{\text{twSch}}(\alpha, r)$ the twisted Schur transform, which maps from this basis to the computational basis.  We can express $U_{\text{twSch}}(\alpha, r)$ as
\begin{align}
   U_{\text{twSch}}(\alpha,r) &= \sum_{\substack{\nu = \alpha + \Box \\ \nu \neq \theta}} \sum_{\xi_{\nu} = \nu - \Box} \sum_{j_{\xi_\nu} = 1}^{d_{\xi_\nu}} \ket{e_\nu, e_{\xi_\nu},e_{j_{\xi_\nu}}} \bra{ f^{\xi_\nu}_{j_{\xi_\nu}}(\alpha, r)}.
\end{align}
In our notation, $\ket{e_\nu, e_{\xi_\nu},e_{j_{\xi_\nu}}} = \ket{e_\nu} \otimes \ket{e_{\xi_\nu}} \otimes \ket{e_{j_{\xi_\nu}}}$ is a computational basis state on three registers, where each register counts the terms in the corresponding sum.  For example, $e_\nu$ takes values from $0$ to the number of $S(n-1)$ irreps $\nu = \alpha + \Box$.  Each of these computational basis states $\ket{e_\nu, e_{\xi_\nu},e_{j_{\xi_\nu}}}$ essentially labels the corresponding row of $U_{\text{twSch}} (\alpha,r)$, $\bra{ f^{\xi_\nu}_{j_{\xi_\nu}}(\alpha, r)}$.  We will use this notation for computational basis states throughout this section. 

These decompositions of $\sqrt{\tilde{\Pi}_i}$ and $\sqrt{\Delta}$ in terms of simple building blocks are the key to implementing port-based teleportation.  From there, we can implement the POVM by finding unitary block-encodings of the building blocks, then use a method inspired by the Linear Combination of Unitaries algorithm to construct the superpositions with coefficients given in Eqs. (\ref{eq:Pidecomp}, \ref{eq:Deltadecomp}). The next step is to find unitary block-encodings $U^c(i)$ of each $\sqrt{\Pi_i}$ operator, using ancillae combined with Naimark's theorem.  This nearly implements the POVM; because we are using block-encodings rather than the measurement operators $\sqrt{\Pi_i}$, the correct POVM will only be performed when we are in the $0$-eigenspace of the ancillae.  The final step is then to amplify the probability that we are in this subspace by using oblivious amplitude amplification.

The rest of this section will show how to find the decompositions given in Eqs. (\ref{eq:Pidecomp}, \ref{eq:Deltadecomp}), as well as describe in more detail how to use them to implement port-based teleportation.  Subsection \ref{sub:repPBT} will explain how the representation theory of the algebra $A^{t_n}_n(d)$ applies to the Kraus operators of port-based teleportation, then subsection \ref{sub:basistransformPBT} will show how to transform them to the $\{ \ket{f} \}$ basis.  Next, subsection \ref{sub:decompsPBT} will show how in the $\{ \ket{f} \}$ basis we can decompose $\sqrt{\tilde{\Pi}_i}$ and $\sqrt{\Delta}$ in terms of the simple building blocks described above to get Eqs. (\ref{eq:Pidecomp}, \ref{eq:Deltadecomp}).  Finally, subsection \ref{sub:naimarksPBT} will describe how we can find the necessary block-encodings and use Naimark's theorem together with amplitude amplification to implement the POVM.
 
\subsection{Applying the representation theory of $A^{t_n}_n(d)$ to the PBT operators}
\label{sub:repPBT}

Recall from Sec.~\ref{sec:prelim} that the Kraus operators for the pretty good measurement are given by 
\begin{equation}
    \sqrt{\Pi_i} = \sqrt{\tilde{\Pi}_i + \Delta}
\end{equation}
with 
\begin{align}
    \tilde{\Pi}_i &= \rho^{-1/2} \rho_i \rho^{-1/2} \\
    \label{eq:rho_algebra2} \rho &= \sum_{i=1}^{n-1} \rho_i = \frac{1}{d^{n-1}} \sum_{i=1}^{n-1} V[(i\ n)]^{t_n} \\
    \Delta &= \frac{1}{n-1}\left(\mathbb{I} - \rho^{-1/2} \rho \rho^{-1/2} \right)
\end{align}

Here we have written the operators $\rho_i$ as partially-transposed permutations, and $\rho$ as a superposition of partially transposed permutations, to see again that $\rho_i, \rho\in A^{t_n}_n(d)$ for all $i$.  Moreover, as these operators are only constructed from elements of $A^{t_n}_n(d)$ that permute and transpose the last system $n$, they are contained in the ideal $M$.  It is challenging to implement the Kraus operators $\sqrt{\Pi_i}$, so our strategy will be to take advantage of this symmetry to simplify the problem.  This will be possible even though $\tilde{\Pi}_i$ and $\sqrt{\Pi_i}$ are not elements of $A^{t_n}_n (d)$ due to the nonlinearity of the square roots.

Recall in the previous section we found that, when representing $A^{t_n}_n(d)$ in $\mathcal{H}_M$, the support space of the ideal $M$, $\mathcal{H}_M$ decomposes as a direct sum of irreps $\mathcal{H}_{\alpha, r}$.  For each of these irreps $\mathcal{H}_{\alpha, r}$, we found an orthonormal basis $\{ \ket{f^{\xi_\nu}_{j_{\xi_\nu}}(\alpha, r)} \}$ that spans it.  Here, recall that $\mathcal{H}_{\alpha,r}$ is divided into blocks corresponding to $S(n-1)$ irreps $\nu = \alpha + \Box \neq \theta$, constructed by adding a box to the Young diagram of the $S(n-2)$ irrep $\alpha$ in all legal ways.  Each $\nu$ block is then subgroup-reduced, that is, divided into $S(n-2)$ irrep blocks labeled by $\xi_\nu = \nu - \Box$, and $j_{\xi_\nu}$ indexes within these blocks.  The dimension of $\mathcal{H}_{\alpha,r}$ is therefore $D_\alpha \equiv \sum_{\nu = \alpha + \Box \neq \theta} d_{\nu}$.

We also found the matrix elements of generators of $A^{t_n}_n(d)$ in the $\{ \ket{f} \}$ basis.  Using the expression for $M_f^\alpha[V[(i\ n)]^{t_n}]$ (\ref{eq:Mf_1}) and Eq. (\ref{eq:rho_algebra2}) to evaluate the matrix elements of $\rho$ in the $\{ \ket{f} \}$ basis, we see \cite{Studzinski22}
\begin{align}
M_f^\alpha[\rho]_{j_{\xi_\omega}j_{\xi_\nu}}^{\xi_\omega\ \xi_\nu}&=\delta^{\omega\nu}\delta^{\xi_\omega\xi_\nu}\delta_{j_{\xi_\omega}j_{\xi_\nu}}\lambda_\nu(\alpha)
\end{align}
where $\lambda_\nu (\alpha)$ are the eigenvalues of $Q(\alpha)$, as given by Eq.~\eqref{eq:lambda_values}. 
The dimension of $M_f^\alpha[\rho]$ is therefore $D_\alpha$, the same as the dimension of $Q(\alpha)$.
Looking at $M_f^\alpha[\rho]$ for all $\alpha$, we then see that the support of $\rho$ is exactly $H_M$.

Going further, the definition of $\tilde{\Pi}_i$ is
\begin{align}
\tilde{\Pi}_i &= \rho^{-1/2} \rho_i \rho^{-1/2},
\end{align}
for $i = 1,..., n-1$, where the multiplicative components $\rho$ and $\rho_i$ both belong to the ideal $M$ of the algebra $A_n^{t_n}(d)$.  In the $\{ \ket{f} \}$ basis, $\rho$ and $\rho_i$ are therefore $(\alpha, r)$-block-diagonalized.  To build up $\tilde{\Pi}_i$ and $\sqrt{\tilde{\Pi}_i}$, we can simply do matrix arithmetic within each $(\alpha, r)$ irrep block, which means that in the same $\{ \ket{f} \}$ basis they will also be $(\alpha,r)$-block-diagonalized.  This implies $\text{supp}(\tilde{\Pi}_i)\subseteq \text{supp}(\rho)=\mathcal{H}_M$.

On the other hand, the formula for $\Delta$, can be rewritten as
\begin{align}
\Delta &= \left(\mathbb{I} - \rho^{-1/2} \rho \rho^{-1/2} \right) \Big/(n-1) \\
&= \left(\mathbb{I} - \mathbb{I}_{H_M} \right) \big/(n-1) \\
\label{eq:Delta_id}&= \mathbb{I}_{H_S}  \big/(n-1),
\end{align}
where in the last equality we use that the orthogonal complement of $\mathcal{H}_M$ in $(\mathbb{C}^d)^{\otimes n}$ is $H_S$.  Therefore, the support of $\Delta$ is exactly $\mathcal{H}_S$, while the support of $\tilde{\Pi}_i$ is entirely contained within its complement, $\mathcal{H}_S$. 

As the supports of $\tilde{\Pi}_i$ and $\Delta$ are contained within the orthogonal spaces $\mathcal{H}_M$ and $\mathcal{H}_S$, respectively, their sum $\Pi_i$ will only act as $\tilde{\Pi}_i$ in $\mathcal{H}_M$ or as $\Delta$ in $\mathcal{H}_S$.  Consequently, the Kraus operators simplify as
\begin{equation}
    \sqrt{\Pi_i} = \sqrt{\tilde{\Pi}_i + \Delta} = \sqrt{\tilde{\Pi}_i} + \sqrt{\Delta}.
\end{equation}
This means that $\sqrt{\Pi_i}$ will be $(\alpha,r)$-block-diagonalized in the basis $\{|f^{\xi_\nu}_{j_{\xi_\nu}}(\alpha, r)\rangle\}$ of $\mathcal{H}_M$, and will be proportional to $\mathbb{I}_{H_S}$ in any arbitrary basis spanning $\mathcal{H}_S$.  This arbitrary basis provides an extension of the $|f\rangle$ basis to the space $(\mathbb{C}^d)^{\otimes n}$. From here on, when we refer to the $\ket{f}$ basis, we are referring to this arbitrary extension to $(\mathbb{C}^d)^{\otimes n}$.

This is a major simplification of the Kraus operators, which will allow us to construct implementations of $\sqrt{\tilde{\Pi}_i}$ and $\sqrt{\Delta}$, individually, in their respective spaces, before easily combining them.  In particular, we will be able to work in the $\ket{f}$ basis of $\mathcal{H}_M$, where $\sqrt{\tilde{\Pi}_i}$ is $(\alpha,r)$-block-diagonalized, to implement it one irrep block at a time.

\subsection{Basis transformation of $\sqrt{\Pi_i}$}
\label{sub:basistransformPBT}

In the previous subsection, we saw that we can implement the Kraus operators for the pretty good measurement by implementing $\sqrt{\tilde{\Pi}_i}$ and $\sqrt{\Delta}$ individually.  Our strategy will be to transform to the $\ket{f}$ basis, in which $\sqrt{\tilde{\Pi}_i}$ is $(\alpha,r)$-block-diagonalized with matrix elements composed from simple building blocks.  In this subsection, we will show how to implement the basis transformation into each $(\alpha,r)$ irrep using the transformation matrix
\begin{align}
   U_{\text{twSch}}(\alpha,r) &= \sum_{\substack{\nu = \alpha + \Box \\ \nu \neq \theta}} \sum_{\xi_{\nu} = \nu - \Box} \sum_{j_{\xi_\nu} = 1}^{d_{\xi_\nu}} \ket{e_\nu, e_{\xi_\nu},e_{j_{\xi_\nu}}} \bra{ f^{\xi_\nu}_{j_{\xi_\nu}}(\alpha, r)}.
\end{align}
Notice that $U_{\text{twSch}}(\alpha,r)$ is not a square matrix, as it transforms from $(\mathbb{C}^d)^{\otimes n}$ to a single irrep $\mathcal{H}_{\alpha,r}$; instead it is of order $D_\alpha\times d^n$. 
 
We will also use the matrix elements of $\tilde{\Pi}_i$ in the $\ket{f}$ basis, which we found in Section~\ref{sec:reptheory}, to find the matrix elements of $\sqrt{\tilde{\Pi}_i}$.
The next subsection will show how these matrix elements decompose into elementary building blocks, decompositions that are ultimately the source of our algorithm's efficiency.

\subsubsection{Transforming $\sqrt{\tilde{\Pi}_i}$}

Using the fact that $\sqrt{\tilde{\Pi}_i}$ is $(\alpha,r)$-block-diagonal in the $\ket{f}$ basis, we can transform it into the $\ket{f}$ basis one irrep block at a time.  We have, 
\begin{align}
    \sqrt{\tilde{\Pi}_i} = \sum_{\alpha, r} U_{\text{twSch}}^\dagger (\alpha, r) \cdot M^\alpha_f [\sqrt{\tilde{\Pi}_i}] \cdot U_{\text{twSch}}(\alpha,r),
    \label{eq:rootPimatrix}
\end{align}
where each summand is a matrix product.  Here $M^\alpha_f [\sqrt{\tilde{\Pi}_i}]$ denotes the representation of $\sqrt{\tilde{\Pi}_i}$ in the $\ket{f}$ basis for any copy $r$ of the irrep $\alpha$.  It can be written out using computational basis states as 
\begin{equation}
    M^\alpha_f [\sqrt{\tilde{\Pi}_i}] =  \sum_{\substack{\nu, \omega = \alpha + \Box \\ \nu, \omega \neq \theta}} \sum_{\substack{\xi_{\nu}, \xi_\omega \\ j_{\xi_\nu}, j_{\xi_\omega}}} \ket{e_\omega, e_{\xi_\omega},e_{j_{\xi_\omega}}} M_f^\alpha[\sqrt{\tilde{\Pi}_i}]_{j_{\xi_\omega}j_{\xi_\nu}}^{\xi_\omega\ \xi_\nu} \bra{e_\nu, e_{\xi_\nu},e_{j_{\xi_\nu}}}.
\end{equation}
where
\begin{equation}
M_f^\alpha[\sqrt{\tilde{\Pi}_i}]_{j_{\xi_\omega}j_{\xi_\nu}}^{\xi_\omega\ \xi_\nu} \equiv \langle f^{\xi_\omega}_{j_{\xi_\omega}}(\alpha, r) | \sqrt{\tilde{\Pi}_i} | f^{\xi_\nu}_{j_{\xi_\nu}}(\alpha, r) \rangle.
\end{equation}
Using these expressions, together with the definition of $U_{\text{twSch}}(\alpha,r)$, Eq. \eqref{eq:rootPimatrix} can be easily verified by substitution.

\subsubsection{Finding the matrix elements of $\sqrt{\tilde{\Pi}_i}$ in the $\ket{f}$ basis}

In order for the transformation of $\sqrt{\tilde{\Pi}_i}$ into the $\ket{f}$ basis to be useful, we need to have an efficiently implementable expression for $M^\alpha_f [\sqrt{\tilde{\Pi}_i}]$.  To this end, we can first use the tools of Section~\ref{sec:reptheory} to calculate $M^\alpha_f [\tilde{\Pi}_i]$.  Explicitly, using Eq. (\ref{eq:Mf_1}), (\ref{eq:rho_algebra2}) and (\ref{eq:rho_i_algebra}), we have \cite{Studzinski22}
\begin{align}
M_f^\alpha[\tilde{\Pi}_i]_{j_{\xi_\omega}j_{\xi_\nu}}^{\xi_\omega\ \xi_\nu}&=\frac{1}{n-1}\frac{\sqrt{d_\nu d_\omega}}{d_\alpha}\sum_{k_\alpha}\psi_{R\ j_{\xi_\omega}k_\alpha}^{\omega\ \xi_\omega\ \alpha}[(i\ n-1)]\psi_{R\ k_\alpha j_{\xi_\nu}}^{\nu\ \alpha\ \xi_\nu}[(i\ n-1)].
\end{align}
It then follows from direct multiplication that
\begin{align}
M_f^\alpha[\tilde{\Pi}_i] M_f^\alpha[\tilde{\Pi}_i] = \left(1-\frac{d_\theta}{(n-1)d_\alpha}\right)M_f^\alpha[\tilde{\Pi}_i]\quad i=1,\dots,n-1,
\end{align}
which shows $M_f^\alpha[\tilde{\Pi}_i]$ is a projector when $d_\theta=0$ and a pseudoprojector when $d_\theta\neq 0$.
We can then find $M^\alpha_f [\sqrt{\tilde{\Pi}_i}]$ by scaling by the eigenvalues of a (pseudo)projector, which gives
\begin{align}
M_f^\alpha[\sqrt{\tilde{\Pi}_i}]_{j_{\xi_\omega}j_{\xi_\nu}}^{\xi_\omega\ \xi_\nu}&=\frac{1}{\sqrt{(n-1)d_\alpha-d_\theta}}\frac{\sqrt{d_\nu d_\omega}}{\sqrt{(n-1)d_\alpha}}\sum_{k_\alpha}\psi_{R\ j_{\xi_\omega}k_\alpha}^{\omega\ \xi_\omega\ \alpha}[(i\ n-1)]\psi_{R\ k_\alpha j_{\xi_\nu}}^{\nu\ \alpha\ \xi_\nu}[(i\ n-1)].
\label{eq:rootPimelements}
\end{align}

\subsubsection{Transforming $\sqrt{\Delta}$}

Finally, although $\sqrt{\Delta}$ has no support in $\mathcal{H}_M$, we can use the expression
\begin{equation}
    \Delta = \left(\mathbb{I} - \mathbb{I}_{\mathcal{H}_M} \right) \big/(n-1)
\end{equation}
to also write it in terms of the basis transformation matrices $\{ U_{\text{twSch}}(\alpha,r) \}$  as
\begin{align}
\sqrt{\Delta} &= \mathbb{I} \big/\sqrt{n-1} - \mathbb{I}_{\mathcal{H}_M} \big/\sqrt{n-1} \\
&=\label{eq:deltabasis2}\mathbb{I} \big/\sqrt{n-1} - 1\big/\sqrt{n-1}\cdot
\sum_{\alpha,r} U^\dagger_{\text{twSch}}(\alpha,r) \cdot U_{\text{twSch}}(\alpha,r).
\end{align}
Using this decomposition, a specification of a basis for $\mathcal{H}_S$ is no longer necessary, as we only work with the identity on the full space and $\mathcal{H}_M$. Furthermore, this decomposition allows us to reuse the building block $U_{\text{twSch}}(\alpha,r)$ in the circuit implementation, which is a significant simplification.

\subsection{Decomposition of $\sqrt{\Pi_i}$ into Schur transforms}
\label{sub:decompsPBT}

Putting together the results of Sec.~\ref{sub:repPBT} and Sec.~\ref{sub:basistransformPBT}, we have seen that each Kraus operator $\sqrt{\Pi_i}$ of port-based teleportation can be written in terms of the basis transformation matrix $U_{\text{twSch}}(\alpha,r)$ and the matrix elements of $\sqrt{\tilde{\Pi}_i}$ in irrep basis $\{ \ket{f} \}$ as
\begin{align}
\sqrt{\Pi_i}&= \sqrt{\tilde{\Pi}_i} + \sqrt{\Delta} \\
&=\sum_{\alpha, r}U^\dagger_{\text{twSch}}(\alpha,r)\cdot M_f^\alpha[\sqrt{\tilde{\Pi}_i}]\cdot U_{\text{twSch}}(\alpha,r) \nonumber\\
& + \mathbb{I} \big/\sqrt{n-1} - 1\big/\sqrt{n-1}\cdot \sum_{\alpha, r} U^\dagger_{\text{twSch}}(\alpha,r)\cdot U_{\text{twSch}}(\alpha,r).
\end{align}
In this subsection, we will show that $M_f^\alpha[\sqrt{\tilde{\Pi}_i}]$ and $U_{\text{twSch}}(\alpha,r)$ can both be implemented by leveraging the efficient quantum circuit for the Schur transform on $(n-1)$ and $(n-2)$ qubits, as well as a few permutations.  These decompositions into simple building blocks will make it possible to implement port-based teleportation efficiently.

\subsubsection{Reviewing the $\ket{f}$ basis}

Before we begin, recall that the elements of the vector basis $\ket{f(\alpha,r)}$ are defined as 
\begin{align}
\label{eq:f_basis_new}|f^{\xi_\nu}_{j_{\xi_\nu}}(\alpha, r)\rangle &= \frac{1}{\sqrt{\lambda_\nu}}\sum_{k=1}^{n-1}\sum_{k_\alpha=1}^{d_\alpha} z_R(\alpha)_{k_\alpha j_{\xi_\nu}}^{k\ \xi_\nu}|\psi^k_{k_\alpha}(\alpha, r)\rangle\quad \nu=\alpha+\Box\neq\theta,\\
|\psi^k_{k_\alpha}(\alpha, r)\rangle &= \sqrt{d}V(\pi_k) \ket{r, \alpha, k_\alpha}^{(n-2)}_{\text{Sch}} |\phi_+\rangle_{n-1n}.
\end{align}
Here, $\{ \ket{f(\alpha,r)} \}$ span a single irrep $\mathcal{H}_{\alpha,r}$ which is divided into blocks corresponding to $S(n-1)$ irreps $\nu = \alpha + \Box\neq \theta$ ($\dim \mathcal{H}_{\alpha,r} = D_\alpha = \sum_{\nu = \alpha + \Box\neq\theta} d_\nu$).  The $\nu$ blocks are subgroup-reduced into $S(n-2)$ irrep blocks labeled by $\xi_\nu = \nu - \Box$, and $j_{\xi_\nu}$ indexes within these blocks.  As a result, there is one basis element for every tuple $(\xi_\nu, j_{\xi_\nu})$.  

As we saw in Section~\ref{sec:reptheory}, the vectors $\{ \ket{\psi^k_{k_\alpha}(\alpha,r)} \}$ naturally arise as spanning vectors of the irreps $\mathcal{H}_{\alpha,r}$.  This is because the generators of $M$ vanish outside of the subspace spanned by $ \{ \ket{\phi_i}_{\overline{kn}} \ket{\phi_+}_{kn}\}$ where $\phi_i$ is a bitstring of length $n-2$ and $\ket{\phi_+}_{kn}$ is the maximally entangled state between qudits $k$ and $n$.  Each of these states can also be written as $V(\pi_k) \ket{\phi_i}\ket{\phi_+}_{n-1 n}$, where $\pi_k$ is the two cycle $(k\ n-1)$ and $V$ is the natural representation on $n-1$ qudits.  Transforming the computational basis states $\ket{\phi_i}$ to the Schur basis on $n-2$ qubits $\ket{q_\alpha^d\equiv r, \alpha, p_\alpha\equiv k_\alpha}^{(n-2)}_{\text{Sch}}$ gave us the states $ \ket{\psi^k_{k_\alpha}(\alpha, r)} $, while also decomposing $\mathcal{H}_M$ into the irreps $\mathcal{H}_{\alpha,r}$.

While the $\{ \ket{\psi^k_{k_\alpha}(\alpha,r)} \}$ vectors span the irreps $\mathcal{H}_{\alpha,r}$, they are not necessarily linearly independent.  We orthonormalize them by using the diagonalization matrix $z_R$ and scaling with the eigenvalue $\lambda_\nu$; this gives us the $\ket{f}$ basis.

Absorbing the scaling factor $1/\sqrt{\lambda_\nu}$ into the diagonalization $z_R(\alpha)$, it is defined as
\begin{align}
\label{eq:zblock}(\tilde{z}_R(\alpha)^k)_{k_\alpha j_{\xi_\nu}}^{\ \ \ \xi_\nu}\equiv \frac{1}{\sqrt{\lambda_\nu}} z_R(\alpha)_{k_\alpha j_{\xi_\nu}}^{k\xi_\nu} = 
\frac{1}{\sqrt{(n-1)d_\alpha}}\sqrt{\frac{d_\nu}{\lambda_\nu}}\psi_R^\nu[\pi_k]^{\alpha\xi_\nu}_{k_\alpha j_{\xi_\nu}}.
\end{align}
Here, $\psi^\nu_R [\pi_k]$ is the permutation $\pi_k$ in the partially reduced irreducible representation of $S(n-1)$ labeled by $\nu$.  As the output of the Schur transform is PRIRs labeled by Young diagrams with height $\leqslant d$, we can find $\pi_k$ in the representation $\psi^\nu_R$ by simple evaluating the matrix elements of $\pi_k$ in the Schur basis for $(n-1)$ qudits.  Choosing a fixed but arbitrary copy $r_\nu$ of the irrep $\nu$, we have 
\begin{equation}
    \psi_R^\nu[\pi_k]^{\alpha\xi_\nu}_{k_\alpha j_{\xi_\nu}} = \bra{r_\nu, \nu, \alpha, k_\alpha}^{(n-1)}_{\text{Sch}} V(\pi_k) \ket{r_\nu, \nu, \xi_\nu, j_{\xi_\nu}}^{(n-1)}_{\text{Sch}}
\end{equation}
Here, $r_\nu$ denotes the fixed copy of the $S(n-1)$ irrep $\nu$, while $\xi_\omega = \omega - \square$ labels the $S(n-2)$ irreps composing $\nu$, and $j_{\xi_\nu}$ indexes within the $S(n-2)$ irreps.  In other words, we label the Schur basis in a way that makes the partially reduced nature of the $\nu$ explicit.

\subsubsection{Decomposing $\tilde{z}_R(\alpha)$}

To decompose the twisted Schur transform $U_{\text{twSch}} (\alpha, r)$, which transforms from the computational basis to the $\ket{f}$ basis, in terms of simple building blocks, we will need to decompose the $\ket{f}$ basis.  We'll start with the scaled diagonalization matrix 
\begin{equation}
    \label{eq:zSchur}(\tilde{z}_R(\alpha)^k)_{k_\alpha j_{\xi_\nu}}^{\ \ \ \xi_\nu} = \frac{1}{\sqrt{(n-1)d_\alpha}}\sqrt{\frac{d_\nu}{\lambda_\nu}} \bra{r_\nu, \nu, \alpha, k_\alpha}^{(n-1)}_{\text{Sch}} V(\pi_k) \ket{r_\nu, \nu, \xi_\nu, j_{\xi_\nu}}^{(n-1)}_{\text{Sch}}
\end{equation}
which we first divide into $(n-1)$ block rows $\tilde{z}_R(\alpha)^k$ as
\begin{align}
    \tilde{z}_R(\alpha)\equiv
    \begin{pmatrix}
        \tilde{z}_R(\alpha)^1\\
        \tilde{z}_R(\alpha)^2\\
        \vdots\\
        \tilde{z}_R(\alpha)^{n-1}
    \end{pmatrix}.
\end{align}
This separation out of the $k$ index makes sense because it refers to which of the $(n-1)$ quantum systems is permuted with the $(n-1)$th system in Eq.~\eqref{eq:zSchur}, as opposed to the other indices, which index within the $S(n-2)$ irreps composing $\mathcal{H}_{\alpha,r}$.

As our expression for $\tilde{z}_R(\alpha)$ in Eq.~\eqref{eq:zSchur} depends on Schur basis states for $n-1$ qudits, we will be able to decompose $\tilde{z}_R(\alpha)$ in terms of the $(n-1)$-qudit Schur transform.  To do this, we define two matrices composed of rows that are elements of the Schur basis on $n-1$ qudits
 \begin{align}
\label{eq:Ualpha} U_{\alpha} &\equiv \sum_{\substack{\omega = \alpha + \Box \\ \omega \neq \theta}} \sum_{\xi_{\omega} = \omega - \Box} \sum_{j_{\xi_\omega} = 1}^{d_{\xi_\omega}} \ket{e_\omega, e_{\xi_{\omega}}, e_{j_{\xi_\omega}}} \bra{r_\omega, \omega, \xi_{\omega}, j_{\xi_\omega}}^{(n-1)}_{\text{Sch}} \\
\label{eq:Unualpha} U_{\nu,\alpha} &\equiv \sum_{k_\alpha = 1}^{d_\alpha} \ket{e_{k_\alpha}} \bra{r_\nu, \nu, \alpha, k_\alpha}^{(n-1)}_{\text{Sch}} .
\end{align}
Here $r_\omega$ and $r_\nu$ are again referring to fixed, but arbitrary, copies of $\omega$ and $\nu$, respectively.  Notice that in $U_{\nu, \alpha}$, we are restricting to the part of the $S(n-1)$ irrep $\nu$ labeled by the $S(n-2)$ irrep $\alpha$.

Using these matrices, we can then rewrite each block row of $\tilde{z}_R$ as
\begin{align}
\tilde{z}_R(\alpha)^k=\frac{1}{\sqrt{(n-1) d_\alpha}}\sum_{\nu=\alpha+\Box\neq\theta} \sqrt{\frac{d_\nu}{\lambda_\nu}}\ U_{\nu,\alpha}\cdot V(\pi_k)\cdot U^\dagger_{\alpha}.
\end{align}
This expression can be quickly verified by substituting in our expressions for $U_\alpha$ and $U_{\nu,\alpha}$:
\begin{align}
    &\frac{1}{\sqrt{(n-1) d_\alpha}}\sum_{\nu=\alpha+\Box\neq\theta} \sqrt{\frac{d_\nu}{\lambda_\nu}}\ U_{\nu,\alpha}\cdot V(\pi_k)\cdot U^\dagger_{\alpha} \\
    =& \frac{1}{\sqrt{(n-1) d_\alpha}} \sum_{\substack{ \nu, \omega = \alpha + \Box \neq \theta \\ \xi_\omega = \omega - \Box, j_{\xi_\omega}}} \sum_{k_\alpha = 1}^{d_\alpha} \sqrt{\frac{d_\nu}{\lambda_\nu}} \ket{e_{k_\alpha}} \bra{r_\nu, \nu, \alpha, k_\alpha}^{(n-1)}_{\text{Sch}} V(\pi_k) \ket{r_\omega, \omega, \xi_{\omega}, j_{\xi_\omega}}^{(n-1)}_{\text{Sch}} \bra{e_\omega, e_{\xi_{\omega}}, e_{j_{\xi_\omega}}} \\
    =& \sum_{\substack{ \nu = \alpha + \Box \neq \theta \\ \xi_\nu = \nu - \Box, j_{\xi_\nu}}} \sum_{k_\alpha = 1}^{d_\alpha} \tilde{z}_R(\alpha)^{k \xi_\nu}_{k_\alpha j_{\xi_\nu}} \ket{e_{k_\alpha}} \bra{e_\nu, e_{\xi_{\nu}}, e_{j_{\xi_\nu}}} = \tilde{z}_R(\alpha)^k
\end{align}
where the sum over $\omega$ vanishes in the third line because the subspaces for the different $S(n-1)$ irreps are orthogonal.  The final equality holds because the left-hand side is simply an expression for $\tilde{z}_R(\alpha)$ with components labeled by computational basis states.

\subsubsection{Decomposing $U_{\text{twSch}}(\alpha,r)$}

Equipped with this decomposition of $\tilde{z}_R(\alpha)$, in terms of submatrices of the Schur transform on $(n-1)$-qudits, we are ready to decompose the remaining contributions to the $\ket{f}$ basis to find the twisted Schur transform 
\begin{align}
   U_{\text{twSch}}(\alpha,r) &= \sum_{\substack{\nu = \alpha + \Box \\ \nu \neq \theta}} \sum_{\xi_{\nu} = \nu - \Box} \sum_{j_{\xi_\nu} = 1}^{d_{\xi_\nu}} \ket{e_\nu, e_{\xi_\nu},e_{j_{\xi_\nu}}} \bra{ f^{\xi_\nu}_{j_{\xi_\nu}}(\alpha, r)}.
\end{align}

In addition to the scaled diagonalization matrix, the $\ket{f}$ basis also depends on the maximally entangled state and the $S(n-2)$ irrep basis, which can be obtained in practice by applying the Schur transform to the computational basis $\{\ket{e_k}\in (\mathbb{C}^d)^{\otimes (n-2)}\}$.  We can capture these contributions in a single matrix 
\begin{align}
    \Phi(\alpha,r) \equiv \sum_{k_\alpha = 1}^{d_\alpha}
     \ket{ e_{k_\alpha}} \bra{r,\alpha,k_\alpha}^{(n-2)}_{\text{Sch}} \sqrt{d}\bra{\phi_+}_{n-1n}.
\end{align}
Using this matrix, we can rewrite $U_{\text{twSch}}(\alpha,r)$ as 
\begin{align}
    \label{eq:U_genSch_ini}U_{\text{twSch}}(\alpha,r)&=\sum_{k=1}^{n-1}\ (\tilde{z}_R(\alpha)^k)^\dagger\cdot \Phi(\alpha,r)\cdot V_L(\pi_k),
\end{align}
where $V_L:S(n-1)\rightarrow Hom[(\mathbb{C}^d)^{\otimes n}]$ is the natural representation of $S(n-1)$ on $n$ qudits.  This expression can be verified by substituting in resolutions of $\Phi (\alpha,r)$ and $\tilde{z}_R (\alpha)^k$ into components, giving 
\begin{align}
    &\sum_{k=1}^{n-1}\ (\tilde{z}_R(\alpha)^k)^\dagger\cdot \Phi(\alpha,r)\cdot V_L(\pi_k) \\
    =& \sum_{\substack{ \nu = \alpha + \Box \neq \theta \\ \xi_\nu = \nu - \Box, j_{\xi_\nu}}} \sum_{k_{\alpha}, l_\alpha = 1}^{d_\alpha} \sum_{k=1}^{n-1}\ 
    \ket{e_\nu, e_{\xi_\nu}, e_{j_{\xi_\nu}} }(\tilde{z}_R (\alpha)^\dagger)^{\xi_\nu k}_{j_{\xi_\nu} k_{\alpha} } \langle e_{k_\alpha} | e_{l_\alpha} \rangle \bra{r,\alpha, l_{\alpha}}^{(n-2)}_{\text{Sch}} \sqrt{d} \bra{\phi_+}_{n-1,n} V[\pi_k] \\
    =& \sum_{\substack{ \nu = \alpha + \Box \neq \theta \\ \xi_\nu = \nu - \Box, j_{\xi_\nu}}} \sum_{k_{\alpha}=1}^{d_\alpha} \sum_{k=1}^{n-1}\ 
    \ket{e_\nu, e_{\xi_\nu}, e_{j_{\xi_\nu}} }(\tilde{z}_R (\alpha)^\dagger)^{\xi_\nu k}_{j_{\xi_\nu} k_{\alpha} }  \bra{\psi^k_{k_\alpha} (\alpha, r)} \\
    =& \sum_{\substack{ \nu = \alpha + \Box \neq \theta \\ \xi_\nu = \nu - \Box, j_{\xi_\nu}}} \ 
    \ket{e_\nu, e_{\xi_\nu}, e_{j_{\xi_\nu}} } \bra{f^{\xi_\nu}_{j_{\xi_\nu}} (\alpha, r)} = U_{\text{twSch}}(\alpha,r)
\end{align}

Putting this result together with our decomposition of $\tilde{z}_R(\alpha)^k$, we have found 
\begin{align}
\label{eq:U_genSch}U_{\text{twSch}}(\alpha,r)&=[(n-1)d_\alpha]^{-\frac{1}{2}}\sum_{\nu=\alpha+\Box\neq\theta} \sqrt{\frac{d_\nu}{\lambda_\nu}}\ U_{\alpha}\cdot V(\pi_k)\cdot U^\dagger_{\nu,\alpha} \cdot\Phi(\alpha,r)\cdot V_L(\pi_k).
\end{align}
We can quickly check the dimension of this expression by multiplying the dimensions of the matrices involved, giving $\text{mul}(D_\alpha\times d^{n-1},d^{n-1}\times d^{n-1},d^{n-1}\times d_\alpha,d_\alpha\times d^n,d^n\times d^n)=D_\alpha\times d^n$, which is what we expect as $U_{\text{twSch}}$ transforms from the computational basis to the $\{ \ket{f} \}$ basis.  Most importantly, as this expression only depends on the permutations $V(\pi_k)$ and $V_L(\pi_k)$ and submatrices of the traditional Schur transform on $(n-1)$ and $(n-2)$ qudits, it will be straightforward to implement as a quantum circuit.

\subsubsection{Decomposing $M_f^\alpha[\sqrt{\tilde{\Pi}_i}]$}

Now that we have decomposed $U_{\text{twSch}}$, we can decompose $M_f^\alpha[\sqrt{\tilde{\Pi}_i}]$ using permutations and the Schur transform in much the same way.
Rewriting Eq. \eqref{eq:rootPimelements} in terms of Schur basis states, we have
\begin{align}
    M_f^\alpha[\sqrt{\tilde{\Pi}_i}]_{j_{\xi_\omega}j_{\xi_\nu}}^{\xi_\omega\ \xi_\nu}=\frac{1}{\sqrt{(n-1)d_\alpha-d_\theta}}\frac{\sqrt{d_\nu d_\omega}}{\sqrt{(n-1)d_\alpha}}\sum_{k_\alpha} & \bra{r_\omega, \omega, \xi_\omega, j_{\xi_\omega}}^{(n-1)}_{\text{Sch}} V(\pi_i) \ket{r_\omega, \omega, \alpha, k_\alpha}^{(n-1)}_{\text{Sch}} \\ &\times \bra{r_\nu, \nu, \alpha, k_\alpha}^{(n-1)}_{\text{Sch}} V(\pi_i) \ket{r_\nu, \nu, \xi_\nu, j_{\xi_\nu}}^{(n-1)}_{\text{Sch}}.
\end{align}

Just as we did when decomposing $\tilde{z}_R$, we can find matrix elements of the permutations $\pi_i$ in the Schur basis where they appear in the product
\begin{align}
    U_{\nu,\alpha} \cdot V(\pi_i) \cdot U_\alpha^\dagger &= \sum_{ \xi_\nu, j_{\xi_\nu}} \sum_{k_\alpha} \ket{e_{k_\alpha}} \bra{r_\nu, \nu, \alpha, k_\alpha}^{(n-1)}_{\text{Sch}} V(\pi_i) \ket{r_\nu, \nu, \xi_{\nu}, j_{\xi_\nu}}^{(n-1)}_{\text{Sch}} \bra{e_\nu, e_{\xi_{\nu}}, e_{j_{\xi_\nu}}}.
\end{align}
It then follows that $M_f^\alpha[\sqrt{\tilde{\Pi}_i}]$ can be written as
\begin{align}
\label{eq:Mf_sqrtPi}M_f^\alpha[\sqrt{\tilde{\Pi}_i}]=\frac{1}{\sqrt{(n-1)d_\alpha-d_\theta}}\frac{1}{\sqrt{(n-1)d_\alpha}}\sum_{\nu_1,\nu_2=\alpha+\Box\neq\theta}\sqrt{d_{\nu_1}d_{\nu_2}}\ U_\alpha\cdot V(\pi_i)\cdot U^\dagger_{\nu_1,\alpha} \cdot U_{\nu_2,\alpha} \cdot V(\pi_i)\cdot U^\dagger_\alpha.
\end{align}
Checking the dimensions again by multiplying the constituent matrices, we have $\text{mul}(D_\alpha\times d^{n-1},d^{n-1}\times d^{n-1},d^{n-1}\times d_\alpha,d_\alpha\times d^{n-1},d^{n-1}\times d^{n-1},d^{n-1}\times D_\alpha)=D_\alpha\times D_\alpha$, as expected.

\subsubsection{Decomposing $\sqrt{\Pi_i}$}

Combining our decompositions of $M_f^\alpha[\sqrt{\tilde{\Pi}_i}]$ and $U_{\text{twSch}}$ in Eq. \eqref{eq:U_genSch}, we have found an expression for $\sqrt{\tilde{\Pi}_i}$ only involving the $U_{\nu,\alpha}$ and $U_\alpha$ matrices derived from the Schur transform, and the permutation operators $V(\pi_k)$ and $V_L(\pi_k)$.
\begin{align}
\label{eq:second_line}\sqrt{\tilde{\Pi}_i}=\sum_{\substack{\alpha,r \\ \nu_1,\nu_2,\nu_l,\nu_r=\alpha+\Box\neq\theta \\ k_l,k_r=1,\dots,n-1}}& [(n-1)d_\alpha-d_\theta]^{-\frac{1}{2}} [(n-1)d_\alpha]^{-\frac{3}{2}} \sqrt{d_{\nu_1}d_{\nu_2}} \sqrt{\frac{d_{\nu_l}d_{\nu_r}}{\lambda_{\nu_l}\lambda_{\nu_r}}}\cdot\nonumber\\ 
&V_L(\pi_{k_l}) \cdot \Phi^\dagger(\alpha,r)\cdot U_{\nu_l,\alpha}\cdot V(\pi_{k_l})\cdot U^\dagger_{\alpha}\cdot U_\alpha \cdot V(\pi_i)\cdot U^\dagger_{\nu_1,\alpha}\cdot \nonumber \\
& U_{\nu_2,\alpha} V(\pi_i)\cdot U^\dagger_\alpha\cdot U_{\alpha} \cdot V(\pi_{k_r})\cdot U^\dagger_{\nu_r,\alpha} \cdot\Phi(\alpha,r)\cdot V_L(\pi_{k_r}).
\end{align}
We can simplify this expression by making two observations. First, consider replacing the $U_\alpha$ in Eq. (\ref{eq:second_line}) by the unitary matrix $U_{\text{Sch}}^{(n-1)}$. It is easy to see that the resulting factor $U_{\nu_l,\alpha}\cdot V(\pi_{k_l})\cdot U_{\text{Sch}}^{(n-1)\dagger}$ is simply a larger matrix where all of the additional entries are zero.  The same thing happens for the other three factors involving $U_\alpha$, and so $\sqrt{\tilde{\Pi}_i}$ is unchanged. But as $U_{\text{Sch}}^{(n-1)}$ is unitary, the two instances of $U_{\text{Sch}}^{(n-1)\dagger}U_{\text{Sch}}^{(n-1)}=\mathbb{I}$ that now appear are canceled out. Second, we notice that unless $\nu_l$ equals $\nu_1$ in the sum, $U_{\nu_l,\alpha}$ and $U_{\nu_1,\alpha}$ act on orthogonal spaces and the product $U_{\nu_l,\alpha}\cdot V(\pi_{k_l}) \cdot V(\pi_i)\cdot U^\dagger_{\nu_1,\alpha}$ will vanish. Similarly, we find $\nu_2=\nu_r$. The equation for $\sqrt{\tilde{\Pi}_i}$ therefore becomes
\begin{align}
\label{eq:sqrt_tilde_Pi_ini}\sqrt{\tilde{\Pi}_i}=\sum_{\substack{\alpha,r \\ \nu_l,\nu_r=\alpha+\Box\neq\theta \\ k_l,k_r=1,\dots,n-1}}& [(n-1)d_\alpha-d_\theta]^{-\frac{1}{2}} [(n-1)d_\alpha]^{-\frac{3}{2}} \frac{d_{\nu_l}d_{\nu_r}}{\sqrt{\lambda_{\nu_l}\lambda_{\nu_r}}} V_L(\pi_{k_l}) \cdot \Phi^\dagger(\alpha,r)\nonumber\\ 
&\cdot U_{\nu_l,\alpha}\cdot V(\pi_{k_l})\cdot V(\pi_i)\cdot U^\dagger_{\nu_l,\alpha}\cdot U_{\nu_r,\alpha} \cdot V(\pi_i)\cdot V(\pi_{k_r})\cdot U^\dagger_{\nu_r,\alpha} \cdot\Phi(\alpha,r)\cdot V_L(\pi_{k_r}).
\end{align}

In exactly the same way, plugging Eq. \eqref{eq:U_genSch} for $U_{\text{twSch}}$ into Eq. \eqref{eq:deltabasis2}, $\sqrt{\Delta}$ can be decomposed as
\begin{align}
\label{eq:sqrt_Delta_ini}\sqrt{\Delta}=\mathbb{I}\big/\sqrt{n-1}-1\big/\sqrt{n-1}\cdot\sum_{\substack{\alpha,r \\ \nu_l=\alpha+\Box\neq\theta \\ k_l,k_r=1,\dots,n-1}} & [(n-1)d_\alpha]^{-1} \frac{d_{\nu_l}}{\lambda_{\nu_l}} V_L(\pi_{k_l}) \cdot \Phi^\dagger(\alpha,r)\cdot\nonumber\\ 
& U_{\nu_l,\alpha}\cdot V(\pi_{k_l})\cdot V(\pi_{k_r})\cdot U^\dagger_{\nu_l,\alpha} \cdot\Phi(\alpha,r)\cdot V_L(\pi_{k_r}).
\end{align}

\subsection{Naimark's theorem}
\label{sub:naimarksPBT}

Armed with the decomposition of each measurement operator $\sqrt{\Pi}_i$ of the pretty-good measurement as a linear combination of products of Schur transform submatrices and permutations, it is straightforward to efficiently implement port-based teleportation.  

Explicitly, we can do this using Naimark's theorem, which can be used to implement a POVM using a combination of a unitary transformation and a projective measurement.  For our case, we will need to construct a unitary matrix $U$ satisfying
\begin{align}
\label{eq:U_def}U \ket{0}_I \ket{\psi}_{AA_n} &= \sum_{i=1}^{n-1} \ket{i}_I \sqrt{\Pi_i} \ket{\psi}_{AA_n}.
\end{align}
where $I$ is an ancilla system that Alice holds of $\text{log}\lceil n-1 \rceil_{2^\bcdot}$ qubits, $\lceil n-1 \rceil_{2^\bcdot}$ denoting the least upper bound of $n-1$ that is a power of $2$.  Recall that $A$ is the set of Alice's ports, while $A_n$ is the quantum system that will hold the state to be teleported.   

Acting on the initial state of our system 
\begin{equation}
    \rho_{\text{ini}} \equiv\ket{0}\bra{0}_{I}\otimes \Phi_{AB} \otimes \eta_{A_n}, 
\end{equation}
where $\Phi_{AB}$ is the joint entangled resource state on Alice and Bob's ports and $\eta_{A_n}$ is the state to be teleported, the Naimark unitary $U$ maps it to the target state
\begin{align}
\rho_G &\equiv \left( U \otimes \mathbb{I}_B \right) \rho_{\text{ini}} \left( U^\dagger \otimes \mathbb{I}_B \right) \\
&=\sum_{j,k=1}^{n-1} \ket{j}\bra{k}_{I} \otimes \left( \sqrt{\Pi_j}  \otimes \mathbb{I}_B \right) \Phi_{AB} \otimes \eta_{A_n} \left( \sqrt{\Pi_k}  \otimes \mathbb{I}_B \right).
\end{align}
Alice can then finish performing the POVM by projectively measuring the ancilla $I$ in the computational basis.  Outcome $i$ will occur with probability 
\begin{align}
    p(i) &= \Tr \left[ (\ket{i}\bra{i}_I \otimes \mathbb{I}_{AA_nB}) \rho_G \right] \\
    &= \Tr\left[ \left( \Pi_i  \otimes \mathbb{I}_B \right) \Phi_{AB} \otimes \eta_{A_n} \right],
\end{align}
as expected~\cite{Chuang10}.  All that remains to implement port-based teleportation is for Alice to classically send the outcome $i$ to Bob, and for Bob to throw away all of the ports except $B_i$.  These are constant time operations\textemdash as is the projective measurement\textemdash so efficiently implementing PBT comes down to being able to efficiently implement the unitary $U$.

\subsubsection{Constructing $U$ using block-encodings}

One strategy to implement the Naimark unitary $U$ in practice is to break it down into two conditional operations.  The first is a unitary matrix $U_0$ which we use to create a superposition over the relevant ancilla states.  Defined by it's first row, 
\begin{align}
U_0\ket{0}_I = \frac{1}{\sqrt{n-1}}\sum_{i=1}^{n-1}\ket{i}_I.
\label{eq:u0}
\end{align}
The second step involves conditioning the Kraus operators $\sqrt{\Pi_i}$ on the ancilla states, using the operator 
\begin{align}
\label{eq:tildeU^c}\tilde{U}^c=\sum_{i=1}^{n-1} \ket{i}\bra{i}_I \otimes \sqrt{\Pi_i} + \sum_{i=n}^{\lceil n-1 \rceil _{2^\bcdot}} \ket{i}\bra{i}_I \otimes \mathbb{I},
\end{align}
Applying these operators in succession would reproduce the action of $U$,
\begin{equation}
    \tilde{U}^c (U_0 \otimes \mathbb{I}_{A A_n}) \ket{0}_I \ket{\psi}_{A A_n} = \frac{1}{\sqrt{n-1}} \sum_{i=1}^{n-1} \ket{i}_I \sqrt{\Pi_i} \ket{\psi}_{AA_n} = \frac{1}{\sqrt{n-1}} U \ket{0}_I \ket{\psi}_{AA_n},
\end{equation}
however, this procedure is not possible in reality because the operator $\tilde{U}^c$ as defined is not unitary.  The non-unitary nature of $\tilde{U}^c$ originates in the fact that the Kraus operators $\sqrt{{\Pi_i}}$ are Hermitian instead of unitary.  We can get around this problem by substituting them with \textit{unitary block-encodings}.

\subsubsection{Unitary block encodings}

Suppose that for each Kraus operators $\sqrt{\Pi_i}$, we have an $(\alpha,a,\delta)$-block-encoding we call $U^c(i)$.  That is, $U^c(i)$ is a unitary acting on the $n$ qubits in $A A_n$, as well as $a$ ancilla qubits, with $\sqrt{\Pi_i}$ approximately encoded in the upper left-hand block,
\begin{align}
    U^c(i) \approx
    \begin{pmatrix}
    \sqrt{\Pi_i}/\alpha & \cdot \\
    \cdot & \cdot
    \end{pmatrix}
    \Leftrightarrow
    \norm{ \sqrt{\Pi_i}-\alpha\left(\bra{0}^{\otimes a}\otimes \mathbb{I}_{A A_n}\right)U^c(i)\left(\ket{0}^{\otimes a}\otimes \mathbb{I}_{A A_n}\right) } \leqslant \delta.
\end{align}
Here, $\alpha \geqslant\norm{\sqrt{\Pi_i}}$ bounds the spectral norm of the Kraus operators.
By substituting these $(\alpha,a,\delta)$-block-encodings $U^c(i)$ for each $\sqrt{\Pi_i}$, we can construct a unitary version of $\tilde{U}^c$,
\begin{align}
\label{eq:U^c}U^c=\sum_{i=1}^{n-1} \ket{i}\bra{i}_I \otimes U^c(i) + \sum_{i=n}^{\lceil n-1 \rceil _{2^\bcdot}} \ket{i}\bra{i}_I \otimes \mathbb{I}.
\end{align}

However, applying $U^c$ in succession with $U_0$ only implements the Naimark unitary $U$ in the 0-eigenspace of the $a$ ancillae.  To get around this problem, we essentially want to project onto this eigenspace using the orthogonal projector
\begin{align}
    \tilde{\Pi} &= \ket{0}\bra{0}^{\otimes a} \otimes \mathbb{I}_{IAA_n},
\end{align}
which would give 
\begin{align}
    \tilde{\Pi} U^c (U_0 \otimes \mathbb{I}_{A A_n}) \ket{0}_I \ket{0}^{\otimes a} \ket{\psi}_{A A_n} &=  \frac{1}{\sqrt{n-1}} \sum_{i=1}^{n-1} \ket{i}_I \ket{0}^{\otimes a} \bra{0}U^c(i) \ket{0}^{\otimes a} \ket{\psi}_{AA_n}\\ 
    \label{eq:projectUc}&\approx \frac{1}{\alpha \sqrt{n-1}} \ket{0}^{\otimes a} U \ket{0}_I  \ket{\psi}_{AA_n}.
\end{align}
This procedure is equivalent to applying $U$, but only up to the constant $\frac{1}{\alpha \sqrt{n-1}}$.  To eliminate this constant, we can use \textit{oblivious amplitude amplification}.

\subsubsection{Oblivious amplitude amplification}

Putting our problem into the correct form for oblivious amplitude amplification, note that with the addition of the $a$ ancillae needed for the block-encoding, the initial and final states of our system are now
\begin{align}
\rho_{\text{ini}} &\equiv\ket{0}\bra{0}_{I}\otimes \ket{0}\bra{0}^{\otimes a}\otimes \Phi_{AB} \otimes \eta_{A_n} \\
\rho_G &\equiv \sum_{j,k=1}^{n-1} \ket{j}\bra{k}_{I} \otimes  \ket{0}\bra{0}^{\otimes a} \otimes \left( \sqrt{\Pi_j}  \otimes \mathbb{I}_B \right) \Phi_{AB} \otimes \eta_{A_n} \left( \sqrt{\Pi_k}  \otimes \mathbb{I}_B \right).
\end{align}
The operator we wish to implement is 
\begin{align}
    W &=(\mathbb{I}^{\otimes a} \otimes U)\Pi \\
    \Pi &= \ket{0}\bra{0}^{\otimes a} \otimes \ket{0}\bra{0}_{I} \otimes \mathbb{I}_{AA_n},
\end{align}
which is an isometry that maps $\rho_{\text{ini}}$ to $\rho_G$.  We have included the projector $\Pi$ to restrict the action of $U$ to the 0-eigenspace of the $I$ ancilla, which contains $\rho_{\text{ini}}$, where its behavior is uniquely specified by Eq. \eqref{eq:U_def}. 

We can then immediately read off from Eq. \eqref{eq:projectUc} that
\begin{align}
\label{eq:W_PiUUPi}\norm{\frac{1}{\alpha\sqrt{n-1}} W - \tilde{\Pi} U^c U_0 \Pi} \leqslant \varepsilon
\end{align}
where $\norm{\cdot}$ is the matrix spectral norm induced from the vector $l^2$-norm $\norm{\ket{\cdot}}$, which we will use to quantify error throughout this paper.  Here, the error $\varepsilon$ originates from the error $\delta$ in the block-encodings of the Kraus operators, as well as the implementation error of $U_0$.

This relationship is the starting point for oblivious amplitude amplification~\cite{Gilyen18,Berry14,Scott12}.  We can adjust $\varepsilon$ so that we can replace $\frac{1}{\alpha \sqrt{n-1}}$ with $\sin{\frac{\pi}{2 m}}$, where $m$ is a positive, odd natural number.  It is then possible to construct a unitary matrix $\tilde{V}$ that is an amplified version of $U^c U_0$, such that
\begin{align}
\norm{W - \tilde{\Pi} \tilde{V} \Pi} \leqslant 2m\varepsilon.
\end{align}

The unitary $\tilde{V}$ will consist of an alternating phase modulation sequence \cite{Gilyen18}
\begin{align}
\tilde{V} = (-1)^{\frac{m-1}{2}} e^{i\phi_1(2\tilde{\Pi} - \mathbb{I})} V \prod_{j=1}^{(m-1)/2} (e^{i\phi_{2j}(2\Pi - \mathbb{I})} V^\dagger e^{i\phi_{2j+1}(2\tilde{\Pi} - \mathbb{I})} V),
 \end{align}
where $V = U^c U_0$ and $\phi_1=(1-m)\pi/2$, $\phi_2,\dots,\phi_m=\pi/2$.  It can be implemented by using the fact $e^{i\phi(2\Pi - \mathbb{I})}=C_{\Pi}\text{NOT}(\mathbb{I}\otimes e^{-i\phi \sigma_z}) C_{\Pi}\text{NOT}$, where the phase gate $e^{-i\phi \sigma_z}$ acts on an ancilla qubit and $C_{\Pi}\text{NOT}\equiv \Pi\otimes X + (\mathbb{I}-\Pi)\otimes \mathbb{I}$.  The circuit therefore uses one ancilla qubit and performs $m$ $V$, $V^\dagger$, $C_{\Pi}\text{NOT}$, $C_{\tilde{\Pi}}\text{NOT}$ and $e^{i\phi\sigma_z}$ gates in total.

Notice that in the case $\varepsilon =0$, $\tilde{\Pi} \tilde{V} \Pi$ is equivalent to W, so it must be an isometry that maps from $\text{img}(\Pi)$ to $\text{img}(\tilde{\Pi})$, like $W$. 
 As a result, $\tilde{V}$ must also be an isometry that maps into $\text{img}(\tilde{\Pi})$, the $0$-eigenspace of the $a$ ancillae, otherwise, the projector $\tilde{\Pi}$ would act non-trivially and $\tilde{\Pi} \tilde{V} \Pi$ wouldn't be an isometry.
This means that applying $\tilde{V}$ leaves the $a$ ancillae unentangled from the rest of the system, despite the fact that it is constructed from the operator $U^c$, which in general can be entangling. 
We then have
 \begin{align}
\rho_G=W\rho_{\text{ini}} W^\dagger \approx \tilde{\Pi} \tilde{V} \rho_{\text{ini}} \tilde{V}^\dagger \tilde{\Pi}  \approx \tilde{V} \rho_{\text{ini}} \tilde{V}^\dagger,
\end{align}
 where the approximation comes from relaxing the condition $\varepsilon = 0$, which means that $\tilde{V}$ alone is an implementation of $W$, and thus of the Naimark unitary.

 So, in summary, we have seen that we can implement the pretty-good measurement by block-encoding the Kraus operators $\sqrt{\Pi_i}$ to construct the unitary $U^c$, which we can then amplify to find the unitary $\tilde{V}$ that approximates the Naimark unitary.  The efficiency of this implementation depends entirely on being able to find efficient block-encodings $U^c(i)$ of each $\sqrt{\Pi_i}$.  In the following subsection, we will see how the representation theory, in particular the  decompositions Eq.~\eqref{eq:Pidecomp} and Eq.~\eqref{eq:Deltadecomp}, enable us to do exactly this.
 
 \subsection{Block-encoding of $\sqrt{\Pi_i}$}

In this subsection, we will discuss how the decompositions Eq.~\eqref{eq:Pidecomp} and Eq.~\eqref{eq:Deltadecomp} make it possible to find efficient unitary block encodings of the Kraus operators $\sqrt{\Pi_i}$, and therefore efficiently-implement port-based teleportation.   

Recall that the decomposition of $\sqrt{\tilde{\Pi}_i}$ is given by 
\begin{align}
\sqrt{\tilde{\Pi}_i}=\sum_{\substack{\alpha,r \\ \nu_l,\nu_r=\alpha+\Box\neq\theta \\ k_l,k_r=1,\dots,n-1}}& C(\alpha, \nu_l) C(\alpha, \nu_r) V_L(\pi_{k_l}) \cdot \Phi^\dagger(\alpha,r)\nonumber\\ 
&\cdot U_{\nu_l,\alpha}\cdot V(\pi_{k_l})\cdot V(\pi_i)\cdot U^\dagger_{\nu_l,\alpha}\cdot U_{\nu_r,\alpha} \cdot V(\pi_i)\cdot V(\pi_{k_r})\cdot U^\dagger_{\nu_r,\alpha} \cdot\Phi(\alpha,r)\cdot V_L(\pi_{k_r}),
\end{align}
where we have consolidated the coefficients
\begin{align}
    C(\alpha,\nu_l)&\equiv [(n-1)d_\alpha-d_\theta]^{-\frac{1}{4}} [(n-1)d_\alpha]^{-\frac{3}{4}} \frac{d_{\nu_l}}{\sqrt{\lambda_{\nu_l}}}.
\end{align}
The building blocks of this decomposition $U_{\nu, \alpha}$ and $\Phi (\alpha,r)$ are manifestly non-unitary because they encode small submatrices of the unitary $(n-1)$- and $(n-2)$-qudit Schur transforms, respectively.  As a result, the Kraus operator is then a superposition of non-unitary terms, weighted by the coefficients $C(\alpha, \nu_l)$. 

Our strategy to block-encode $\sqrt{\tilde{\Pi}_i}$ will be to use a method inspired by the Linear Combination of Unitaries (LCU) algorithm~\cite{Gilyen18, Berry15, Childs17}.  This procedure allows us to encode the ``inner'' portion of the summand using the entire $(n-1)$-qudit Schur transform, rather than non-unitary submatrices, while weighting the different portions of the basis with $C(\alpha,\nu)$.  By handling both the $(n-1)$-qudit Schur transform and the coefficients, this method performs the bulk of the work of the block-encoding.  It may also be of more general interest because it enjoys the additional efficiency, over the LCU algorithm, of using the Schur-transformed permutation operator $U_{\text{Sch}}^{(n-1)} \cdot V(\pi_i)\cdot V(\pi_{k_r})\cdot U_{\text{Sch}}^{\dagger (n-1)}$ only once, rather than repeating the Schur transform to extract the various irrep blocks.

\subsubsection{Method inspired by the LCU algorithm}

Sketching out the method, our goal is to first block encode $\sum_{\alpha=1}^{n_\alpha} \ket{e_\alpha}\bra{e_\alpha}\otimes O(\alpha, k,i)$, where
\begin{align}
    \label{eq:O_small}O(\alpha,k_l,i) &\equiv \sum_{\nu_l=\alpha+\Box\neq\theta} C(\alpha,\nu_l) U_{\nu_l,\alpha}\cdot V(\pi_{k_l})\cdot V(\pi_i)\cdot U^\dagger_{\nu_l,\alpha} \\
    &= \sum_{\nu_l=\alpha+\Box\neq\theta} \sum_{k_\alpha, k_\alpha' = 1}^{n_{d_\alpha}} C(\alpha,\nu_l) \ket{e_{k_\alpha}} \bra{r_{\nu_l},\nu_l,\alpha,k_\alpha}^{(n-1)}_{\text{Sch}} V(\pi_{k_l}) \cdot V(\pi_i)  \ket{r_{\nu_l},\nu_l,\alpha,k'_\alpha} ^{(n-1)}_{\text{Sch}} \bra{e_{k_\alpha'}}.
\end{align}
Here we have switched to using the practical form of the Schur transform, where the size of each register in the Schur basis is padded by ancilla qubits to the minimum size necessary for any Schur basis state.  The constants $n_{r_\nu}$, $n_\nu$,  $n_\alpha$ and $n_{d_{\alpha}}$ denote the padded number of basis states contained in the $\ket{q_{n-1} = r}$, $\ket{\lambda_{n-1}=\nu}$, $\ket{\lambda_{n-2}=\alpha}$ and $\ket{p_{n-2}=k_\alpha}$ registers, respectively.  $O(\alpha,k_l,i)$ acts on the $\ket{k_\alpha}$ register, and is a $n_{d_\alpha}\times n_{d_\alpha}$ matrix, and for the block-encoding we introduce two ancilla registers $\ket{\text{Anc}_0} = \ket{r} \ket{\nu}$, named for their roles when combining this block-encoding with the rest of $\sqrt{\tilde{\Pi}_i}$.

The block-encoding uses four controlled unitary matrices: $P_1$, $P_2$, $P_{L}[x]$ and $P_{R}[x]$.  To define them, first enumerate the set of $S(n-1)$ irreps created by adding a box to $\alpha$ as $\nu^\alpha=\alpha+\Box\neq\theta \equiv\{\nu_1^\alpha, \nu_2^\alpha,\dots,\nu_{n_{\alpha+\Box}}^\alpha\}$ where $n_{\alpha+\Box}$ is the size of the set.  $P_{L}[x]$ and $P_{R}[x]$ are matrices designed to introduce the $C(\alpha, \nu)$ coefficients, and are defined as 
\begin{align}
P_L \equiv \sum_{\alpha} P_L[x](\alpha) \otimes \ket{e_\alpha} \bra{e_\alpha}_\alpha,\quad\quad P_R \equiv \sum_{\alpha} P_R[x](\alpha) \otimes \ket{e_\alpha} \bra{e_\alpha}_\alpha
\end{align}
where $P_L[x](\alpha)$ and $P_R[x](\alpha)$ are unitaries that only mix the basis states $ \{ |0,e_{\nu_1^\alpha} \rangle_{r \nu}, \linebreak[1] \dots,|0,e_{\nu_{n_{\alpha+\Box}}^\alpha} \rangle_{r \nu}, \linebreak[1] |1,0\rangle_{r \nu }, \linebreak[1] |1,1\rangle _{r \nu} \} $ and leave other basis states in the $\ket{r}$ and $\ket{\nu}$ registers invariant.  Specifying each of them by the first-row of their non-identity submatrix, 
\begin{align}
\label{eq:P_L_el}\bra{0,e_{\nu_1^\alpha}}P_{L/R}[x](\alpha)\ket{0,e_{\nu_j^\alpha}}_{r \nu} &= \frac{1}{x}\sqrt{C(\alpha,\nu_j^\alpha)}\\
\bra{0,e_{\nu_1^\alpha}}P_{L}[x](\alpha)\ket{1,0}_{r \nu} &= \sqrt{C_{\text{rem}}(\alpha)}\\
\bra{0,e_{\nu_1^\alpha}}P_{L}[x](\alpha)\ket{1,1}_{r \nu} &= 0\\
\bra{0,e_{\nu_1^\alpha}}P_{R}[x](\alpha)\ket{1,0}_{r \nu} &= 0\\
\bra{0,e_{\nu_1^\alpha}}P_{R}[x](\alpha)\ket{1,1}_{r \nu} &= \sqrt{C_{\text{rem}}(\alpha)}\\
\label{eq:cons1}C_{\text{rem}}(\alpha)\equiv 1-\sum_{j=1}^{n_{\alpha+\Box}}\frac{1}{x^2} C&(\alpha,\nu_j^\alpha)\geqslant 0\quad \forall\alpha,
\end{align}
where $x\in\mathbb{R}^+$ is a scaling parameter that satisfies the constraints Eq. \eqref{eq:cons1}, and $C_{\text{rem}}$ is defined to ensure each of these matrices is unitary.  Notice that we put the $r$ register into the $\ket{1}_r$ state to mark where the $\sqrt{C_{\text{rem}}}$ terms appear; generically there could be some $S(n-1)$ irrep copy such that $\ket{e_r, e_\nu}_{r \nu} = \ket{1,0}_{r \nu}$ or $\ket{e_r, e_\nu}_{r \nu} = \ket{1,1}_{r \nu}$ but it can be easily checked that this does not cause any issues. 

We have also selected the computational basis state labeling the fixed irrep $\nu_1^\alpha \in v^\alpha$ to trigger $P_{L/R}[x]$ to introduce the coefficients for all of the $S(n-1)$ irrep states $\ket{0, e_{\nu^\alpha_j}}_{r\nu}$.  The controlled matrix $P_2$ will play the role of putting the $\nu$ ancilla in this state to activate $P_{L/R}[x]$.  It is defined as 
\begin{align}
    P_2 &\equiv \sum_{\alpha} \left( \ket{0,e_{\nu_1^\alpha}} \bra{0,0}_{r \nu} + \ket{0,0} \bra{0,e_{\nu_1^\alpha}}_{r \nu} + \sum_{e_\nu \neq e_{\nu_1^\alpha}, e_\nu \neq 0} \ket{0,e_\nu}\bra{0,e_\nu}_{r \nu} + \sum_{e_r\neq 0} \ket{e_r}\bra{e_r}_r \right) \otimes \ket{e_\alpha} \bra{e_\alpha}_\alpha.
\end{align}
Finally, $P_1$ is defined to put the $r$ register in the correct state for transforming to the Schur basis, and is given by
\begin{align}
    P_1 &\equiv  \sum_{\nu} \left( \ket{e_{r_{\nu}}} \bra{0}_{r} + \ket{0} \bra{e_{r_{\nu}}}_{r} + \sum_{e_r \neq e_{r_{\nu}}, e_r \neq 0} \ket{e_r}\bra{e_r}_{r} \right) \otimes \ket{e_\nu} \bra{e_\nu}_\nu 
\end{align}

Using these matrices, the block encoding $U[x^2] (k,i)$ of $\sum_{\alpha=1}^{n_\alpha} \ket{e_\alpha}\bra{e_\alpha}\otimes O(\alpha, k,i)$ is given by 
\tikzset{
leftinternal/.append style={teal, xshift=-0.1cm, scale=0.95},
rightinternal/.append style={teal, xshift=0.1cm, scale=0.95}
}

\tikzscale{0.73}{\begin{quantikz}[row sep={0.7cm,between origins}]
\lstick{$\ket{\text{Anc}_0}$} & \gate[wires=3][2cm]{U[x^2](k,i)} & \rstick{$\ket{\text{Anc}_0}$}\qw \\
\lstick{$\ket{\alpha}$} & & \rstick{$\ket{\alpha}$}\qw\\
\lstick{$\ket{k_\alpha}$} & & \rstick{$\ket{k_\alpha}$}\qw
\end{quantikz}$\cong$}

\vspace{-1cm}
\tikzscale{0.73}{\begin{quantikz}[row sep={0.7cm,between origins}, column sep={0.15cm}]
\lstick[2]{$\ket{\text{Anc}_0}$} & \octrl{1} & \gate[wires=2][1.4cm]{P_R^\dagger[x](\alpha)} & \gate{P_1(\nu)} & \gate[wires=4][2.8cm]{U_\text{Sch}^{\dagger(n-1)}} \gateinput{$\ket{r_{n-1}}$} \gateoutput{$\ket{\text{Pad}}$}  & \qw & \qw & \gate[wires=4][2.8cm]{U_\text{Sch}^{(n-1)}}\gateoutput{$\ket{r_{n-1}}$} \gateinput{$\ket{\text{Pad}}$}  &\gate{P_1(\nu)} & \gate[wires=2][1.4cm]{P_L[x](\alpha)} & \octrl{1} & \rstick[2]{$\ket{\text{Anc}_0}$}\qw \\
 & \gate{P_2(\alpha)} & & \ocontrol[rectangle]{}\vqw{-1} & \gateinput{$\ket{\nu}$}\gateoutput{$\ket{1}$} & \gate[wires=3][2.2cm]{V(\pi_i)} & \gate[wires=3][2.2cm]{V(\pi_k)} &\gateoutput{$\ket{\nu}$}\gateinput{$\ket{1}$} & \ocontrol[rectangle]{}\vqw{-1} & & \gate{P_2(\alpha)} &\qw \\
\lstick{$\ket{\alpha}$} & \ocontrol[rectangle]{}\vqw{-1} & \ocontrol[rectangle]{}\vqw{-1} & \qw & \gateinput{$\ket{\alpha}$}\gateoutput{\vdots} & & & \gateoutput{$\ket{\alpha}$}\gateinput{\vdots} & \qw & \ocontrol[rectangle]{}\vqw{-1} & \ocontrol[rectangle]{}\vqw{-1} & \rstick{$\ket{\alpha}$}\qw\\
\lstick{$\ket{k_\alpha}$} & \qw & \qw & \qw & \gateinput{$\ket{k_\alpha}$}\gateoutput{$\ket{n-1}$} & & & \gateoutput{$\ket{k_\alpha}$}\gateinput{$\ket{n-1}$} & \qw & \qw & \qw & \rstick{$\ket{k_\alpha}$}\qw
\end{quantikz}}
We can verify that this is a block-encoding by post-selecting for the $\ket{0,0}_{r\nu}$ state on the ancilla $\ket{\text{Anc}_0}$.  This gives
\begin{align}
&\bra{0,0,e_\alpha, e_{k_\alpha}} U[x^2](k,i) \ket{0,0,e_\alpha, e_{k'_\alpha}} \\
&= \bra{0,0,e_\alpha, e_{k_\alpha}} P_2 \cdot P_L[x] \cdot P_1 \cdot U_{\text{Sch}}^{(n-1)} \cdot V(\pi_k) \cdot V(\pi_i) \cdot U^{\dagger (n-1)}_{\text{Sch}} \cdot P_1 \cdot P^{\dagger}_R[x] \cdot P_2  \ket{0,0,e_\alpha, e_{k'_\alpha}} \\
&= \bra{0,e_{\nu_1^\alpha},e_\alpha, e_{k_\alpha}} P_L[x] \cdot P_1 \cdot U_{\text{Sch}}^{(n-1)} \cdot V(\pi_k) \cdot V(\pi_i) \cdot U^{\dagger (n-1)}_{\text{Sch}} \cdot P_1 \cdot P^{\dagger}_R[x] \ket{0,e_{\nu_1^\alpha},e_\alpha, e_{k'_\alpha}}
\end{align}
where $P_2(\alpha)$ has acted on each side to put the $\ket{\nu}$ register of the ancilla in the computational basis state labeling $\nu_1^\alpha \in \nu^\alpha$.  The first rows of $P_L[x](\alpha)$ and $P_R[x](\alpha)$ then introduce the coefficients $C(\alpha, \nu_j)$.  After $P_1$ has acted on the $r$ register, this gives
\begin{align}
    &\bra{0,0,e_\alpha, e_{k_\alpha}} U[x^2](k,i) \ket{0,0,e_\alpha, e_{k'_\alpha}} \\
    &= \Bigg(\sqrt{C_{\text{rem}}(\alpha)} \bra{1,0,e_\alpha, e_{k_\alpha}}  + \sum_{\nu_j\in \nu^\alpha}\frac{1}{x} \sqrt{C(\alpha,\nu_j)} \bra{e_{r_{\nu_j}},e_{\nu_j},e_\alpha, e_{k_\alpha}} \Bigg)  U_{\text{Sch}}^{(n-1)} \\
    &\times  V(\pi_k) \cdot V(\pi_i) \cdot U^{\dagger (n-1)}_{\text{Sch}}  \Bigg(  \sqrt{C_{\text{rem}}(\alpha)} \ket{1,1,e_\alpha, e_{k_\alpha'}} + \sum_{\nu_m\in \nu^\alpha}\frac{1}{x} \sqrt{C(\alpha,\nu_m)} \ket{e_{r_{\nu_m}},e_{\nu_m},e_\alpha, e_{k_\alpha'}}\Bigg),
\end{align}
assuming $(1,0)$ or $(1,1)$ do not equal $(e_{r_\nu}, e_\nu)$ for any $\nu$ (not necessarily $\nu \in \nu^\alpha$).  In this case, we see no cross-terms proportional to $\sqrt{C_{\text{rem}}}$ will survive.  In the case $(1,0)$ or $(1,1)$ \textit{does} correspond to some $(e_{r_\nu}, e_\nu)$, $P_1$ will have acted to put the $r$ register of the term(s) proportional to $\sqrt{C_{\text{rem}}}$ in the $\ket{0}$ state, so that the cross-terms will still vanish.

Now we can perform the Schur transform, which is defined as 
\begin{equation}
    U^{(n-1)}_{\text{Sch}} = \sum_{r_\omega} \sum_{\omega \vdash n-1} \sum_{\xi_\omega = \omega - \Box} \sum_{j_{\xi_\omega} = 1}^{d_{\xi_\omega}} \ket{e_{r_\omega}, e_\omega, e_{\xi_\omega}, e_{j_{\xi_\omega}}} \bra{r_\omega, \omega, \xi_\omega, j_{\xi_\omega}}^{(n-1)}_{\text{Sch}}.
\end{equation}
As the permutation $V(\pi_k) \cdot V(\pi_i) \in S(n-1)$ is $\nu$-block-diagonalized in the Schur basis, all off-diagonal blocks will vanish.  We then see 
\begin{align}
    &\bra{0,0,e_\alpha, e_{k_\alpha}} U[x^2](k,i) \ket{0,0,e_\alpha, e_{k'_\alpha}} \\
    =& \sum_{\nu_j\in \nu^\alpha}\frac{1}{x^2} C(\alpha,\nu_j) \bra{r_{\nu_j},\nu_j,\alpha, k_\alpha}^{(n-1)}_{\text{Sch}} V(\pi_k) \cdot V(\pi_i) \ket{r_{\nu_j},\nu_j,\alpha, k_\alpha'}^{(n-1)}_{\text{Sch}} \\
    \approx & \frac{1}{x^2} \bra{e_{k_\alpha}} O(\alpha, k, i) \ket{e_{k_\alpha'}}
\end{align}
The block-encoding is approximate because the numerical Schur transform, as well as the numerical operators $P_L[x]$ and $P_R[x]$, can only be implemented up to target accuracies, which trades-off with efficiency.  Calling this error $\varepsilon_2$, this means $U[x^2](k,i)$ is therefore an $(x^2, \text{log}\ n_{r_\nu} + \text{log}\ n_\nu, \varepsilon_2)$-encoding of $\sum_{\alpha=1}^{n_\alpha} \ket{e_\alpha}\bra{e_\alpha}\otimes O(\alpha, k,i)$.

\subsubsection{Block-encoding the rest of $\sqrt{\Pi_i}$}

From here it is relatively straightforward to build up a block-encoding of the rest of the Kraus operator,
\begin{align}
\sqrt{\tilde{\Pi}_i}=\sum_{\substack{\alpha,r \\ k_l,k_r=1,\dots,n-1}}& V_L(\pi_{k_l}) \cdot \Phi^\dagger(\alpha,r) 
\cdot O(\alpha, k_l, i) \cdot O(\alpha, k_r, i) \cdot\Phi(\alpha,r)\cdot V_L(\pi_{k_r}).
\end{align}
The first step is to take advantage of the fact that we had block-encoded $\sum_{\alpha=1}^{n_\alpha} \ket{e_\alpha}\bra{e_\alpha}\otimes O(\alpha, k,i)$, which each entangles $O(\alpha, k,i)$ with the $\alpha$ register.  We can do this by rewriting the Kraus operator as 
\begin{align}
    \label{eq:start}\sqrt{\tilde{\Pi}_i} = \sum_{k_l,k_r=1,\dots,n-1} V_L(\pi_{k_l})\cdot \Phi^\dagger \cdot O_{\text{cen}}(i,k_l,k_r) \cdot \Phi \cdot V_L(\pi_{k_r}),
\end{align}
where 
\begin{align}
    \label{eq:O_cen}O_{\text{cen}}(i,k_l,k_r) &\equiv \sum_{r = 1}^{n_r}\sum_{\alpha = 1}^{n_\alpha} \ket{e_r, e_\alpha} \bra{e_r, e_\alpha} \otimes O(\alpha,k_l,i)\cdot O^\dagger (\alpha,k_r,i) \\
    \Phi &\equiv \sum_{r = 1}^{n_r} 
    \sum_{\alpha = 1}^{n_\alpha} \ket{e_r, e_\alpha} \otimes \Phi(\alpha, r).
\end{align}

We can then use the essential fact that the product of block-encoded matrices faithfully represents the product of the encoded matrices, with only additive error.  More precisely, if $U$ is an $(\alpha, a,\delta)$-block-encoding of an $s$-qubit operator $A$, and $V$ is an $(\beta, b,\varepsilon)$-block-encoding of an $s$-qubit operator $B$, then $(\mathbb{I}_{2^b}\otimes U)(\mathbb{I}_{2^a}\otimes V)$, is an $(\alpha\beta, a+b,\alpha\varepsilon+\beta\delta)$-block-encoding of $AB$~\cite{Gilyen18}.  This means that multiplying $U[x^2](k_l, i)$ and $U^\dagger[x^2](k_r, i)$ immediately gives us a $(x^4,2\text{log}\ n_{r_{\nu}} + 2\text{log}\ n_\nu, 2x^2\varepsilon_2)$-block-encoding of $O_{\text{cen}}(i,k_l,k_r)$.

It also means that by block-encoding $\Phi$, we can build up the entire summand multiplicatively.  As it turns out, it is much easier to block-encode $\Phi$, which contains the $(n-2)$-qudit Schur basis states, than it was to block-encode $\sum_{\alpha=1}^{n_\alpha} \ket{e_\alpha}\bra{e_\alpha}\otimes O(\alpha, k,i)$.  This is because, in the earlier case, the different parts of the $(n-1)$-qudit Schur basis were weighted by different coefficients.  In contrast, for $\Phi$, we can simply use the entire $(n-2)$-qudit Schur transform (as well as a unitary to contribute the maximally entangled state).  Block-encoding the entire summand is then relatively straightforward, and entails padding the factors so that the matrices are of the appropriate size, multiplying the block-encodings, and keeping track of the ancillas involved.  The details can be found in Appendix~\ref{app:circuit}.

As $\sqrt{\Delta}$ depends only on the twisted Schur transform and the identity, it can be block-encoded in exactly the same way.  Then, armed with a block-encoding of the summand, we can generate the superposition over $k_l$ and $k_r$ in Eq.~\eqref{eq:start} with a simple unitary circuit, and add on the block-encoded contribution from $\sqrt{\Delta}$ at the same time, to find a block-encoding of $\sqrt{\Pi_i}$.  This circuit can also be found in Appendix~\ref{app:circuit}.

In summary, we have seen that the decompositions Eq.~\eqref{eq:Pidecomp} and Eq.~\eqref{eq:Deltadecomp} made it possible to find unitary block-encodings of each $\sqrt{\Pi_i}$.  We first block-encoded the portion that depends on the $(n-1)$-qudit Schur basis, incorporating the coefficients weight the different parts of the basis using a method-inspired by the LCU algorithm, then built up the decompositions multiplicatively.   
As we have seen in the previous subsections, when used with the Naimark extension and oblivious amplitude amplification, these block-encodings make it possible to implement port-based teleportation.  

Moreover, this implementation is manifestly efficient: the decomposition in Eq.~\eqref{eq:start} has a number of terms that is polynomial in $n$, and the block-encoding of each term can be built up multiplicatively with a short-depth unitary circuit that depends on the $(n-1)$- and $(n-2)$-qudit Schur transforms.  Implementing this block-encoding of the Kraus operator therefore inherits the efficiency of the Schur transform, and is polynomial in $n$ and $d$, as is the Naimark extension.  The scaling factors of the block-encodings remain $\poly{n,d}$ throughout the multiplicative construction of the block-encodings, meaning that the oblivious amplitude amplification, which is needed to amplify them out, will also be $\poly{n,d}$.  All together, this means that port-based teleportation can be performed efficiently.  

Finally, the error of this algorithm can be made constant in $n$ and $d$ by setting high enough target accuracies for each block-encoding as well as for the numerical implementations of matrices like the Schur transform.  This process merely has a $\polylog{n,d}$ overhead in time and space complexity, so doesn't affect the efficiency.  The next section will summarize all of the results about complexities and errors.

%% file: sections/complexity.tex
\section{Circuit complexity}
\label{sec:complexity}
In a port-based teleportation protocol with $n$ $d$-dimensional ports on the sender's side, our algorithm is efficient with a $\poly{n,d}$ time complexity, an $O(n^{2d})$ space complexity (including $O[d^2\log(n)]$ ancilla qubits), 
and a constant error in spectral norm independent of the dimensional parameters $n$ and $d$. The space cost originates from the $O(n^{d})$ irreps, and while prohibitive, is still polynomial for fixed $d$. The time and space complexity for a single shot are determined primarily by the cost of the $(n-1)$- and $(n-2)$-qudit Schur transforms, and to successfully perform the amplitude amplification the $O(n^{5/2}d^3)$ shots are required.

In Appendix~\ref{app:complexity}, we analyze the complexity and error propagation in detail in two stages. In the first stage, the time and space complexity and error of the entire circuit $\tilde{V}$ (or U) is expressed in terms of $n$, $d$, and the gate accuracies. Using the Solovay-Kitaev theorem~\cite{Kitaev97, Dawson05}---each one-or-two-qubit gate can be approximated to an accuracy $\epsilon$ in time $\polylog{1/\epsilon}$ by a product of $\polylog{1/\epsilon}$ gates from a fixed and finite set---the time and space complexity would be measured by the number of gates from a universal set, e.g., the set of Clifford operators combined with a $\pi/8$ rotation, and the gate accuracies would be the difference norm between the exact and implemented one-or-two-qubit gates. In the second stage, we then require the total error to be constant in $n$ and $d$, and parameterize the target gate accuracies by $n$ and $d$. We find that this requires $\poly{n,d}^{-1}$ gate accuracies, giving a $\polylog{1/\epsilon}=\polylog{n,d}$ overhead in both runtime and circuit size, and therefore not affecting the asymptotic complexities.

%% file: sections/discussion.tex
\section{Discussion}
\label{sec:discussion}
We have presented a $\operatorname{poly}(d,n)$ time algorithm for port-based teleportation of a single qudit system using $n$ maximally entangled states of $n$ qudits as a resource. The algorithm amounts to performing the pretty good measurement on a family of states, one associated with each possible port. The states $\rho_i$ are elements of the partially transformed permutation algebra $A_n^{(k)}(d)$.  Their sum $\rho$ is additionally symmetric under permutations of the $n-1$ untransposed systems. We therefore construct a $\operatorname{poly}(d,n)$ algorithm which we call the twisted Schur transform for transforming to an irrep basis of the algebra $A_n^{(k)}(d)$ which is furthermore reduced with respect to the restriction to $S(n-1)$, thereby diagonalizing $\rho$. After transforming to that basis, the pretty good measurement can be implemented by a combination of block-encodings, linear combinations and oblivious amplitude amplification.

In addition to the applications of port-based teleportation discussed in section~\ref{subsec:applications}, we anticipate that the twisted Schur transform could have further applications to the implementation of unitarily equivariant quantum channels and entanglement detection.
As we show in our companion paper, we can generalize our algorithm for the twisted Schur transform to efficiently perform an \textit{even-more-twisted} Schur transform to the irrep basis of $A^{(k)}_n(d)$ and its commutant $\mathbb{C}\big\{\chi^{(k)}(U)=U^{\otimes n-k}\otimes U^{*\otimes k}\big\}$, again focusing on the ideal $\mathcal{H}_M$ that supports all elements of $A^{(k)}_n(d)$ that nontrivially permute and transpose all of the last $k$ systems~\cite{inducedRepPrep}.

\medskip
\noindent\textbf{Unitarily equivariant quantum channels.}  As pointed out in~\cite{Grinko23}, the even-more-twisted Schur transform provides an avenue for the generic implementation of all unitarily equivariant quantum channels. This is enabled by the channel-state correspondence, where each unitarily equivariant quantum channel is represented by a $U^{\otimes n-k} \otimes U^{*\otimes k}$-invariant Choi state. Transforming to the Kraus or Stinespring representation, for example, could make the algorithm concrete~\cite{Buhrman22}.

Spelling this out, the correspondence~\cite{Choi75,Jamiolkowski72} declares that each linear map $\Lambda$ from $\text{End}(\mathbb{C}^{d_\text{in}})$ to $\text{End}(\mathbb{C}^{d_\text{out}})$ is isomorphic to a Choi operator $\rho_\Lambda\in \text{End}(\mathbb{C}^{d_\text{out}}\otimes \mathbb{C}^{d_\text{in}})$ as
\begin{align}
    \label{eq:CJ_iso}\rho_\Lambda = \frac{1}{d}\sum_{i,j=1}^{d_\text{in}}   \Lambda(\ket{i}\bra{j})  \otimes \ket{i}\bra{j},\quad \Lambda(\sigma) = \Tr_2[\rho_\Lambda(\mathbb{I}_{d_\text{out}} \otimes \sigma^T)].
\end{align}
Here, $\Lambda$ being a completely positive (CP) and trace-preserving (TP) quantum channel corresponds to $\rho_\Lambda$ being a positive semidefinite (PSD) quantum state that obeys $\Tr_1[\rho_\Lambda]=\mathbb{I}_{d_\text{in}}/d$. Setting $\mathbb{C}^{d_\text{in}} = (\mathbb{C}^{d})^{\otimes k}$ and $\mathbb{C}^{d_\text{out}} = (\mathbb{C}^{d})^{\otimes n-k}$, a unitarily equivariant channel satisfies
\begin{align}
    \label{eq:equivariance}\Lambda(U^{\otimes k}\sigma U^{\dagger\otimes k}) = U^{\otimes n-k}\Lambda(\sigma) U^{\dagger\otimes n-k}
\end{align}
for any $U\in U(d)$. It is then easy to see from Eq.~\eqref{eq:CJ_iso} that $[\rho_\Lambda, U^{\otimes n-k} \otimes U^{*\otimes k}]=0$.

Unitarily equivariant channels play an important role in quantum information theory because the optimal channels for many figures of merit are unitarily equivariant. For example, the fully symmetrized channel $\tilde{\Lambda}(\sigma)=\int dU(d) U^{\dagger\otimes n-k}\Lambda(U^{\otimes k}\sigma U^{\dagger\otimes k})U^{\otimes n-k}$ has an entanglement fidelity no worse than that of $\Lambda$ owing to the concavity of fidelity, and therefore the optimal channel should be some  unitarily equivariant $\Lambda$ that equals its symmetrization $\tilde{\Lambda}$. Diamond norm distance favors this channel, too.

An instance of an optimal unitarily equivariant channel is that for port-based teleporting a qudit state using Bell pairs as the resource state, as defined in Eq.~\eqref{eq:channel_PBT}. The corresponding Choi state is $\sum_i \Tr_{\overline{B_i}}(E^i_{B_i \overline{B_i}A_n})$, where $\{E^i\}$ is the POVM used to execute the teleportation. So, essentially, the optimality of the channel determines that the POVM $\{E^i\}$ must be $U^{n-1}\otimes U^{*}$-invariant. Other examples include the optimal channels for asymmetric quantum cloning~\cite{Nechita23,Studzinski_cloning14,Cerf00}, quantum majority vote~\cite{Buhrman22}, quantum purification~\cite{Keyl01}, quantum neural network~\cite{Nguyen22} and encoding in quantum error correction~\cite{Kong22_Qec}. Semidefinite programming can be used to find such optimal channels~\cite{Grinko23} by parametrizing all Choi states~\cite{Buhrman22,Nguyen22} using the representation theory of $U^{\otimes n-k}\otimes U^{*\otimes k}$ and restricting them to be PSD variables with maximally mixed second states. The twisted Schur transform can in turn be used to implement those channels, as just discussed.

Moreover, there is a natural generalization of port-based teleportation that is applicable when the input state consists of $k$ registers~\cite{Kopszak2021multiportbased,studzinski2022efficient}. \textit{Multiport-based teleportation} transfers the state to $k$ of the output ports instead of a single one. This procedure can achieve higher fidelities than standard port-based teleportation, albeit with similar parametric scaling. 
The algebra relevant to multiport-teleportation is $A^{(k)}_n(d)$, the algebra of permutation operators where now the last $k$ systems are transposed as compared to just the last system in standard port-based teleportation. Our even-more-twisted Schur transform, given in the companion paper~\cite{inducedRepPrep}, should therefore provide an efficient implementation of multiport-based teleportation. 
\medskip
\medskip

\noindent\textbf{Entanglement detection.}  Partial transposition and unitary conjugation symmetries play a central role in the theory of entanglement detection. A necessary condition for bipartite states to be separable, as opposed to entangled, is that they are positive under partial transposition. This condition is moreover sufficient~\cite{Peres96,Horodecki96} in the cases of $\mathbb{C}^2\otimes \mathbb{C}^2$ and $\mathbb{C}^2\otimes \mathbb{C}^3$. Nonpositivity of the partial transpose is also a necessary and sufficient condition for states to contain ``1-distillable'' entanglement in $\mathbb{C}^2\otimes \mathbb{C}^d$ for all $d \geq 2$~\cite{dur2000distillability}.

More generally, every entangled state can be detected by at least one positive (P), but not completely positive (CP), map that sends it to negative operator, although finding such a map is NP-hard. Recent progress~\cite{Bardet20} shows that the search space can be hugely constrained: for each $t\in \mathbb{Z}^+$, there exists a complete sequence of unitarily covariant $t$-positive maps that can detect every $t$-entangled state at some point of the sequence; and that the study of each unitarily covariant map can be reduced to studying a set of $(a_j,b_j)$-unitarily equivariant constituents, defined by replacing the right-hand side of Eq.~\eqref{eq:equivariance} by $U^{\otimes a_j} \otimes U^{*\otimes b_j}  \Lambda(\sigma) U^{\dagger\otimes a_j}\otimes U^{T\otimes b_j}$ and whose Choi operator instead has a $[\rho_\Lambda, U^{\otimes a_j} \otimes U^{*\otimes b_j+k}]=0$ symmetry.

Another central means of detecting entangled states is by entanglement witnesses, which measure expectation values and therefore facilitate experiments. In fact, the isomorphism in Eq.~\eqref{eq:CJ_iso} exactly relates~\cite{Horodecki09} each P (but not CP) map $\Lambda$ to an entanglement witness $\rho_\Lambda$ that is block-positive (but not PSD). Huber~\cite{Huber21} and Balanz\'o-Juand\'o et al.~\cite{Maria22} respectively exploited this duality to obtain entanglement winesses from unitarily equivariant and from $(a,b)$-unitarily equivariant maps. Just notice that they rewrote Eq.~\eqref{eq:CJ_iso} as $\Lambda(\sigma)=\Tr_2[\rho_\Lambda^{T_2}( \mathbb{I}_{d_\text{out}} \otimes \sigma)]$ and therefore acquired $U^{\otimes n}$- and $U^{\otimes a+k} \otimes U^{*\otimes b}$-invariant witnesses $\rho_\Lambda^{T_2}$. The twisted Schur transform, again, serves to implement such witnesses.

%% file: sections/appendixblock.tex
\appendix

\section{Block-encoding $\sqrt{\Pi_i}$\label{app:circuit}}

In this appendix, we will detail how to construct $(\alpha,a,\varepsilon)$-block-encodings $U^c(i)$ of each of the Kraus operators $\sqrt{\Pi_i} = \sqrt{\tilde{\Pi}_i} + \sqrt{\Delta}$ where
\begin{align}
    \sqrt{\tilde{\Pi}_i} &= \sum_{k_l,k_r=1,\dots,n-1} V_L(\pi_{k_l})\cdot \Phi^\dagger \cdot O_{\text{cen}}(i,k_l,k_r) \cdot \Phi \cdot V_L(\pi_{k_r}), \\
    \label{eq:start2}\sqrt{\Delta} &= \frac{\mathbb{I}}{\sqrt{n-1}}-\frac{1}{\sqrt{n-1}} \sum_{k_l,k_r=1,\dots,n-1} V_L(\pi_{k_l})\cdot \Phi^\dagger \cdot O'_{\text{cen}}(k_l,k_r) \cdot \Phi \cdot V_L(\pi_{k_r}).
\end{align}
Here we have rewritten the decomposition of $\sqrt{\Delta}$ given in Eq.~\eqref{eq:sqrt_Delta_ini} in a form that parallels our expression for $\sqrt{\tilde{\Pi}_i}$ by defining
\begin{align}
\label{eq:O_p_cen}O'_{\text{cen}}(k_l,k_r) &\equiv \sum_{r = 1}^{n_r} \sum_{\alpha = 1}^{n_\alpha} \ket{e_r, e_\alpha}\bra{e_r, e_\alpha} \otimes  O'(\alpha,k_l,k_r)\\
O'(\alpha,k_l,k_r) &\equiv \sum_{\nu_l=\alpha+\Box\neq\theta} C'(\alpha,\nu_l) U_{\nu_l, \alpha} \cdot V(\pi_{k_l}) \cdot V(\pi_{k_r}) \cdot U^\dagger_{\nu_l, \alpha}  \\
\label{eq:end}C'(\alpha,\nu_l)&\equiv [(n-1)d_\alpha]^{-1} \frac{d_{\nu_l}}{\lambda_{\nu_l}}.
\end{align}
These block-encodings will be used together to implement $U^c$, the Naimark extension of our POVM, as shown in Eq. \eqref{eq:U^c}.
Our strategy will be to block-encode the smallest components of our decomposition of $\sqrt{\tilde{\Pi}_i}$ and $\sqrt{\Delta}$, then use these components to build up $U^c(i)$. We have already shown in detail how to block-encode $O(\alpha,k_l,i)$ in the main text, so here we will begin with $O_{\text{cen}}(i,k_l,k_r)$, and likewise for $\sqrt{\Delta}$.  To assemble these constituent pieces, we will take advantage of the fact that there are straightforward ways to add and multiply block-encoded matrices that faithfully respect the arthimetic of the encoded matrices.  A summary of the results is shown in Table~\ref{table:block}.

Throughout this section, all quantum wires in circuit diagrams are bundled unless marked with slash-$1$.

\begin{table}[H]
\caption{Summary of the unitary block-encoding.  For each matrix listed in the first column, we build up a $(\beta, b, \gamma)$-block-encoding, which we name in the second column.  We list the scaling factor $\beta$ of each encoding, as well as the number of ancilla qubits used $b$, and the error $\gamma$.  For the final encoding of $\sqrt{\Pi_i}$, the scaling factor is $\alpha = (n-1)^2dx^4 + (n-1)^{3/2}dx'^2 + (n-1)^{-1/2}$ and the error $\delta$ is found in Appendix~\ref{app:complexity}.}\label{table:block}
\begin{center}
\begin{tabular}{@{} p{0.29\textwidth}@{}@{}p{0.17\textwidth}@{}@{}p{0.08\textwidth}@{}@{}p{0.3\textwidth}@{}@{}p{0.15\textwidth}@{}} 
\toprule
\thd{Matrix} & \thd{Block-encoding} & \thd{Scaling} & \thd{Num. of Ancilla qubits} & \thd{Error}\\
\midrule
\addlinespace
\thd{$\sum_{\alpha=1}^{n_\alpha} \ket{e_\alpha}\bra{e_\alpha}\otimes O(\alpha, k,i)$} & \thd{$U[x^2](k,i)$} & \thd{$x^2$} & \thd{$\log n_{r_\nu} + \log n_\nu$} & \thd{$\varepsilon_2$}\\
\thd{$O_\text{cen}(i, k_l, k_r)$} & \thd{$U_2[x^4](i,k_l, k_r)$} & \thd{$x^4$} & \thd{$2\log n_{r_\nu} + 2\log n_\nu$} & \thd{$2 x^2 \varepsilon_2$} \\
\thd{$\tilde{\Phi}$} & \thd{$U^c_{\Phi}$} & \thd{$\sqrt{d}$} & \thd{$1$} & \thd{$\varepsilon_3$} \\
\thd{$\tilde{O}_\text{cen}(i, k_l, k_r)$} & \thd{$U^c_{\text{cen}}[x^4](i,k_l, k_r)$} & \thd{$x^4$} & \thd{$2\log n_{r_\nu} + 2\log n_\nu$\\$ - 2 \log \lceil d \rceil _{2^\bcdot} + 2$} & \thd{$2 x^2 \varepsilon_2$} \\
\thd{$V_L(\pi_{k_l}) \Phi^\dagger O_{\text{cen}}(i,k_l,k_r)\cdot$\\$\cdot \Phi V_L(\pi_{k_r})$} & \thd{$U^c
(i,k_l, k_r)$} & \thd{$dx^4$} & \thd{$2\text{log}\ n_{r_{\nu}} + 2\text{log}\ n_\nu$\\$ - 2\text{log} \lceil d\rceil _{2^\bcdot} + 4$} & \thd{$2dx^2\varepsilon_2$\\$+ 2\sqrt{d} x^4\varepsilon_3$} \\
\thd{$\sqrt{\Pi_i}$} & \thd{$U^c
(i)$} & \thd{$\alpha$} & \thd{$2\text{log}\ n_{r_{\nu}} + 2\text{log}\ n_\nu - 2\text{log} \lceil d\rceil _{2^\bcdot}$\\$+ 2\text{log} \lceil n-1\rceil _{2^\bcdot} + 6$} & \thd{$\delta$} \\
\addlinespace
\bottomrule
\end{tabular}
\end{center}
\end{table}

\subsection{Block encoding $O_{\text{cen}}(i,k_l,k_r)$}

To find a block-encoding of $O_{\text{cen}}(i,k_l,k_r)$, we can first rewrite it as 
\begin{align}
    O_{\text{cen}}(i,k_l,k_r)=\mathbb{I}_{n_r}\otimes \left(\sum_{\alpha=1}^{n_\alpha} | e_\alpha \rangle \langle e_\alpha | \otimes O(\alpha,k_l,i) \right) \cdot \left(\sum_{\alpha=1}^{n_\alpha} | e_\alpha \rangle \langle e_\alpha | \otimes O^\dagger (\alpha,k_r,i) \right).
\end{align}
In this form, we see that we can take advantage of the fact that the product of block-encodings, faithfully represents the product of the encoded matrices.  We have already found a $(x^2, \text{log}\ n_{r_\nu} + \text{log}\ n_\nu, \varepsilon_2)$-block-encoding of the matrix $\sum_{\alpha=1}^{n_\alpha} \ket{e_\alpha}\bra{e_\alpha}\otimes O(\alpha, k,i)$, which we call $U[x^2](k,i)$, in Section~\ref{sec:circuit}.  Passing through the $\ket{r}$ register and multiplying it by $U^\dagger[x^2](k_r,i)$ then gives a $(x^4,2\text{log}\ n_{r_{\nu}} + 2\text{log}\ n_\nu, 2x^2\varepsilon_2)$-block-encoding of $O_{\text{cen}}(i,k_l,k_r)$.  Calling it $U_2[x^4](i,k_l,k_r)$, we have

\tikzscale{0.75}{\begin{quantikz}[row sep={0.6cm,between origins}]
\lstick{$\ket{\text{Anc}}$} & \gate[wires=4][2.4cm]{U_2[x^4](i,k_l,k_r)} & \rstick{$\ket{\text{Anc}}$}\qw\\
\lstick{$\ket{r}$} & & \rstick{$\ket{r}$}\qw\\
\lstick{$\ket{\alpha}$} & & \rstick{$\ket{\alpha}$}\qw\\
\lstick{$\ket{k_\alpha}$} & & \rstick{$\ket{k_\alpha}$}\qw
\end{quantikz}
$\cong$
\begin{quantikz}[row sep={0.6cm,between origins}, column sep={0cm}]
\lstick[2]{$\ket{\text{Anc}}$} & \qw & \gate[style={draw=red!0},swap][0.05cm]{} & \qw & \qw & \qw & \gate[style={draw=red!0},swap][0.05cm]{} & \qw &\rstick[2]{$\ket{\text{Anc}}$}\qw \\ [-0.3cm]
 & \gate[style={draw=red!0},swap][0.05cm]{} & &\qw & \gate[style={draw=red!0},swap][0.05cm]{} & \qw &  & \gate[style={draw=red!0},swap][0.05cm]{}&\qw\\[-0.1cm]
\lstick{$\ket{r}$} & & \qw & \gate[wires=3][2cm]{U^\dagger[x^2](k_r,i)}\gateinput{$\ket{\text{Anc}_0}$}\gateoutput{$\ket{\text{Anc}_0}$} & & \gate[wires=3][2cm]{U[x^2](k_l,i)}\gateinput{$\ket{\text{Anc}_0}$}\gateoutput{$\ket{\text{Anc}_0}$} & \qw &\qw &\rstick{$\ket{r}$} \\
\lstick{$\ket{\alpha}$} & \qw & \qw & & \qw & & \qw &\qw & \qw \rstick{$\ket{\alpha}$}\\
\lstick{$\ket{k_\alpha}$} & \qw & \qw & & \qw & & \qw &\qw & \rstick{$\ket{k_\alpha}$}\qw
\end{quantikz}}
Here we have used $\ket{\text{Anc}}$, a larger ancilla that holds two copies of $\ket{\text{Anc}_0}$. 

To obtain an analogous block-encoding for $O'_{\text{cen}}(k_l, k_r)$, defined in Eq. (\ref{eq:O_p_cen}), we first need to block-encode $\sum_{\alpha=1}^{n_\alpha} \ket{e_\alpha} \bra{e_\alpha} \otimes O'(\alpha, k_l, k_r)$.  We can do this using the same method inspired by the Linear Combination of Unitaries algorithm outlined in the main text, and simply replacing $C(\alpha,\nu_j^\alpha)$ with $C'(\alpha,\nu_j^\alpha)$ in the construction of $P'_{L/R}[x']$.  Doing so produces the constraints
\begin{align}
\label{eq:cons2}C'_{\text{rem}}(\alpha)\equiv 1-\sum_{j=1}^{n_{\alpha+\Box}}\frac{1}{x'^2} C'(\alpha,\nu_j^\alpha)\geqslant 0\quad \forall\alpha.
\end{align}
No multiplication is necessary to construct $O'_{\text{cen}}(k_l,k_r)$ from $O'(\alpha, k_l, k_r)$, so we then immediately find a $(x'^2,2\text{log}\ n_{r_{\nu}} + 2\text{log}\ n_\nu, \varepsilon_2')$-block-encoding of $O'_{\text{cen}}(k_l,k_r)$ we call $U_1[x'^2](k_l, k_r)$.

\tikzscale{0.75}{\begin{quantikz}[row sep={0.55cm,between origins}]
\lstick{$\ket{\text{Anc}}$} & \gate[wires=4][2.4cm]{U_1[x'^2](k_l,k_r)} & \rstick{$\ket{\text{Anc}}$}\qw\\
\lstick{$\ket{r}$} & & \rstick{$\ket{r}$}\qw\\
\lstick{$\ket{\alpha}$} & & \rstick{$\ket{\alpha}$}\qw\\
\lstick{$\ket{k_\alpha}$} & & \rstick{$\ket{k_\alpha}$}\qw
\end{quantikz}
$\cong$
\begin{quantikz}[row sep={0.7cm,between origins}, column sep={0cm}]
\lstick[2]{$\ket{\text{Anc}}$} & \qw & \qw & \qw &\rstick[2]{$\ket{\text{Anc}}$}\qw \\ [-0.4cm]
 & \gate[style={draw=red!0},swap][0.05cm]{} & \qw & \gate[style={draw=red!0},swap][0.05cm]{} & \qw\\[-0.2cm]
\lstick{$\ket{r}$} & & \gate[wires=3][2cm]{U[x'^2](k_l,kr)}\gateinput{$\ket{\text{Anc}_0}$}\gateoutput{$\ket{\text{Anc}_0}$} & &\rstick{$\ket{r}$} \\
\lstick{$\ket{\alpha}$} & \qw & & \qw & \qw\rstick{$\ket{\alpha}$}\\
\lstick{$\ket{k_\alpha}$} & \qw &  & \qw &\rstick{$\ket{k_\alpha}$}\qw
\end{quantikz}}
\vspace{0.6cm}

\subsection{Block-encoding $\Phi$}

First, notice that we can write $\Phi$ as 
\begin{equation}
    \Phi = \sqrt{d} U^{(n-2)}_{\text{Sch(exact)}} \otimes \bra{\phi_+}_{n-1 n},
\end{equation}
where $U^{(n-2)}_{\text{Sch(exact)}}$ is the exact Schur transform on $(n-2)$-qudits, and the registers have been padded.  $\Phi$ performs the Schur transform on the first $(n-2)$-qudits, but only when qudits $n-1$ and $n$ are maximally entangled.  Notice that it is a rectangular matrix: if we hadn't padded the registers for the Schur transform, it would have had dimension $d^{n-2} \times d^n$.  With the padding, it has dimension $(n_r n_\alpha n_{d_\alpha}) \times (n_r n_\alpha n_{d_\alpha}) \lceil d\rceil^2_{2^\bcdot}$ where $\lceil d\rceil_{2^\bcdot}$ is the minimum number of qubits needed to encode a qudit.  

To block-encode $\Phi$, we begin by first promoting it to a square matrix
\begin{align}
\tilde{\Phi}\equiv \sqrt{d} \left( U_{\text{Sch(exact)}}^{(n-2)} \otimes \ket{0}\bra{0}_{n-1n} U_s \right).
\end{align}
Here, $U_S$ is defined such that its first row satisfies $\bra{0}_{n-1n} U_S = \bra{\phi_+}_{n-1n}$.  To use this version of $\Phi$, we also need to update our definition of $O_{\text{cen}}$ to
\begin{align}
\tilde{O}_{\text{cen}}(i,k_l,k_r) \equiv O_{\text{cen}}(i,k_l,k_r) \otimes \ket{0}\bra{0}_{n-1n},
\end{align}
We will still be able to use the block-encoding of $O_{\text{cen}}$ that we have already found though: as we will show shortly, it is possible to easily lift $U_2[x^4](i,k_l,k_r)$ to a block-encoding of $\tilde{O}_{\text{cen}}(i,k_l,k_r)$.

Together, the product of these now-square matrices gives
\begin{align}
\tilde{\Phi}^\dagger \tilde{O}_{\text{cen}}(i,k_l,k_r) \tilde{\Phi} = \Phi^\dagger O_{\text{cen}}(i,k_l,k_r) \Phi.
\end{align}

\subsubsection{Block-encoding $\Tilde{\Phi}$}

We can block-encode $\tilde{\Phi}$ as
\begin{align}
U^c_\Phi = \mathbb{I}_{A_2} \otimes U_{\text{Sch}}^{(n-2)} \otimes \ket{0}\bra{0}_{n-1n} U_s + X_{A_2} \otimes U_{\text{Sch}}^{(n-2)} \otimes \sum_{j=1}^{\lceil d\rceil _{2^\bcdot}^2}\ket{j}\bra{j}_{n-1n} U_s,
\end{align}
where $\ket{A_2}$ is the necessary ancilla qubit.  Represented as a circuit, this is
\tikzscale{0.75}{\begin{quantikz}[row sep={0.37cm,between origins}]
\lstick{$\ket{A_2}$} & \gate[wires=6][2.6cm]{U^c_{\Phi} }  &\qw \rstick{$\ket{A_2}$}\\
\lstick{$\ket{\text{Pad}}$} & & \rstick{$\ket{n-1}$}\qw\\
\lstick{$\ket{1}$} & & \rstick{$\ket{n}$}\qw\\
\lstick{$\ket{2}$} & & \rstick{$\ket{r}$}\qw\\
\lstick{\vdots\space} & & \rstick{$\ket{\alpha}$}\qw\\
\lstick{$\ket{n}$} & & \rstick{$\ket{k_\alpha}$}\qw
\end{quantikz}
$\cong$
\begin{quantikz}[row sep={0.35cm,between origins}, column sep={0.15cm}]
\lstick{$\ket{A_2}$} & \qwbundle{1} & \gate{X} & \gate{X}  &  \rstick{$\ket{A_2}$}\qw\\[0.15cm]
\lstick{$\ket{\text{Pad}}$} & \gate[4][2.1cm]{U_{\text{Sch}}^{(n-2)}}& &&\\
\lstick{$\ket{1}$} &  &\qw & \qw & \rstick{$\ket{r}$}\qw\\
\lstick{\vdots\space} &  &\qw & \qw & \rstick{$\ket{\alpha}$}\qw\\
\lstick{$\ket{n-2}$} &  &\qw & \qw & \rstick{$\ket{k_\alpha}$}\qw\\[0.42cm]
\lstick{$\ket{n-1}$} & \gate[2][1cm]{U_s}  &\qw & \octrl{-5} & \rstick{$\ket{n-1}$}\qw\\
\lstick{$\ket{n}$} &   &\qw & \octrl{-1} & \rstick{$\ket{n}$}\qw\\
\end{quantikz}}
where $\ket{\text{Pad}}$ is a register of dimension $(n_r n_{\alpha} n_{d_\alpha}) / \lceil d\rceil_{2^\bcdot}^{n-2}$, necessary because the Schur basis registers $\ket{r}\ket{\alpha}\ket{k_\alpha}$ are padded.  Note that, because $\Phi$ and $\tilde{\Phi}$ are defined in terms of the exact, padded Schur transform, the $\ket{\text{Pad}}$ register is not part of the ancilla register used for the block-encoding.
We can quickly check that $U^c_{\Phi}$ is a block-encoding by projecting onto the $0$-eigenspace of the $\ket{A_2}$ ancilla, which gives
\begin{align}
\left(\bra{0}_{A_2} \otimes \mathbb{I}_{r\alpha k_\alpha}\otimes \mathbb{I}_{n-1n} \right)& U^c_{\Phi} \left(\ket{0}_{A_2} \otimes \mathbb{I}_{\text{Pad}}\otimes \mathbb{I}_{[n]} \right)
=U_{\text{Sch}}^{(n-2)} \otimes \ket{0}\bra{0}_{n-1n} U_S
\approx\frac{1}{\sqrt{d}} \tilde{\Phi}.
\end{align}
Introducing $\varepsilon_3$ to describe the numerical error from the implementation of the $(n-2)$-qudit Schur transform, this means that $U^c_\Phi$ is a $(\sqrt{d},1,\varepsilon_3)$-block-encoding of $\tilde{\Phi}$.

\subsubsection{Block-encoding $\Tilde{O}_{\text{cen}}(i, k_l, k_r)$}

To lift the block-encoding of $O_{\text{cen}}(i, k_l, k_r)$ to a block-encoding of $\tilde{O}_{\text{cen}}(i, k_l, k_r)$, we will need to introduce the registers $\ket{n}$ and $\ket{n-1}$, as well as some additional ancilla qubits.  We'll play a clever trick here, using the $\ket{n}$ and $\ket{n-1}$ registers to reduce the total number of ancilla qubits we need.

Explicitly, we are going to store the ancilla register $\ket{\text{Anc}}$, which was originally used to block-encode $O_{\text{cen}}(i, k_l, k_r)$, in $\ket{n}$ and $\ket{n-1}$ and an ancilla $\ket{A_{13}}$. Because $\ket{\text{Anc}}$ is of dimension $(n_{r_\nu} n_\nu)^2$, $\ket{A_{13}}$ will be of dimension $(n_{r_\nu} n_\nu)^2 / \lceil d \rceil^2_{2^\bcdot}$.  We also introduce two additional ancilla qubits we call $\ket{A_{11}}$ and $\ket{A_{12}}$, and collect all three in the register $\ket{A_1} = \ket{A_{11}} \ket{A_{12}} \ket{A_{13}}$.

This may seem counterintuitive: to retrieve $\tilde{O}_{\text{cen}}(i, k_l, k_r)$, we need to project onto the $0$-eigenspace of $\ket{\text{Anc}}$, but $\ket{n}$ and $\ket{n-1}$ are not ancillary and will not be involved in the projection.  It works by defining the unitary as
\begin{align}
U^c_{\text{cen}}[x^4](i,k_l,k_r) =& \left[ \mathbb{I}_{A_{11}} \otimes \ket{0}\bra{0}_{n-1n} + X_{A_{11}} \otimes \sum_{j=1}^{\lceil d\rceil _{2^\bcdot}^2} \ket{j}\bra{j}_{n-1n} \right] U_2[x^4](i,k_l,k_r) \nonumber\\
&\times \left[ \mathbb{I}_{A_{12}} \otimes \ket{0}\bra{0}_{n-1n} + X_{A_{12}} \otimes \sum_{j=1}^{\lceil d\rceil _{2^\bcdot}^2} \ket{j}\bra{j}_{n-1n} \right].
\end{align} 
\tikzscale{0.75}{\begin{quantikz}[row sep={0.5cm,between origins}]
\lstick{$\ket{A_1}$} & \gate[wires=6,disable auto height][2.8cm]{\begin{array}{c} U^c_{\text{cen}}[x^4](i,k_l,k_r) \\ \textcolor{magenta}{U^{c'}_{\text{cen}}[x'^2](k_l,k_r) }\end{array}}  &\qw \rstick{$\ket{A_1}$}\\
\lstick{$\ket{n-1}$} & & \rstick{$\ket{n-1}$}\qw\\
\lstick{$\ket{n}$} & & \rstick{$\ket{n}$}\qw\\
\lstick{$\ket{r}$} & & \rstick{$\ket{r}$}\qw\\
\lstick{$\ket{\alpha}$} & & \rstick{$\ket{\alpha}$}\qw\\
\lstick{$\ket{k_\alpha}$} & & \rstick{$\ket{k_\alpha}$}\qw
\end{quantikz}
$\cong$
\begin{quantikz}[row sep={0.4cm,between origins}, column sep={0.15cm}]
\lstick[3]{$\ket{A_1}$} & \qw & \qw & \qwbundle{1} & \gate{X} \qw & \gate{X} \qw  &  \rstick[3]{$\ket{A_1}$}\qw\\
& \gate{X}&\gate{X}& \qwbundle{1} & \qw &\qw &\qw\\
& \qw &\qw & \gate[wires=6,disable auto height][2.8cm]{\begin{array}{c} U_2[x^4](i,k_l,k_r) \\ \textcolor{magenta}{U_1[x'^2](k_l,k_r)}\end{array}}\gateinput[3]{$\ket{\text{Anc}}$}\gateoutput[3]{$\ket{\text{Anc}}$}& \qw &\qw &\qw\\
\lstick{$\ket{n-1}$} &\qw &\octrl{-2} & & \octrl{-3} &\qw & \rstick{$\ket{n-1}$}\qw\\
\lstick{$\ket{n}$} &\qw &\octrl{-1} & & \octrl{-1} &\qw & \rstick{$\ket{n}$}\qw\\
\lstick{$\ket{r}$} &\qw &\qw & & \qw &\qw & \rstick{$\ket{r}$}\qw\\
\lstick{$\ket{\alpha}$} &\qw &\qw & & \qw &\qw & \rstick{$\ket{\alpha}$}\qw\\
\lstick{$\ket{k_\alpha}$} &\qw &\qw & & \qw &\qw & \rstick{$\ket{k_\alpha}$}\qw
\end{quantikz}}
\vspace{0.6cm}
Then, by direct computation, we can see that
\begin{align}
\big(\bra{0}_{A_1} \otimes &\mathbb{I}_{n-1 n r\alpha k_\alpha} \big) U^c_{\text{cen}}[x^4](i,k_l,k_r) \left(\ket{0}_{A_1} \otimes \mathbb{I}_{n-1n r\alpha k_\alpha} \right) \\
&= \left(\bra{0}_{A_{13}} \otimes \ket{0}\bra{0}_{n-1n} \otimes \mathbb{I}_{r\alpha k_\alpha}\right) U_2[x^4](i,k_l,k_r) \left(\ket{0}_{A_{13}} \otimes \ket{0}\bra{0}_{n-1n} \otimes \mathbb{I}_{r\alpha k_\alpha}\right)\\
&= \ket{0}\bra{0}_{n-1n} \left(\bra{0}_{A_{13}} \otimes \bra{0}_{n-1n} \right) U_2[x^4](i,k_l,k_r) \left(\ket{0}_{A_{13}} \otimes \ket{0}_{n-1n} \right)\\
&= \ket{0}\bra{0}_{n-1n} \bra{0} U_2[x^4](i,k_l,k_r) \ket{0}_{\text{Anc}} \approx \frac{1}{x^4} \tilde{O}_{\text{cen}}(i,k_l,k_r),
\end{align}
verifying that $U^c_{\text{cen}}[x^4](i,k_l,k_r)$ is an $(x^4,  2\text{log}\ n_{r_{\nu}} + 2\text{log}\ n_\nu - 2\text{log} \lceil d\rceil _{2^\bcdot} +2, 2x^2\varepsilon_2)$-encoding of $\tilde{O}_{\text{cen}}(i,k_l,k_r)$. 

Similarly, we can block-encode $\tilde{O}'_{\text{cen}}(k_l,k_r)$ by replacing $U_2[x^4](i,k_l,k_r)$ with $U_1[x'^2](k_l,k_r)$.  This give a $(x'^2,  2\text{log}\ n_{r_{\nu}} + 2\text{log}\ n_\nu - 2\text{log} \lceil d\rceil _{2^\bcdot} +2, \varepsilon_2')$-encoding of $\tilde{O}'_{\text{cen}}(k_l,k_r)$, which we call $U^{c'}_{\text{cen}}[x'^2](k_l,k_r)$, and which is shown in the diagram above in pink.

\subsection{Block-encoding $V_L(\pi_{k_l})\cdot \Phi^\dagger \cdot O_{\text{cen}}(i,k_l,k_r) \cdot \Phi \cdot V_L(\pi_{k_r})$}

We are now ready to multiply together all of the block-encodings we have constructed to get a single summand $V_L(\pi_{k_l})\cdot \Phi^\dagger \cdot O_{\text{cen}}(i,k_l,k_r) \cdot \Phi \cdot V_L(\pi_{k_r})$ making up $\sqrt{\Tilde{\Pi}_i}$ Eq. (\ref{eq:start}).

We have already seen that the product of square matrices $\tilde{\Phi}^\dagger \tilde{O}_{\text{cen}}(i,k_l,k_r) \tilde{\Phi}$ encodes $\Phi^\dagger O_{\text{cen}}(i,k_l,k_r) \Phi$, and we have encoded $\tilde{\Phi}$ in $U^c_\Phi$ and $\tilde{O}_{\text{cen}}(i,k_l,k_r)$ in $U^c_{\text{cen}}[x^4](i,k_l,k_r)$. Block-encoding $\Phi^\dagger O_{\text{cen}}(i,k_l,k_r) \Phi$ then merely requires multiplying $U^c_{\Phi}$, $U^c_{\text{cen}}[x^4](i,k_l,k_r)$ and $U^{c\dagger}_{\Phi}$ and managing the ancillae. The permutations $V_L(\pi_{k_l})$ and $V_L(\pi_{k_r})$ don't require block-encoding because they are unitary, and so can also be multiplied on.

The following circuit $U^c(i,k_l,k_r)$ is then an $(d x^4,  2\text{log}\ n_{r_{\nu}} + 2\text{log}\ n_\nu - 2\text{log} \lceil d\rceil _{2^\bcdot} + 4, 2dx^2\varepsilon_2 + 2\sqrt{d} x^4\varepsilon_3)$-encoding of one summand in Eq. (\ref{eq:start}).

\tikzscale{0.73}{\begin{quantikz}[row sep={0.47cm,between origins}]
\lstick{$\ket{A_U}$} & \gate[wires=6,disable auto height][2.9cm]{\begin{array}{c} U^c(i,k_l,k_r) \\ \textcolor{magenta}{U^{c'}(k_l,k_r) }\end{array} } & \rstick{$\ket{A_U}$}\qw\\
\lstick{$\ket{\text{Pad}}$} & & \rstick{$\ket{\text{Pad}}$}\qw\\
\lstick{$\ket{1}$} & & \rstick{$\ket{1}$}\qw\\
\lstick{$\ket{2}$} & & \rstick{$\ket{2}$}\qw\\
\lstick{\vdots\space} & & \rstick{\space\vdots}\qw\\
\lstick{$\ket{n}$} & & \rstick{$\ket{n}$}\qw
\end{quantikz}
$\cong$}

\vspace{-1cm}
\tikzscale{0.73}{\begin{quantikz}[row sep={0.48cm,between origins}, column sep={0cm}]
\lstick{$\ket{A_3}$} &[0.6cm] \qwbundle{1} &[0.7cm] \qw & \qw & \qw & \gate[style={draw=red!0},swap][0.05cm]{} & \qw & \qw &[0.7cm] \qw &[0.6cm] \rstick{$\ket{A_2}$}\qw\\
\lstick{$\ket{A_1}$} & \qw & \qw & \gate[style={draw=red!0},swap][0.05cm]{} & \qw & & \gate[style={draw=red!0},swap][0.05cm]{} & \qw & \qw & \rstick{$\ket{A_1}$}\qw\\
\lstick{$\ket{A_2}$} & \qwbundle{1} & \gate[wires=6][3.2cm]{U^c_{\Phi} }\gateinput{$\ket{A_2}$}\gateoutput{$\ket{A_2}$} &  &[-2cm] \gate[wires=6,disable auto height][2.7cm]{\begin{array}{c} U^c_{\text{cen}}[x^4](i,k_l,k_r) \\ \textcolor{magenta}{U^{c'}_{\text{cen}}[x'^2](k_l,k_r) }\end{array}} & \qw &  & \gate[wires=6][3.2cm]{U^{c\dagger}_{\Phi} }\gateoutput{$\ket{A_3}$}\gateinput{$\ket{A_3}$} & \qw &\rstick{$\ket{A_3}$}\qw\\
\lstick{$\ket{\text{Pad}}$} & \qw & \gateinput{$\ket{\text{Pad}}$}\gateoutput{$\ket{n-1}$} & \qw & & \qw & \qw & \gateoutput{$\ket{\text{Pad}}$}\gateinput{$\ket{n-1}$} & \qw &\rstick{$\ket{\text{Pad}}$}\qw\\
\lstick{$\ket{1}$} & \gate[wires=4][2.2cm]{V_L(\pi_{k_r}) } & \gateinput{$\ket{1}$}\gateoutput{$\ket{n}$} & \qw & & \qw & \qw & \gateoutput{$\ket{1}$}\gateinput{$\ket{n}$} & \gate[wires=4][2.2cm]{V_L(\pi_{k_l}) } &\rstick{$\ket{1}$}\qw\\
\lstick{$\ket{2}$} & & \gateinput{$\ket{2}$}\gateoutput{$\ket{r}$} & \qw & & \qw & \qw & \gateoutput{$\ket{2}$}\gateinput{$\ket{r}$} & &\rstick{$\ket{2}$}\qw\\
\lstick{\vdots\space} & & \gateinput{\vdots}\gateoutput{$\ket{\alpha}$} & \qw & & \qw & \qw & \gateoutput{\vdots}\gateinput{$\ket{\alpha}$} & &\rstick{ \space\space\vdots}\qw\\
\lstick{$\ket{n}$} & & \gateinput{$\ket{n}$}\gateoutput{$\ket{k_\alpha}$} & \qw & & \qw & \qw & \gateoutput{$\ket{n}$}\gateinput{$\ket{k_\alpha}$} & &\rstick{$\ket{n}$}\qw
\end{quantikz}}
\vspace{0.6cm}

Similarly, $U^{c'}(k_l,k_r)$ is an $(d x'^2, 2\text{log}\ n_{r_{\nu}} + 2\text{log}\ n_\nu - 2\text{log} \lceil d\rceil _{2^\bcdot} + 4, d\varepsilon_2'+2\sqrt{d}x'^2\varepsilon_3)$-encoding of one summand in Eq. (\ref{eq:start2}).  In both cases, the combined register $\ket{A_U}$ contains the indicated ancilla state.

\subsubsection{Block-encoding $\sqrt{\Pi_i}$}

Armed with the block-encodings of each summand, we are finally ready to encode the Kraus operator
\begin{align}
\sqrt{\Pi_i}=& \sqrt{\tilde{\Pi}_i} + \sqrt{\Delta} \\
=& \sum_{k_l,k_r=1,\dots,n-1} V_L(\pi_{k_l})\cdot \Phi^\dagger \cdot O_{\text{cen}}(i,k_l,k_r) \cdot \Phi \cdot V_L(\pi_{k_r}) \\
&-1\big/\sqrt{n-1}\cdot \sum_{k_l,k_r=1,\dots,n-1} V_L(\pi_{k_l})\cdot \Phi^\dagger \cdot O'_{\text{cen}}(k_l,k_r) \cdot \Phi \cdot V_L(\pi_{k_r}) + \mathbb{I}\big/\sqrt{n-1}.
\end{align}
Recall that we will use this block-encoding, which we call $U^c(i)$, to execute the POVM by controlling it with the $\ket{I}$ register, as in Eq.~\eqref{eq:U^c} which we restate here:
\begin{align}
U^c=\sum_{i=1}^{n-1} \ket{i}\bra{i}_I \otimes U^c(i) + \sum_{i=n}^{\lceil n-1 \rceil _{2^\bcdot}} \ket{i}\bra{i}_I \otimes \mathbb{I}.\nonumber
\end{align}

To implement $U^c(i)$, we will need to add the contributions from $\sqrt{\Tilde{\Pi}_i}$ and $\sqrt{\Delta}$ in appropriate ratios, as well as execute the superpositions over $k_l$ and $k_r$.  To perform the first task, we'll use a two-qubit ancilla we call $\ket{A_4}$, and to perform the second task we'll use two ancilla registers we call $\ket{k_l}$ and $\ket{k_r}$, each of dimension $\lceil n-1 \rceil_{2^\bcdot}$. 

The first task is performed by using controlled unitary matrices $U_l$ and $U_r$ to introduce coefficients to scale the terms.  These matrices are very similar to the $P_{L/R}$ matrices used to introduce coefficients in the block-encoding of $\sum_{\alpha = 1}^{n_\alpha} \ket{e_\alpha} \bra{e_\alpha} \otimes O(\alpha, k, i)$.  They are controlled by the $\ket{I}$ register: when any state $\ket{i \in [n-1]}_I$ is encountered, the first row of these matrices will be given by
\begin{align}
\bra{00} U_{l/r} \ket{00}_{A_4} &= a_{l/r} \\
\bra{00} U_{l/r} \ket{01}_{A_4} &= b_{l/r} \\
\bra{00} U_{l/r} \ket{10}_{A_4} &= c_{l/r} \\
\bra{00} U_{l/r} \ket{11}_{A_4} &= 0.
\end{align}
In other words, $U_l$ or $U_r$ map $\ket{00}_{A_4}$ from a superposition over different $\ket{A_4}$ states with the real coefficients defined by 
\begin{align}
\begin{cases}
a_l:b_l:c_l=(n-1)^{5/4}\sqrt{d}x^2:(n-1)\sqrt{d}x':1 \\
a_r:b_r:c_r=(n-1)^{5/4}\sqrt{d}x^2:(1-n)\sqrt{d}x':1 \\
a_{l/r}^2 + b_{l/r}^2 + c_{l/r}^2 = 1\\
c_{l/r} \geqslant 0
\end{cases}\\
\Rightarrow c_{l/r} = \left[ (n-1)^{5/2}dx^4 + (n-1)^{2}dx'^2 + 1 \right]^{-1/2}.
\end{align}
Here, $a_{l/r}$, which are tied to the $\ket{00}_{A_4}$ state, will be the coefficients for $U^c(i, k_l, k_r)$, i.e. the $\sqrt{\Tilde{\Pi}_i}$ portion of the Kraus operator, which will also be tied to $\ket{00}_{A_4}$.  Similarly, $b_{l/r}$ will be the coefficients for $U^{c'}(k_l, k_r)$, the superposition portion of $\sqrt{\Delta}$, and $c_{l/r}$ will be the coefficients for the identity portion of $\sqrt{\Delta}$.

The second task is performed using the unitary $U_k$, which can be used to put the $\ket{k_l}$ or $\ket{k_r}$ registers in a uniform superposition over $k \in [n-1]$.  That is,
\begin{align}
\bra{0} U_k \ket{k}_{k_l/k_r} &=
\begin{cases}
\frac{1}{\sqrt{n-1}} & k=1,\dots,n-1 \\
0 & k=n,\dots, \lceil n-1 \rceil _{2^\bcdot}
\end{cases}
\end{align}

Armed with $U_l$, $U_r$, and $U_k$ to put the ancilla qubits in a carefully chosen weighted superposition, all we need to do to construct the block-encoding is to condition $U^c(i, k_l, k_r)$, $U^{c'}(k_l, k_r)$, and the identity on the appropriate ancilla states.  Explicitly, we define
\begin{align}
    &U^c(i) = \left(U_l \otimes \mathbb{I}_{k_l,k_r, A_U, 
    \text{in}}\right) \nonumber\\
    &\left[ \ket{00}\bra{00}_{A_4}\otimes \tilde{U}(i) + \ket{01}\bra{01}_{A_4}\otimes \tilde{U}'(i) + \left(\ket{10}\bra{10}+\ket{11}\bra{11}\right)_{A_4}\otimes \mathbb{I}_{k_l,k_r,A_U, [n]}  \right] \left(U_r^\dagger \otimes \mathbb{I}_{k_l,k_r,A_U, \text{in}}\right)
\end{align}
where $\ket{\text{in}} \equiv \ket{\text{Pad}} \ket{1,\dots,n}$ is the input register for $U^c(i)$.  In this expression, the term conditioned on $\ket{00}_{A_4}$ is the operator $U^c(i,k_l,k_r)$ controlled by the $\ket{k_l}$ and $\ket{k_r}$ ancillae, and the term conditioned on $\ket{01}_{A_4}$ is the same for $U^{c'}(k_l, k_r)$.  Written out, we have 
\begin{align}
    &\tilde{U}(i) \equiv \left(U_k \otimes U_k \otimes \mathbb{I}_{A_U,\text{in}}\right) \nonumber\\
    &\left[ \sum_{k_l,k_r=1}^{n-1}\ket{k_l}\bra{k_l}_{k_l} \otimes \ket{k_r}\bra{k_r}_{k_r} \otimes U^c(i,k_l,k_r) + \sum_{k_l,k_r=n}^{\lceil n-1 \rceil _{2^\bcdot}} \ket{k_l}\bra{k_l}_{k_l} \otimes \ket{k_r}\bra{k_r}_{k_r} \otimes \mathbb{I}_{A_U,\text{in}} \right] 
    \left( U_k^\dagger \otimes U_k^\dagger \otimes \mathbb{I}_{A_U, \text{in}}\right).\\
    &\tilde{U}'(i) \equiv \left(U_k \otimes U_k \otimes \mathbb{I}_{A_U, \text{in}}\right) \nonumber\\
    &\left[ \sum_{k_l,k_r=1}^{n-1}\ket{k_l}\bra{k_l}_{k_l} \otimes \ket{k_r}\bra{k_r}_{k_r} \otimes U^{c'}(k_l,k_r) + \sum_{k_l,k_r=n}^{\lceil n-1 \rceil _{2^\bcdot}} \ket{k_l}\bra{k_l}_{k_l} \otimes \ket{k_r}\bra{k_r}_{k_r} \otimes \mathbb{I}_{A_U,\text{in}} \right] 
    \left( U_k^\dagger \otimes U_k^\dagger \otimes \mathbb{I}_{A_U, \text{in}}\right).
\end{align}

\tikzscale{0.73}{\begin{quantikz}[row sep={0.52cm,between origins}, column sep={0.3cm}]
\lstick{$\ket{I}$} &  \ocontrol[rectangle]{}\vqw{1} &\rstick{$\ket{I}$} \qw\\
\lstick{$\ket{\text{Anc}_{\text{tot}}}$} & \gate[wires=3][1.8cm]{U^c(i)} & \rstick{$\ket{\text{Anc}_{\text{tot}}}$}\qw\\
\lstick{$\ket{\text{Pad}}$} & & \rstick{$\ket{\text{Pad}}$}\qw\\
\lstick{$\ket{1,\dots,n}$} & & \rstick{$\ket{1,\dots,n}$}\qw
\end{quantikz}
\hspace{0.3cm}
$\cong$
\hspace{0.3cm}
\begin{quantikz}[row sep={0.51cm,between origins}]
\lstick{$\ket{I}$} &[-0.1cm] \ocontrol[rectangle]{}\vqw{1} & \qw & [-0.2cm]\ocontrol[rectangle]{}\vqw{1} & [-0.2cm]\qw & [-0.1cm] \ocontrol[rectangle]{}\vqw{1} & \rstick{$\ket{I}$}\qw\\
\lstick{$\ket{A_4}$} & \gate{U_r^\dagger} & \qwbundle{2} & \ocontrol[rectangle]{}\vqw{1} & \qw & \gate{U_l} & \rstick{$\ket{A_4}$}\qw\\
\lstick{$\ket{k_l}$} & \qw & \gate{U_k^\dagger} & \ocontrol[rectangle]{}\vqw{1} & \gate{U_k} & \qw & \rstick{$\ket{k_l}$}\qw\\[0.3cm]
\lstick{$\ket{k_r}$} & \qw & \gate{U_k^\dagger} & \ocontrol[rectangle]{}\vqw{1} & \gate{U_k} & \qw & \rstick{$\ket{k_r}$}\qw\\
\lstick{$\ket{A_U}$} & \qw & \qw & \gate[wires=5,disable auto height][2.7cm]{\begin{array}{c} U^c(i,k_l,k_r) \\ \textcolor{magenta}{U^{c'}(k_l,k_r) }\end{array}  } & \qw & \qw & \rstick{$\ket{A_U}$}\qw\\
\lstick{$\ket{\text{Pad}}$} & \qw & \qw & & \qw & \qw & \rstick{$\ket{\text{Pad}}$}\qw\\
\lstick{$\ket{1}$} & \qw & \qw & & \qw & \qw & \rstick{$\ket{1}$}\qw\\
\lstick{\vdots\space} & \qw & \qw & & \qw & \qw & \rstick{\space\vdots}\qw\\
\lstick{$\ket{n}$} & \qw & \qw & & \qw & \qw & \rstick{$\ket{n}$}\qw
\end{quantikz}}

To see that $U^c(i)$, defined in this way, is a block-encoding of $\sqrt{\Pi_i}$, we just need to project onto the $0$-eigenspace of the total ancilla $\ket{\text{Anc}_{\text{tot}}} \equiv \ket{A_4} \ket{k_l} \ket{k_r} \ket{A_U}$. In the first step, the matrices $U_{l/r}$ create a superposition over $\ket{A_4}$ that scales each term of $U^c(i)$ by different coefficients:
\begin{align}
    &\left( \bra{0}_{\text{Anc}_{\text{tot}}} \otimes \mathbb{I}_{\text{in}} \right) U^c(i) \left( \ket{0}_{\text{Anc}_{\text{tot}}} \otimes \mathbb{I}_{\text{in}} \right)\\
    &=\left( \bra{0}_{k_l,k_r,A_U} \otimes \mathbb{I}_{\text{in}} \right) \left[ a_l a_r \tilde{U}(i) - b_l b_r \tilde{U}'(i) + c_l c_r\mathbb{I}_{k_l,k_r,A_U, \text{in}} \right] \left( \ket{0}_{k_l,k_r,A_U} \otimes \mathbb{I}_{\text{in}} \right)
\end{align}
In the first two terms, $U_k$ then turns the controlled matrices into the superpositions over $k_l, k_r \in [n-1]$ we are looking for.  For example,
\begin{align}
    &a_l a_r \left( \bra{0}_{k_l,k_r,A_U} \otimes \mathbb{I}_{\text{in}} \right) \tilde{U}(i) \left( \ket{0}_{k_l,k_r,A_U} \otimes \mathbb{I}_{\text{in}} \right) \\
    &= \frac{a_l a_r}{(n-1)^2} \sum_{k_l k_r = 1}^{n-1} \left( \bra{0}_{A_U} \otimes \mathbb{I}_{\text{in}} \right) U^c(i, k_l,k_r) \left( \ket{0}_{A_U} \otimes \mathbb{I}_{\text{in}} \right) \\
    &\approx \frac{a_l a_r}{dx^4 (n-1)^2} \sum_{k_l k_r = 1}^{n-1} V_L(\pi_{k_l})\cdot \Phi^\dagger \cdot O_{\text{cen}}(i,k_l,k_r) \cdot \Phi \cdot V_L(\pi_{k_r})
\end{align}
Here the second term in $\Tilde{U}(i)$ has vanished because $U_k$ only creates a superposition over $\ket{k \in [n-1]}$.  Notice that, as $\frac{a_l a_r}{c_l c_r} = (n-1)^{5/2} dx^4$, the prefactor for this term can be rewritten as $c_l c_r \sqrt{n-1}$.  In other words, we have chosen $a_{l/r}$ to cancel the scaling factor of the block-encoding $U^c(i, k_l, k_r)$.  The coefficients $b_{l/r}$ were also chosen this way, so that overall we find
\begin{align}
    &\left( \bra{0}_{\text{Anc}_{\text{tot}}} \otimes \mathbb{I}_{\text{in}} \right) U^c(i) \left( \ket{0}_{\text{Anc}_{\text{tot}}} \otimes \mathbb{I}_{\text{in}} \right) \\
    &\approx c_l c_r \sqrt{n-1} \Biggl[ \sum_{k_l,k_r=1}^{n-1} V_L(\pi_{k_l})\cdot \Phi^\dagger \cdot O_{\text{cen}}(i,k_l,k_r) \cdot \Phi \cdot V_L(\pi_{k_r}) + \mathbb{I}_{\text{in}}\big/ \sqrt{n-1} \nonumber\\
    &\quad\quad\quad\quad\quad - 1 \big/ \sqrt{n-1}\cdot \sum_{k_l,k_r=1}^{n-1} V_L(\pi_{k_l})\cdot \Phi^\dagger \cdot O'_{\text{cen}}(k_l,k_r) \cdot \Phi \cdot V_L(\pi_{k_r}) \Biggr] = c_l c_r \sqrt{n-1} \sqrt{\Pi_i}
\end{align}

Therefore, we have shown that $U^c(i\in[n])$ is an $(\alpha,a,\delta)$-block-encoding of $\sqrt{\Pi_i}$, where
\begin{align}
\label{eq:alpha_a}\begin{cases}
\alpha &= (n-1)^2dx^4 + (n-1)^{3/2}dx'^2 + (n-1)^{-1/2} \\
a &= 2\text{log}\ n_{r_{\nu}} + 2\text{log}\ n_\nu - 2\text{log} \lceil d\rceil _{2^\bcdot} + 2\text{log} \lceil n-1\rceil _{2^\bcdot} + 6\\
\end{cases}
\end{align}
and $\delta$ is the error that will be calculated in the next appendix. The $\ket{\text{Pad}}$ register adds on
\begin{align}
    \# \text{Pad} = \text{log}\ n_r + \text{log}\ n_\alpha + \text{log}\ n_{d_\alpha} - (n-2)\text{log} \lceil d\rceil _{2^\bcdot}
\end{align}
qubits.

%% file: sections/appendixcomplexity.tex
\section{Complexity analysis\label{app:complexity}}

In this appendix, we will detail how to calculate the complexity and error propagation of our circuit. A summary of the results is shown in Tables~\ref{table:gate_accuracy},~\ref{table:parameters} and~\ref{table:complexities}.

\begin{table}[H]
\caption{Accuracies of the one-or-two-qubit gates that implement the numerical operators. Each of those gates are then approximated, to the same accuracy, by a product of gates from a finite universal gate set. These accuracies (in violet) are set to ensure that the error $2m\varepsilon$ of the entire circuit $\tilde{V}$ (or U) is $O(1)$, when expressed in powers of $n$ and $d$.}\label{table:gate_accuracy}
\begin{center}
\begin{tabular}{@{} p{0.15\textwidth}@{}@{}p{0.11\textwidth}@{}@{}p{0.1\textwidth}@{}@{}p{0.11\textwidth}@{}@{}p{0.1\textwidth}@{}@{}p{0.1\textwidth}@{}@{}p{0.12\textwidth}@{}@{}p{0.1\textwidth}@{}@{}p{0.13\textwidth}@{}@{} @{}} 
\toprule
\thd{Operator} & \thd{$U_0$} & \thd{$P_{L/R}$} & \thd{$P'_{L/R}$} & \thd{$U_{\text{Sch}}^{(n-1)}$} & \thd{$U_S$} & \thd{$U_{\text{Sch}}^{(n-2)}$} & \thd{$U_k$} & \thd{$U_{l/r}$}\\
\midrule
\thd{Gate Accuracy} & \thd{$\epsilon_0$} & \thd{$\epsilon_1$} & \thd{$\epsilon_1'$} & \thd{$\epsilon_2$} & \thd{$\epsilon_3$} & \thd{$\epsilon_4$} & \thd{$\epsilon_5$} & \thd{$\epsilon_6$}\\
\thd{\textcolor{violet}{Setting $\Theta[\cdot]$}} & \thd{\textcolor{violet}{$ (n^{9/2} d^3)^{-1}$}} & \thd{\textcolor{violet}{$ (n^{3} d^{5})^{-1}$}} & \thd{\textcolor{violet}{$ (n^{5/2} d^4)^{-1}$}} & \thd{\textcolor{violet}{$ (n d^3)^{-1}$}} & \thd{\textcolor{violet}{$ (n^{3} d^{5})^{-1}$}} & \thd{\textcolor{violet}{$ (n^{4} d^{13/2})^{-1}$}} & \thd{\textcolor{violet}{$ (n^{5/2} d)^{-1}$}} & \thd{\textcolor{violet}{$ (n^{9/4} d^{5/2})^{-1}$}}\\
\addlinespace
\bottomrule
\end{tabular}
\end{center}
\end{table}

\begin{table}[H]
\caption{Summary of the block-encoding parameters, including the number of ancilla qubits, the scaling factors and the accuracies (normed errors) of the operators.}\label{table:parameters}
\begin{center}
\begin{tabular}{@{} p{0.06\textwidth}@{}@{}p{0.27\textwidth}@{}@{}p{0.68\textwidth} @{}} 
\toprule
\thd{ID} & \thd{Encoding Parameter} & \thd{Value (In Terms of $n$, $d$ and the Gate Accuracies)}\\
\midrule
\addlinespace
\thd{$1$} & \thd{$\#$ of ancilla qubits $a$} & \thd{$O[d^2\log(n)]$} \\
& \thd{$\#\text{Pad}$} & \thd{$O[d^2\log(n)]$} \\
\thd{$2$} & \thd{the scaling factor\\$x$ in $P_{L/R}$} & \thd{$\sqrt{d}$}\\
& \thd{the scaling factor\\$x'$ in $P'_{L/R}$} & \thd{$\sqrt{d}$}\\
\thd{$3$} & \thd{the scaling factor $\alpha\sqrt{n-1}$\\in amplitude amplification} & \thd{$O(n^{5/2} d^3)$}\\
\thd{$4$} & \thd{the error $\varepsilon_2$ in\\$O_\text{cen}(i,k_l,k_r)$} & \thd{$d[k^*+(k^*+1)2(n-1)d^3\epsilon_2]$ where $k^*=2\sqrt{d}(d+2)\epsilon_1+d(d+2)^2\epsilon_1^2$}\\
\thd{$5$} & \thd{the error $\varepsilon_2'$ in\\$O'_\text{cen}(k_l,k_r)$} & \thd{$d[{k'}^*+({k'}^*+1)2(n-1)d^3\epsilon_2]$ where ${k'}^*=2\sqrt{d}(d+2)\epsilon_1'+d(d+2)^2{\epsilon_1'}^2$}\\
\thd{$6$} & \thd{the error $\varepsilon_3$ of $\tilde{\Phi}$} & \thd{$d^2\epsilon_3 + \sqrt{d}d^3(n-2)\epsilon_4$}\\
\thd{$7$} & \thd{the error $\delta$ of the \\block-encoding of $\sqrt{\Pi_i}$} & \thd{$\sqrt{d}\ \text{poly}_2\Big\{ \left[ (n-1)^{5/2}d^3 + (n-1)^{2}d^2 + 1 \right]^{1/2}\epsilon_6 \Big\}$\\$+  (n-1)^2 \left( 2d^2\varepsilon_2+2\sqrt{d}d^2\varepsilon_3 \right)$\\$ + (n-1)^{3/2} \left( d\varepsilon_2'+2\sqrt{d}d\varepsilon_3 \right) + \left[\sqrt{d}+1/\sqrt{n-1}\right]\text{poly}_4\left[\epsilon_5(n-1)^{3/2}\right] $}\\
\thd{$8$} & \thd{the error $2m\varepsilon$ of\\the entire circuit $\tilde{V}$ (or $U$)} & \thd{$\sqrt{d}(n-1)\pi\cdot \delta + \alpha(n-1)^2\pi\epsilon_0$}\\
\addlinespace
\bottomrule
\end{tabular}
\end{center}
\end{table}

\begin{table}[H]
\caption{Time and space complexity (both in terms of gates in the universal set) of our circuit. The computations of the representation-theoretic coefficients in $P\textcolor{magenta}{'}_{L/R}$ are performed on a classical Turing machine, paralleled and stored by matrix element. Note that in this table, we have made $n-1$ a power of $2$, so that the $n-1$ states in $U_0$ and $U_k$ are equally superposed using $\log(n-1)$ Hadamard gates, eliminating the unessential $\poly{n}$ costs. The final complexities (in violet) are calculated by setting the gate accuracies high enough to make the total error $O(1)$, as in Table~\ref{table:gate_accuracy}.}\label{table:complexities}
\begin{center}
\begin{tabular}{@{} p{0.12\textwidth}@{}@{}p{0.43\textwidth}@{}@{}p{0.43\textwidth} @{}} 
\toprule
\thd{Operator} & \thd{Time Complexity (T)} & \thd{Space Complexity (S)}\\
\midrule
\addlinespace
\thd{$U_0$} & \thd{$\log{(n)}\polylog{\frac{1}{\epsilon_0}}$} & \thd{$\log{(n)}\polylog{\frac{1}{\epsilon_0}}$} \\
\thd{$P\textcolor{magenta}{'}_{L/R}$ \\ $\textcolor{magenta}{\epsilon_1 \rightarrow \epsilon_1'}$} & \thd{$\poly{d}\polylog{\frac{1}{\epsilon_1}}$} & \thd{$\#\alpha\ O[(d+2)^2]\polylog{\frac{1}{\epsilon_1}}$\\$=O(n^d)\polylog{\frac{1}{\epsilon_1}}$} \\
\thd{$U_{\text{Sch}}^{(n-1)}$} & \thd{$n\ \text{poly}[d,\log{(n)},\log{(\frac{1}{\epsilon_2})}]$} & \thd{$O(n^{2d})\polylog{\frac{1}{\epsilon_2}}$} \\
\thd{$U_S$} & \thd{$\poly{d}\polylog{\frac{1}{\epsilon_3}}$} & \thd{$O(d^4)\polylog{\frac{1}{\epsilon_3}}$}\\
\thd{$U_{\text{Sch}}^{(n-2)}$} & \thd{$n\ \text{poly}[d,\log(n),\log{(\frac{1}{\epsilon_4})}]$} & \thd{$O(n^{2d})\polylog{\frac{1}{\epsilon_4}}$} \\
\thd{$U_k$} & \thd{$\log{(n)}\polylog{\frac{1}{\epsilon_5}}$} & \thd{$\log{(n)}\polylog{\frac{1}{\epsilon_5}}$} \\
\thd{$U_{l/r}$} & \thd{$\polylog{d,n,\frac{1}{\epsilon_6}}$} & \thd{$\polylog{\frac{1}{\epsilon_6}}$} \\
\thd{$U^c$} & \thd{$T\big\lbrace\text{max}[2P_L+2P_R, P_L'+P_R']  +4 U_{\text{Sch}}^{(n-1)} $\\$ + 2 \text{max}[U_{\text{Sch}}^{(n-2)},U_S] +2U_k + U_l + U_r\big\rbrace $} & \thd{$S\big\lbrace 2P_L + 2P_R +P_L'+P_R' + 4 U_{\text{Sch}}^{(n-1)} $\\$ + 2 U_{\text{Sch}}^{(n-2)} + 2U_S + 4U_k + U_r+U_l \big\rbrace$}\\
\thd{Total} & \thd{$\poly{n}\polylog{\frac{1}{d\epsilon_1}}$ (compute coeffs. in $P\textcolor{magenta}{'}_{L/R}$)\\$ +O(n^{5/2}d^3)  T\big\lbrace U^c \big\rbrace + O(n^{5/2}d^3) T\big\lbrace U_0 \big\rbrace $\\\textcolor{violet}{$=\poly{n,d}$}} & \thd{$O(n^{5/2}d^3)  S\big\lbrace U^c\big\rbrace + O(n^{5/2}d^3) S\big\lbrace U_0 \big\rbrace $\\\textcolor{violet}{$=O(n^{2d})$}} \\
\addlinespace
\bottomrule
\end{tabular}
\end{center}
\end{table}

\subsection{Unitary complementation and ``qubitizing'' unitary matrices \label{sec:qubitization}}

A main theme of our algorithm is to quantumly implement numerical matrices, which are complemented from a given row/column vector that incorporate appropriate weights for a superposition. This is done by the following three steps:
\begin{enumerate}
    \item Use singular value decomposition (SVD) to unitarily complement $k$ orthonormal vectors $\mathbb{C}^m$ ($k<m$) to a unitary matrix. The unitary matrix is obtained by stacking the right null row space of the $k\times m$ matrix onto the matrix itself. This takes $O(mk^2)\text{polylog}(1/\epsilon_1)$ elementary operations \cite{Vinita17} and gives matrix elements of accuracies $\epsilon_1$.
    \item Given an $m\times m$ unitary matrix with known elements, use a classical Turing machine to decompose it in time $\poly{m}\polylog{1/\epsilon_2}$ into $O(m^2)$ one-or-two-qubit gates in sequence, with gate accuracies $\epsilon_2$ (in spectral norm) \cite{Zeilinger94, Barenco95}.
    \item Each of these one-or-two-qubit gates can then be decomposed to an accuracy $\epsilon_3$ in time $\polylog{1/\epsilon_3}$ into $\polylog{1/\epsilon_3}$ gates from a fixed and finite set, complexity following the Solovay-Kitaev theorem~\cite{Kitaev97, Dawson05}.
\end{enumerate}

To make the best use of these accuracies, we let $\epsilon_1 / m = \epsilon_2=\epsilon_3$. This is because norm errors add up when unitaries are multiplied, so the spectral error of the $m\times m$ unitary is $O(m^2 \epsilon_2)\leqslant \epsilon_F = \sqrt{m^2 \epsilon_1^2}=m\epsilon_1$, where $F$ denotes the Frobenius norm; and because $\epsilon_3$ need not be less than $\epsilon_2$.

In all, a complemented $m\times m$ numerical matrix will be decomposed into $O(m^2)\polylog{1/\epsilon_2}$ quantum gates from a universal set in time $\poly{m}\polylog{1/\epsilon_2}$. After this \textit{online} classical ``compilation'', the output quantum circuit is built from the indicated gates. Our order of magnitude estimate gives accuracies of an [$m\times m\ \text{matrix}:\text{1-or-2-qubit-gate}:\text{matrix element}= m^2:1:m$].

\subsection{Complexity of the quantum Schur transform\label{app:complexity_Sch}}
\subsubsection{Standard and semistandard tableau}
In Sec.~\ref{sec:sch}, we illustrated that the Schur basis is canonical, i.e., the irrep chains subgroup-reduced down the towers $S(n)\supset S(n-1) \supset\cdots\supset S(1)$ and $U(d)\supset U(d-1)\supset \cdots \supset U(1)$ are all uniquely defined and each of these chains corresponds to a basis element. For the symmetric group portion, they are represented by sequences of one-box removals, for example,
\begin{equation}
    \yng(2,2,1) \rightarrow \yng(2,1,1) \rightarrow \yng(2,1) \rightarrow \yng(1,1) \rightarrow \yng(1).
    \label{eq:Youngdiagramchain}
\end{equation}
is a basis element in the irrep $\lambda = (2,2,1)$ and for $n=5$. The register $\ket{p_\lambda}$ can then be divided into $\ket{\lambda_{n-1}}\ket{\lambda_{n-2}}\cdots \ket{\lambda_1}$ where the valid labels of the Schur basis should satisfy $\lambda_{i-1}=\lambda_i -\Box$.  Notice that, since the irreps $\lambda_i$ of $S(i)$ are in $(\mathbb{C}^d)^{\otimes i}$, $\lambda_i$ should have height $h(\lambda_i)\leqslant d$. More concisely, each chain can be written as a \textit{standard $\lambda$-tableau} with non-repeating entries in $\{1,...,n\}$ which corresponds to the ordering of one-box removals. For example, the basis element given in Eq. \eqref{eq:Youngdiagramchain} can be written as
\begin{equation}
    \young(13,25,4).
\end{equation}
This means that standard tableau have strictly increasing rows and columns.

Similarly, the unitary group basis register $\ket{q_\lambda^d}$ can be divided into $\ket{q_{d-1}}\ket{q_{d-2}}\cdots\ket{q_1}$, where $q_i$ labels an irrep of $U(i)$, and so $h(q_i)\leqslant i$, and the valid chains have $q_{i-1}\lesssim q_i$. We read $\lesssim$ \textit{interlaces} and write $\mu\lesssim \nu$ whenever $\mu$ is a valid diagram obtained from removing zero or one boxes from each column of $\nu$. Each $U(d)$ irrep chain can therefore be written as a \textit{semistandard $\lambda$-tableau}, with entries in $\{1,...,d\}$ that weakly increase across rows and strictly increase down columns. For example, for $d=2$ and $\lambda = (2)$, there are multiple possible semistandard $\lambda$-tableau
\begin{equation}
    \young(11),\ \young(12),\ \young(22)
\end{equation}
and so the corresponding $U(d)$ irrep is three-dimensional.

\subsubsection{The algorithm in~\cite{Harrow05,Harrow06}}
The theoretical underpinning of the efficient algorithm in~\cite{Harrow05,Harrow06} is a duality between the subgroup branching in $S(n)$ and the Clebsch-Gorden (CG) transforms of $U(d)$.  Restricting $S(n)$ on $d$-dimensional systems to $S(k)\times S(n-k)$---permutations that leave the set $\{1,\dots,k\}$ and $\{k+1,\dots,n\}$ invariant---branches an $S(n)$ irrep $\mathcal{P}_\lambda$ into a direct sum over all irreps $\mathcal{P}_\mu\hat{\otimes} \mathcal{P}_\nu$ of $S(k)\times S(n-k)$. Their multiplicities satisfy
\begin{align}
\dim{\text{Hom}_{S_k\times S_{n-k}}}(\mathcal{P}_\mu\hat{\otimes} \mathcal{P}_\nu, \mathcal{P}_\lambda) = \dim{\text{Hom}_{U_d}(\mathcal{Q}^d_\lambda, \mathcal{Q}^d_\mu\otimes \mathcal{Q}^d_\nu})
\end{align}
In particular, when $k=n-1$ the branching in every $\mathcal{P}_\lambda$ is canonical, i.e., $\dim{\text{Hom}_{S_{n-1}}(\mathcal{P}_{\lambda'}\hat{\otimes} \mathcal{P}_\Box, \mathcal{P}_\lambda)} = \dim{\text{Hom}_{U_d}(\mathcal{Q}^d_\lambda, \mathcal{Q}^d_{\lambda'}\otimes \mathcal{Q}^d_\Box})=1$ only when $\lambda = \lambda' + \Box$, zero otherwise. Branching an irrep $\lambda$ of $S(n)$ down the subgroup tower $S(n-1) \supset S(n-2) \supset ... \supset S(1)$ thus gives all standard $\lambda$-tableau. By exploiting the CG dual we obtain an algorithm: starting with $\mathcal{Q}^d_{\Box}$ = $\mathbb{C}^{d}$, iteratively ($i= 0,\dots, n-1$) apply $U_{\text{CG}}^{[d]}(i)$, which merges an additional $\mathbb{C}^{d}$ into all irreps $\lambda'\vdash i$ as
\begin{align}
\mathcal{Q}^d_{\lambda'} \otimes \mathcal{Q}^d_\Box \mycong{$U_d$} \bigoplus_{\lambda = \lambda' +\Box, h(\lambda)\leqslant d} \mathcal{Q}^d_{\lambda},
\end{align}
until finally $(\mathbb{C}^{d})^{\otimes n}$ is reorganized into all $U(d)$ irreps $\mathcal{Q}^d_{\lambda}$ with $\lambda \vdash n$ and $h(\lambda)\leqslant d$. A consistency check is that, each copy of $\mathcal{Q}^d_{\lambda}$ for a fixed $\lambda$ actually comes with an $S(n)$ branching sequence, and thus each basis element of the multiplicity space of $\mathcal{Q}^d_{\lambda}$, $\mathcal{P}_{\lambda}$, can be labeled by a standard $\lambda$-tableau.

The CG transform $U_{\text{CG}}^{[d]}(i)$ conditions on $\lambda'$ and unitarily transforms $\ket{\lambda'}\ket{q_{\lambda'}}\ket{i+1}$ to $\ket{\lambda'}\ket{\lambda}\ket{q_\lambda}$. This uses $U_{\text{CG}}^{[d-1]}(i)$ as a subroutine, and a bunch of $d\times d$ unitary operators $T^{[d]}(\lambda',\lambda'')$, which condition on the labels $\lambda'\vdash i$ and $\lambda'' \lesssim \lambda',h(\lambda'')\leqslant d-1$ and comprise the reduced Wigner coefficients $\hat{T}^{\lambda',\lambda'+e_{j_1},\lambda'',\lambda''+e_{j_2}}, j_1,j_2=1,\dots,d$. Spelling it out, the subgroup-reduced and iterative nature of the algorithm ensures that the register $\ket{q_{\lambda'}}$ can be split up into $\ket{\lambda''}\ket{q_{\lambda''}}$. The algorithm then performs
\begin{align}
    \ket{\lambda'}\ket{\lambda''}\ket{q_{\lambda''}}\ket{i+1} \overset{U_{\text{CG}}^{[d-1]}(i)} {\longrightarrow} \ket{\lambda'}\ket{\lambda''}\ket{\lambda''+e_{j_2}}\ket{q_{\lambda''+e_{j_2}}} \overset{T^{[d]}(\lambda',\lambda'')} {\longrightarrow}  \ket{\lambda'}\ket{\lambda'+e_{j_1}}\ket{\lambda''+e_{j_2}}\ket{q_{\lambda''+e_{j_2}}}
\end{align}
with special care to $\ket{i+1}$ being in the state $\ket{d}$, and renames $\ket{\lambda'+e_{j_1}}$ as $\ket{\lambda}$ and $\ket{\lambda''+e_{j_2}}\ket{q_{\lambda''+e_{j_2}}}$ as $\ket{q_\lambda}$. $U_{\text{CG}}^{[d-1]}(i)$ in turn uses $U_{\text{CG}}^{[d-2]}(i)$ and $T^{[d-1]}(\lambda'',\lambda''')$ as subroutines where $\lambda''' \lesssim \lambda'', h(\lambda''')\leqslant d-2$, et cetera.

\subsubsection{Complexity and error}
Here we restate the time complexity of the algorithm, and calculate its circuit size and error.
Let the gate accuracy for $T^{[d]}(\lambda',\lambda'')$ be $\epsilon$. Since the $T^{[d]}(\lambda',\lambda'')$ for different $(\lambda',\lambda'')$ pairs are quantum paralleled, their runtime and spectral error should be paralleled while their circuit sizes be added up. Therefore, the total error of $U_{\text{Sch}}^{(n)}$ is $nd\cdot O(d^2\epsilon)=O(nd^3\epsilon)$ and the runtime is $n\cdot\poly{d,\log{n},\log{1/\epsilon}}$, with the overhead of classically computing the reduced Wigner coefficients included.

The circuit size, i.e., the number of gates from a universal set, is $O(n^{2d})\polylog{1/\epsilon}$.
\begin{proof}
We consider the most complex operator, $T^{[d]}$ in $U^{[d]}_{\text{CG}}(n-1)$. Induced by a particular $(\lambda',\lambda'')$ pair, $T^{[d]}(\lambda',\lambda'')$ is a $d\times d$ unitary matrix and thus can be implemented by $O(d^2)\polylog{1/\epsilon}$ gates. The remainder is to calculate the number of $(\lambda',\lambda'')$ pairs whose circuits need to be constructed separately.

The interlacing relation tells us that $\#\lambda''=\prod_{j=1}^{d-1}a_j$ where $a_j\equiv\lambda'_{j}-\lambda'_{j+1}+1$. So $
\#\lambda''\leqslant \left(n-1+1\right)^{d-1} \leqslant n^{d-1}.
$

$\# \lambda' \leqslant p(n,1) + p(n,2) + \cdots + p(n,d)$ where $p(n,k)$ denotes the number of valid partitions of $n$ into exactly $k$ parts \cite{Oruc16}.
$
p(n,k)\leqslant \frac{1}{k!}
\begin{pmatrix}
n+\frac{k(k-1)}{2}-1\\
k-1
\end{pmatrix}\\
\leqslant \frac{1}{k!}e^{k-1}\left(\frac{n}{k-1} + \frac{k}{2} - \frac{1}{k-1}\right)^{k-1}
\leqslant e^{2k-2}\left[\frac{2^{k-2}}{k^k}\left(\frac{n}{k-1}\right)^{k-1}+\frac{1}{2k}\right]
=O\left[\left(\frac{\sqrt{2}e}{k-1}\right)^{2k-1} n^{k-1}\right]=O(n^{k-1}).\\
\Rightarrow\#\lambda'\leqslant O(n^{d-1})$.

Collecting the contributions from $T^{[j]}(\lambda',\lambda''),j=1,\dots,d$ and then from $U^{[d]}_{\text{CG}}(i),i=0,\dots,n-1$, the circuit size is $\#\lambda'\#\lambda''\cdot O(nd^3) \polylog{1/\epsilon}=O(n^{2d})\polylog{1/\epsilon}$.
\end{proof}

\subsection{The proofs for Table~\ref{table:parameters}}\label{sec:error}
\begin{enumerate}
\item The ancilla and padded qubits primarily come from the Schur transforms.
There are two sets of registers, respectively for the $(n-2)$-system and $(n-1)$-system transforms: $\ket{q_{n-2}=r}$, $\ket{\lambda_{n-2}=\alpha}$ and $\ket{p_{n-2}=k_\alpha}$; $\ket{q_{n-1}=r_\nu}$, $\ket{\lambda_{n-1}=\nu}$, $\ket{\lambda_{n-2}=\alpha}$ and $\ket{p_{n-2}=k_\alpha}$. To achieve alignment, the labels $\lambda_i$ in $\ket{\lambda_{n-2}=\alpha}$ and in $\ket{k_\alpha}=\ket{\lambda_{n-3}}\ket{\lambda_{n-2}}\cdots \ket{\lambda_1}$ should encode Young diagrams in the same way in both the transforms.

To save qubits, we would encode $\lambda_i,i={n-3},\dots,1$ by $e^i\in[d]$, where $\lambda_{i}=\lambda_{i-1}+\Box$ and the newly added box to $\lambda_i$ is in its $e^i$-th row. For $\lambda_{n-2}$ and $\lambda_{n-1}$, we would encode them in a direct way by $d$ numbers in $\{0,\dots,n-1\}$, the number of boxes in each row, because $P_{1},P_2,P_{L/R}$ are controlled by those labels. So $\log{n_{d_\alpha}}=(n-3)\log{d}$,  $\log{n_\nu}=\log{n_\alpha}=d\log{n}$.

Each $U(d)$ basis vector $\ket{q}=\ket{q_{d-1}}\ket{q_{d-2}}\cdots\ket{q_1}$ is encoded in the algorithm by $d-1$ height-$< d$ Young diagrams where $q_{i-1}\lesssim q_i$, which compose a semistandard tableau, and so by $d^2\log{(n+1)}$ qubits. Thus, $\log{n_r}=d^2\log{(n-1)}$, $\log{n_{r_\nu}}=d^2\log{n}$.

Plugging them into Eq. (\ref{eq:alpha_a}) gives $a=\#\text{Pad}=O(d^2\log{n})$.

\item $x$ and $x'$ should be preset for a given dimensional pair $(n,d)$ before running the circuit. Either for reducing the amplification factors or for reducing the errors, the smaller they are the better. The only constraints that they must satisfy are Eq. (\ref{eq:cons1}) and (\ref{eq:cons2}):
\begin{align}
\left\{
\begin{aligned}
&x^2 \geqslant \sum_{\nu=\alpha + \Box} C(\alpha,\nu) & \forall \alpha\\
&x'^2 \geqslant \sum_{\nu=\alpha + \Box} C'(\alpha,\nu) & \forall \alpha.
\end{aligned}
\right.
\end{align}
In the above,
\begin{align}
\left\{
\begin{aligned}
C(\alpha,\nu) &= [(n-1)d_\alpha-d_\theta]^{-\frac{1}{4}} [(n-1)d_\alpha]^{-\frac{3}{4}} \frac{d_{\nu}}{\sqrt{\lambda_{\nu}}}\leqslant\left[ \frac{1}{(n-1)^3}\frac{d_\nu^3}{d_\alpha^3}\frac{m_\alpha}{m_\nu} \right]^{1/2}\\
C'(\alpha,\nu)&= [(n-1)d_\alpha]^{-1} \frac{d_{\nu}}{\lambda_{\nu}}\leqslant\left[ \frac{1}{(n-1)^2}\frac{d_\nu^2}{d_\alpha^2}\frac{m_\alpha}{m_\nu} \right]
\end{aligned}
\right.
\end{align}
where $\lambda_\nu=(n-1)(m_\nu d_\alpha)/(m_\alpha d_\nu)$.

Notice that
\begin{align}
(n-1)d_\alpha = \sum_{\nu=\alpha+\Box}d_\nu,
\end{align}
because
\begin{align}
\text{Ind}_{S(n-2)}^{S(n-1)}(\alpha) \cong \bigoplus_{\nu=\alpha+\Box}\nu.
\end{align}
Also notice that $m_\alpha\leqslant m_\nu$, considering the number of semistandard tableau. So $C(\alpha,\nu)\leqslant 1$ and $C'(\alpha,\nu)\leqslant 1$. And since the number of ways to add a box to any $\alpha$ is at most $d$,
\begin{align}
\left( x, x' \right) = \left( \sqrt{d}, \sqrt{d} \right)
\end{align}
should satisfy the constraints.

\item This is direct by plugging $x=x'=\sqrt{d}$ into Eq. (\ref{eq:alpha_a}).

\item For a fixed $\alpha$ and $\nu_l\in \nu^\alpha = \alpha+\Box$, denote by $a_{\nu_l}$ and $b_{\nu_l}$ the implemented matrix elements of $P_L[x](\alpha)$ and $P_R[x](\alpha)$, which approximate the exact ones in Eq. (\ref{eq:P_L_el}) as $\Big\lvert a_{\nu_l}-\frac{\sqrt{C(\alpha,\nu_l)}}{x} \Big\rvert \leqslant (n_{\alpha+\Box}+2)\epsilon_1$ and $\Big\lvert b_{\nu_l}-\frac{\sqrt{C(\alpha,\nu_l)}}{x} \Big\rvert \leqslant (n_{\alpha+\Box}+2)\epsilon_1$. Let $k\equiv \Big\lvert C(\alpha,\nu_l)-x^2 a_{\nu_l} b_{\nu_l}^*\Big\rvert$, then we can obtain $k\leqslant 2\sqrt{C(\nu_l,\alpha)}x(n_{\alpha+\Box}+2) \epsilon_1 + x^2(n_{\alpha+\Box}+2)^2\epsilon_1^2\leqslant 2\sqrt{d}(d+2)\epsilon_1+d(d+2)^2\epsilon_1^2\equiv k^*$.
\begin{align}
&\norm{\sum_{\alpha=1}^{n_\alpha} \ket{e_\alpha}\bra{e_\alpha}\otimes O(\alpha, k,i) - x^2 \left( \bra{0}_{r \nu}\otimes \mathbb{I}_{\alpha k_\alpha}\right) U[x^2](k,i) \left( \ket{0}_{r \nu}\otimes \mathbb{I}_{\alpha k_\alpha}\right)}\\
=&\norm{ \sum_{\alpha=1}^{n_\alpha} \ket{e_\alpha}\bra{e_\alpha}\otimes \Biggl[ O(\alpha, k,i) - x^2 \sum_{\nu_l=\alpha+\Box}\sum_{k_\alpha,k'_\alpha=1}^{n_{d_\alpha}} a_{\nu_l} b_{\nu_l}^* \ket{e_{k_\alpha}} \bra{r_{\nu_l},\nu_l,\alpha,k_\alpha} ^{(n-1)}_{\text{Sch}} \nonumber\\
&\hspace{6cm} \times V(\pi_i)\cdot V(\pi_k) \ket{r_{\nu_l},\nu_l,\alpha,k'_\alpha} ^{(n-1)}_{\text{Sch}} \bra{e_{k'_\alpha}}\Biggr] }\\
\leqslant & \text{max}_\alpha \Big\lbrace \sum_{\nu_l=\alpha+\Box} k \norm{ \sum_{k_\alpha,k'_\alpha=1}^{n_{d_\alpha}} \ket{e_{k_\alpha}} \bra{r_{\nu_l},\nu_l,\alpha,k_\alpha} V(\pi_i)\cdot V(\pi_k)  \ket{r_{\nu_l},\nu_l,\alpha,k'_\alpha} ^{(n-1)}_{\text{Sch(exact)}} \bra{e_{k'_\alpha}} } \nonumber\\
&\hspace{1cm} + \Big\lvert x^2 a_{\nu_l} b_{\nu_l}^* \Big\rvert \norm{ \sum_{k_\alpha,k'_\alpha=1}^{n_{d_\alpha}} \ket{e_{k_\alpha}} \Biggl( \bra{r_{\nu_l},\nu_l,\alpha,k_\alpha} ^{(n-1)} V(\pi_i)\cdot V(\pi_k) \ket{r_{\nu_l},\nu_l,\alpha,k'_\alpha} ^{(n-1)}_{\text{Sch(exact)}} \nonumber\\
&\hspace{4cm} -  \bra{r_{\nu_l},\nu_l,\alpha,k_\alpha}  V(\pi_i)\cdot V(\pi_k) \ket{r_{\nu_l},\nu_l,\alpha,k'_\alpha} ^{(n-1)}_{\text{Sch}} \Biggr) \bra{e_{k'_\alpha}} }  \Big\rbrace \\
\leqslant & \text{max}_\alpha \Big\lbrace \sum_{\nu_l=\alpha+\Box} k + \left[k+ C(\alpha,\nu_l) \right] 2(n-1)d^3\epsilon_2\Big\rbrace\\
\leqslant & d[k^*+(k^*+1)2(n-1)d^3\epsilon_2]
\end{align}

\item Similar to 4.

\item \begin{align}
&\norm{\sqrt{d}\left(\bra{0}_{A_2} \otimes \mathbb{I}_{r\alpha k_\alpha}\otimes \mathbb{I}_{n-1n} \right) U^c_{\Phi} \left(\ket{0}_{A_2} \otimes \mathbb{I}_{\text{Pad}}\otimes \mathbb{I}_{[n]} \right) - \tilde{\Phi}}\\
=& \norm{\sqrt{d} U_{\text{Sch(exact)}}^{(n-2)} \otimes \ket{0}\bra{0}_{n-1n} \left( U_S - U_{S\text{(exact)}}\right)} + \norm{\sqrt{d} \left(U_{\text{Sch}}^{(n-2)} - U_{\text{Sch(exact)}}^{(n-2)} \right) \otimes \ket{0}\bra{0}_{n-1n} U_S}\\
\leqslant& d^2\epsilon_3 + \sqrt{d}d^3(n-2)\epsilon_4
\end{align}
In the last line, the error of the first row of $U_S$ has been replaced with $d\sqrt{d}\epsilon_3$.

\item For $i=1,\dots,n-1$,
\begin{align}
&\norm{\left[(n-1)^2dx^4 + (n-1)^{3/2}dx'^2 + (n-1)^{-1/2} \right] \pp{ \bra{0}_{\text{Anc}_{\text{tot}}} \otimes \mathbb{I}_{[n]} } U^c(i)\pp{ \ket{0}_{\text{Anc}_{\text{tot}}}\otimes \mathbb{I}_{[n]} } - \sqrt{\Pi_i} } \\
\leqslant & \norm{\bigg\{ \left[ (n-1)^2dx^4 + (n-1)^{3/2}dx'^2 + (n-1)^{-1/2} \right] c_lc_r -1 \bigg\} \left( \bra{0}_{k_l,k_r,A_U} \otimes \mathbb{I}_{[n]} \right) \left[(n-1)^2dx^4 \tilde{U}(i) \right.\nonumber\\
& \left. +\mathbb{I}_{k_l,k_r,A_U,[n]} /\sqrt{n-1} - (n-1)^{3/2}dx'^2 \tilde{U}'(i) \right] \left( \ket{0}_{k_l,k_r,A_U} \otimes \mathbb{I}_{[n]} \right) } + \norm {\left( \bra{0}_{k_l,k_r,A_U} \otimes \mathbb{I}_{[n]} \right) \nonumber\\
& \left[(n-1)^2dx^4 \tilde{U}(i) +\mathbb{I}_{k_l,k_r,A_U,[n]} /\sqrt{n-1} - (n-1)^{3/2}dx'^2 \tilde{U}'(i) \right] \left( \ket{0}_{k_l,k_r,A_U} \otimes \mathbb{I}_{[n]} \right) - \sqrt{\Pi_i} } \\
\lessapprox & \bigg| \left[ (n-1)^2dx^4 + (n-1)^{3/2}dx'^2 + (n-1)^{-1/2} \right] c_lc_r -1 \bigg| \cdot \norm{\sqrt{\Pi_i}} +  \nonumber\\
&\norm{(n-1)^2dx^4\left( \bra{0}_{k_l,k_r,A_U} \otimes \mathbb{I}_{[n]} \right) \tilde{U}(i) \left( \ket{0}_{k_l,k_r,A_U} \otimes \mathbb{I}_{[n]} \right) - \sqrt{\tilde{\Pi}_i}} + 1/\sqrt{n-1} \norm{(n-1)^2dx'^2\nonumber\\
& \left( \bra{0}_{k_l,k_r,A_U} \otimes \mathbb{I}_{[n]} \right) \tilde{U}'(i) \left( \ket{0}_{k_l,k_r,A_U} \otimes \mathbb{I}_{[n]} \right) - \sum_{k_l,k_r=1}^{n-1} V_L(\pi_{k_l})\cdot \Phi^\dagger \cdot O_{\text{cen}}'(k_l,k_r) \cdot \Phi \cdot V_L(\pi_{k_r}) }
\end{align}

In the first term above, by using $\Big\lvert c_{l/r} - c_{l/r\text{(exact)}} \Big\rvert \leqslant 4\epsilon_6$, we can get $\Big| \left[ (n-1)^{5/2}dx^4 + (n-1)^{2}dx'^2 + 1 \right] \linebreak[1]c_lc_r - 1 \Big|\leqslant  8 \left[ (n-1)^{5/2}d^3 + (n-1)^{2}d^2 + 1 \right]^{1/2}\epsilon_6 + 16 \left[ (n-1)^{5/2}d^3 + (n-1)^{2}d^2 + 1 \right]\epsilon_6^2 \\ \equiv \text{poly}_2\Big\{\left[ (n-1)^{5/2}d^3 + (n-1)^{2}d^2 + 1 \right]^{1/2}\epsilon_6 \Big\}$.

Now we will calculate the second term: Denoting by $a_{k}$ and $b_{k}$ the implemented matrix elements of $U_k$ on the left and right-hand side of the circuit, respectively, it follows that $\Big\lvert a_{k}-1/\sqrt{n-1} \Big\rvert \leqslant \epsilon$ and $\Big\lvert b_{k}-1/\sqrt{n-1} \Big\rvert \leqslant \epsilon$ where the element accuracy $\epsilon=O(n-1)\epsilon_5$. Let $k\equiv \Big\lvert (n-1)^2 a_{k_l} a_{k_r} b_{k_l}^* b_{k_r}^* - 1 \Big\rvert$, it can be seen that $k\leqslant \epsilon^4(n-1)^2+4\epsilon^3(n-1)^{3/2} + 6 \epsilon^2(n-1) +4\epsilon (n-1)^{1/2}\equiv \text{poly}_4\left[\epsilon_5(n-1)^{3/2}\right]$. The second term then obeys
\begin{align}
&\norm{(n-1)^2dx^4 \left( \bra{0}_{k_l,k_r,A_U} \otimes \mathbb{I}_{[n]} \right) \tilde{U}(i) \left( \ket{0}_{k_l,k_r,A_U} \otimes \mathbb{I}_{[n]} \right) - \sqrt{\tilde{\Pi}_i}}\\
\leqslant& \norm{ (n-1)^2 \bra{0}_{k_lk_r} U_k \otimes U_k \sum_{k_l,k_r=1}^{n-1}\ket{k_l k_r}\bra{k_l k_r}_{k_lk_r} \otimes \Big[ dx^4\left( \bra{0}_{A_U} \otimes \mathbb{I}_{[n]} \right) U^c(i,k_l,k_r) \left( \ket{0}_{A_U} \otimes \mathbb{I}_{[n]} \right) \nonumber\\
& - V_L(\pi_{k_l})\cdot \Phi^\dagger \cdot O_{\text{cen}}(i,k_l,k_r) \cdot \Phi \cdot V_L(\pi_{k_r}) \Big] U_k^\dagger \otimes U_k^\dagger \ket{0}_{k_lk_r}}
 + \norm{ (n-1)^2 \bra{0}_{k_lk_r} U_k \otimes U_k  \nonumber\\
& \sum_{k_l,k_r=1}^{n-1}\ket{k_l k_r}\bra{k_l k_r}_{k_lk_r} \otimes\Big[ V_L(\pi_{k_l})\cdot \Phi^\dagger \cdot O_{\text{cen}}(i,k_l,k_r) \cdot \Phi \cdot V_L(\pi_{k_r}) \Big] U_k^\dagger \otimes U_k^\dagger \ket{0}_{k_lk_r} -\sqrt{\tilde{\Pi}_i} } \\
\leqslant & (n-1)^2 \left( 2dx^2\varepsilon_2+2\sqrt{d}x^4\varepsilon_3 \right) + k \norm{ \sqrt{\tilde{\Pi}_i} } \\
\leqslant & (n-1)^2 \left( 2d^2\varepsilon_2+2\sqrt{d}d^2\varepsilon_3 \right) + \sqrt{d}\ \text{poly}_4\left[\epsilon_5(n-1)^{3/2}\right],
\end{align}
where we have used \cite{Studzinski18} $\norm{ \sqrt{\tilde{\Pi}_i} }\leqslant\left[\left(\lambda_{\text{min}}\rho\right)^{-1} \lambda_{\text{max}}\rho_i \right]^{1/2} = \big\lbrace d/\left[ d - \min{(n-2,d-1)} \right] \big\rbrace^{1/2}\leqslant \sqrt{d}$.

Similarly, the third term satisfies
\begin{align}
& \norm{(n-1)^2dx'^2 \left( \bra{0}_{k_l,k_r,A_U} \otimes \mathbb{I}_{[n]} \right) \tilde{U}'(i) \left( \ket{0}_{k_l,k_r,A_U} \otimes \mathbb{I}_{[n]} \right)\nonumber\\& - \sum_{k_l,k_r=1}^{n-1} V_L(\pi_{k_l})\cdot \Phi^\dagger \cdot O_{\text{cen}}'(k_l,k_r) \cdot \Phi \cdot V_L(\pi_{k_r}) }
\leqslant  (n-1)^2 \left( d\varepsilon_2'+2\sqrt{d}d\varepsilon_3 \right) + \text{poly}_4\left[\epsilon_5(n-1)^{3/2}\right].
\end{align}

The desired result is finally obtained by adding up the three terms.

\item 
\begin{align}
&\norm{\frac{1}{\alpha\sqrt{n-1}} W - \tilde{\Pi} V \Pi}\\
\leqslant &\frac{1}{\sqrt{n-1}} \norm{ \sum_{i=1}^{n-1} \ket{i}\bra{0}_{I} \otimes \left[ \frac{1}{\alpha}\sqrt{\Pi_i} - \left(\bra{0}^{\otimes a} \otimes \mathbb{I}_{IAAn} \right) U^c(i) \left( \ket{0}^{\otimes a} \otimes \mathbb{I}_{AAn} \right) \right] } \nonumber\\
& + \norm {U_0 \ket{0}\bra{0}_{I} - \frac{1}{\sqrt{n-1}} \sum_{i=1}^{n-1} \ket{i}\bra{0}_{I}} \\
\leqslant & \frac{n-1}{\alpha\sqrt{n-1}} \norm{\sqrt{\Pi_1}}\delta + (n-1)^{3/2}\epsilon_0 
\leqslant  \frac{\sqrt{n-1}}{\alpha} \sqrt{d}\delta + (n-1)^{3/2}\epsilon_0
\end{align}
Since $\sin(\pi/2m)=1/(\alpha\sqrt{n-1})$ and $\alpha = O(n^{2} d^3)$ according to Eq. \eqref{eq:alpha_a}, we have $\pi/2m \ll 1$ and in turn $2m=\alpha\sqrt{n-1}\pi$. The result directly follows. 
\end{enumerate}